\documentclass[12pt, letterpaper]{extarticle}

\usepackage{subcaption}
\usepackage{fullpage}
\usepackage[switch]{lineno}
\usepackage{amsmath}
\usepackage{amssymb}
\usepackage{rotating}
\usepackage{array}
\usepackage{mathtools}
\usepackage[ruled]{algorithm2e}
\usepackage{algorithmic}
\usepackage{bm}
\usepackage{breqn}
\usepackage{comment}
\usepackage{enumitem}
\usepackage{graphics}
\usepackage{graphicx}
\usepackage{latexsym}
\usepackage{mathrsfs}
\usepackage{morefloats}
\usepackage{nicefrac}
\usepackage{authblk}
\usepackage{pifont}
\usepackage{tabularx}
\usepackage{array}
\usepackage{ltablex, booktabs}
\usepackage{ragged2e}
\usepackage{makecell}
\usepackage{multirow}

\usepackage{colortbl}
\usepackage{color}
\usepackage{xcolor}

\newcolumntype{L}{>{\raggedright\arraybackslash}X}

\definecolor{blue}{HTML}{88CCEE}
\definecolor{red}{HTML}{CC6677}

\newif\ifcomment
\commenttrue

\ifcomment
    \newcounter{HINumberOfComments}
    \stepcounter{HINumberOfComments}

    \newcounter{DJNumberOfComments}
    \stepcounter{DJNumberOfComments}
    \newcommand{\djnote}[1]{\textcolor{red}{\small \bf [DJ\#\arabic{DJNumberOfComments}\stepcounter{DJNumberOfComments}: #1]}}
    
    \newcounter{YZNumberOfComments}
    \stepcounter{YZNumberOfComments}

    \newcommand{\NOTE}[1]
    {
      {\footnotesize\it
        \begin{center}
          \begin{tabular}{|c|}
           \hline
            \parbox{0.85\columnwidth}{
              \medskip
              #1
              \medskip} \\
            \hline
          \end{tabular}
        \end{center}
        }
    }
\else
    \newcommand\djnote[1]{}
    \newcommand\NOTE[1]{}
\fi

\usepackage[hyphens]{url}
\usepackage{hyperref}
\hypersetup{colorlinks=false,breaklinks=true}

\title{TikTok's recommendations skewed towards Republican content during the 2024 U.S. presidential race}

\author[1+]{Hazem Ibrahim}
\author[1+]{HyunSeok Daniel Jang}
\author[1]{Nouar Aldahoul}
\author[2]{\\ Aaron R. Kaufman}
\author[1*]{Talal Rahwan}
\author[1*]{Yasir Zaki}

\affil[1]{\normalsize Computer Science, New York University Abu Dhabi, UAE.}
\affil[2]{\normalsize Political Science, New York University Abu Dhabi, UAE.}
\affil[+]{\footnotesize Joint first author}
\affil[*]{\footnotesize Corresponding authors. E-mail: \{talal.rahwan,yasir.zaki\}@nyu.edu}
\date{}

\renewcommand{\bf}{}
\linenumbers

\begin{document} 

\maketitle 

\begin{abstract}
\noindent 
TikTok is a major force among social media platforms with over a billion monthly active users worldwide and 170 million in the United States. The platform's status as a key news source, particularly among younger demographics, raises concerns about its potential influence on politics in the U.S. and globally. Despite these concerns, there is scant research investigating TikTok's recommendation algorithm for political biases. We fill this gap by conducting 323 independent algorithmic audit experiments testing partisan content recommendations in the lead-up to the 2024 U.S. presidential elections. Specifically, we create hundreds of ``sock puppet'' TikTok accounts in Texas, New York, and Georgia, seeding them with varying partisan content and collecting algorithmic content recommendations for each of them. Collectively, these accounts viewed $\sim$394,000 videos from April 30th to November 11th, 2024, which we label for political and partisan content. Our analysis reveals significant asymmetries in content distribution: Republican-seeded accounts received $\sim$11.8\% more party-aligned recommendations compared to their Democratic-seeded counterparts, and Democratic-seeded accounts were exposed to $\sim$7.5\% more opposite-party recommendations on average. These asymmetries exist across all three states and persist when accounting for video- and channel-level engagement metrics such as likes, views, shares, comments, and followers, and are driven primarily by negative partisanship content. Our findings provide insights into the inner workings of TikTok's recommendation algorithm during a critical election period, raising fundamental questions about platform neutrality.
\end{abstract}

\section*{Introduction} 
TikTok, a social media platform owned by Chinese company ByteDance, has rapidly grown to more than a billion monthly active users worldwide~\cite{titktok_mau}, establishing itself as a major player in the social media space, particularly among younger demographics. In the United States alone, TikTok supports over 170 million monthly active users~\cite{tiktok_us_users}, with one-third of U.S. adults—and a majority of those under 30—using the platform regularly~\cite{pewresearchFactsAbout}. Due to its scale, the platform has emerged as a significant source for news content, with 39\% of adults under 30 and 19\% of those aged 30-49 reporting that they regularly get news from the app~\cite{pewresearchMoreAmericans}. This shift in how Americans consume news has raised concerns about the platform’s potential to shape political narratives and influence the democratic process.

Despite TikTok’s policy prohibiting political advertising~\cite{tiktokPoliticsReligion}, the platform’s role in elections has been widely acknowledged and highly scrutinized in the media. For example, TikTok may have played a pivotal role in the 2024 Romanian presidential election: researchers identified thousands of bot-like accounts engaging in coordinated activity to amplify the campaign of a surprise Pro-Russia candidate, Călin Georgescu~\cite{politicoTikTokSummoned}. In response, Romania’s Prime Minister called for an investigation into the candidate’s TikTok campaign funding and a national regulator suggested suspending TikTok for election interference~\cite{theguardianRomaniaRegulator}. The European Parliament has since summoned TikTok’s CEO to address these allegations, with some lawmakers questioning whether the platform violated the EU’s Digital Services Act~\cite{europaEUsDigital} by failing to curb the spread of disinformation and inauthentic behavior~\cite{politicoTikTokSummoned}. In the U.S., concerns about TikTok’s influence are compounded by national security owing to its Chinese ownership. A 2020 executive order by Donald Trump sought to ban the app unless it was sold to a U.S.\ company~\cite{federalregisterFederalRegister}, though the order was ultimately blocked by a federal judge~\cite{npr_judge} and Trump later reversed his position during to the 2024 U.S. election, advocating against the banning of TikTok~\cite{washingtonPost_halt}. Nevertheless, bipartisan support for a potential TikTok ban has resurfaced, with a Senate-approved bill requiring the app to either divest from its Chinese parent company or face a ban in the United States~\cite{nytimesCongressPassed}. The bill, which passed with a vote of 360-58 in April 2024, exists in a broader geopolitical conflict between the U.S.\ and China. Some arguments supporting the ban emphasize TikTok as a path for Chinese political influence on the U.S., especially considering the Chinese government's heavy hand in shaping corporate strategy within Chinese-owned companies~\cite{cnnJackLoses, reutersGaming}. Moreover, TikTok has a history of politically-motivated censorship: TikTok's internal moderation guidelines, covered by The Guardian in 2019~\cite{theguardianRevealedTikTok}, suggest that the company censored Chinese politically-sensitive topics such as Tiananmen Square or Tibetan independence, and TikTok publicly apologized after a glitch that censored hashtags relating to the Black Lives Matter movement in the wake of George Floyd's death in 2020~\cite{businessinsiderTikTokApologized}. As of January 20th, 2025, after a brief interruption to TikTok's services, the platform resumed activity in the United States~\cite{nbcnews_trump_tiktok}.

Yet, despite its growing influence and the ongoing discussion surrounding its ban, TikTok remains relatively unexplored in the literature on social media recommendation algorithms. There is extensive prior work auditing social media recommendation algorithms and user behavior on platforms such as Facebook~\cite{guess2023reshares, gonzalez2023asymmetric, nyhan2023like, bode2012facebooking, batorski2018three, bakshy2015exposure, quattrociocchi2016echo}, Instagram~\cite{allcott2024effects, lalancette2019power, trevisan2019towards, parmelee2019insta}, YouTube~\cite{ibrahim2023youtube, hosseinmardi2020evaluating, hosseinmardi2021examining, ledwich2019algorithmic, lambrecht2021algorithmic}, X (formerly Twitter)~\cite{balasubramanian2024public, colleoni2014echo, conover2011political, barbera2015tweeting, aragon2013communication, ibrahim2024analyzing}, and Reddit~\cite{cinelli2021echo, morini2021toward, de2021no}, but little on TikTok, potentially due to its relative novelty as a social media platform and tight restrictions on its official API. Nevertheless, some important studies have examined political content on TikTok through different lenses, such as investigating political user demographics~\cite{medina2020dancing}, the characteristics of content that political parties publish on the platform~\cite{cervi2021political, zamora2023securing, moir2023use, guinaudeau2022fifteen}, polarization and toxicity in TikTok political discourse~\cite{vasconcellos2023analyzing}, and the intersection of humor and political engagement~\cite{zeng2023okboomer, brown2024affective}. TikTok specifically is uniquely positioned as a platform of interest due to the manner in which users engage with the service. On YouTube, for instance, while users are offered recommendations (which is impacted by both the user's own preferences, as well as co-viewership networks), the choice of the next video a user watches largely lies with the user (excluding YouTube shorts, or allowing videos to auto-play), which we term as ``direct'' agency. On TikTok, while users can potentially alter their feed by engaging with videos in various ways (e.g. liking, commenting on, or sharing a video), users ultimately only have ``indirect'' agency with regards to what video they watch next on their feed. As such, TikTok's recommendation algorithm plays a far more central role in what users consume relative to other social media services.

Taken together, we believe the literature leaves two important gaps. We know little about the number and partisan distribution of political content creators on TikTok, or their engagement and audiences; we know even less about TikTok's recommendation algorithm. We begin to fill these gaps here.

We conduct a series of systematic audit experiments on TikTok to examine partisan content recommendations in the lead-up to the 2024 U.S. presidential elections. Using controlled sock puppet accounts across three states with varying political demographics (New York, Texas, and Georgia), we analyze over 340,000 videos encountered through 381 experimental runs to understand how TikTok's recommendation algorithm distributes political content based on users' watch history and geographic location. Through a human-validated classification pipeline utilizing an ensemble of large language models, we identify and categorize political videos to track recommendation patterns over a 27-week period from April 30th to November 11th, 2024. Note that our original research design called for collecting data until after the January 2025 inauguration and the U.S. government's regulatory actions against TikTok; however, after November 11th, TikTok increased its bot-detection efforts preventing us from further data collection.

To explore the possibility of political bias in TikTok's recommendation algorithm, we operationalize bias as the difference in partisan recommendation rates for accounts seeded with partisan content for both major U.S.\ political parties. Our findings reveal a significant Republican bias in TikTok's recommendation algorithm. Republican-conditioned accounts received approximately 11.5\% more party-aligned content compared to Democratic-conditioned accounts across all states; Democratic-conditioned accounts are exposed to 8.0\% more opposite-party content than their Republican counterparts on average, particularly content from the main political candidates of each party. These asymmetries cannot be explained by differences in video- or channel-level engagement metrics such as likes, views, shares, comments, or followers. These findings provide insights into the inner workings of TikTok's recommendation algorithm during a critical election period.

\section*{Experimental Setup}
We conduct a series of 323 audit experiments on the platform using sock puppets (i.e., bots), mimicking user activity. Each experiment employs a bot to watch a researcher-controlled sequence of videos seeded with political content, which we call the ``conditioning stage.'' This treatment conditioning teaches TikTok's recommendation algorithm our bot's political preferences. Following the conditioning stage, our bot watches videos on TikTok's recommendation page (the ``For You'' page), which we call the ``recommendation stage.'' In this stage, each bot repeats the process of watching 10 videos, followed by a one-hour pause in which the bot does not engage with TikTok (to circumvent TikTok's bot-detection algorithm). The content of these videos constitutes our outcome data: we infer and analyze the partisan, topical, and misinformation content of these videos, including partisan alignment between the conditioning videos and the recommended ones. In total, each experimental run was conducted over the duration of one week, with one day dedicated to the conditioning stage, and six days dedicated to the recommendation stage (see Supplementary Figure~1 for an illustration of a single experimental run).

Each bot is assigned two treatment attributes: (1) political leaning and (2) geography. A bot's political leaning may be ``Democratic,'' ``Republican,'' or ``Neutral'' (our control group). In the conditioning stage, Democratic bots watched up to 400 Democratic-aligned videos and Republican bots watched up to 400 Republican-aligned videos; Neutral bots bypassed the conditioning stage entirely and moved directly to the recommendation stage to mimic users who do not have a particular interest in politics. Note that our control group receives no conditioning rather than politically neutral conditioning, since identifying politically neutral content is a large and potentially fruitless task in itself~\cite{ackerman1983neutral,mccarthy2014neutral}.

A bot's geography is the location where the bot is virtually present. Each bot is connected to a VPN server in New York, Texas, or Georgia. These were selected due to their projection as a likely Democratic, Republican, or swing state in the 2024 U.S. elections, respectively. In addition to VPN tunneling, each bot used a GPS mocking application to artificially set its location to its given VPN state (for more details, see the \hyperref[methods:experiment_setup:pre_post_protocol]{Pre- and Post-Experiment Protocols} in Methods). 

Each week from April 30th, 2024 to November 4th, 2024, we created 21 new TikTok accounts and assigned them to one of seven experimental conditions. We assigned three bots each to be Democrats in New York, Georgia, and Texas, and three bots each to be Republicans in New York, Georgia, and Texas. We set three final bots to be politically neutral in Georgia, designed to capture how the recommendation algorithm treats apolitical users in a swing state, an electorally critical demographic.

We exclude a number of experimental runs for technical reasons: some bots were shut down by TikTok's bot detection algorithm, while others failed due to internet outages. To account for this and ensure that we compare bots with the same amount of conditioning and the same number of recommended videos, we conduct a pair-matching procedure to match bots of opposite conditioning in each state during a given week. For each pair of bots, we only consider the first \textit{N} recommendations, where \textit{N} is the lesser of the total videos watched by either bot. We also exclude pairs of bots that watched fewer than 150 videos each during the recommendation stage to allow for a sufficient number of political videos to be recommended, bearing in mind that not all recommendations are political in nature. Of the 567 experimental runs conducted, 323 met our inclusion criteria. Supplementary Table~1 lists the number of successful experimental runs per condition over time.

\begin{figure}
    \centering
    \includegraphics[width=\linewidth]{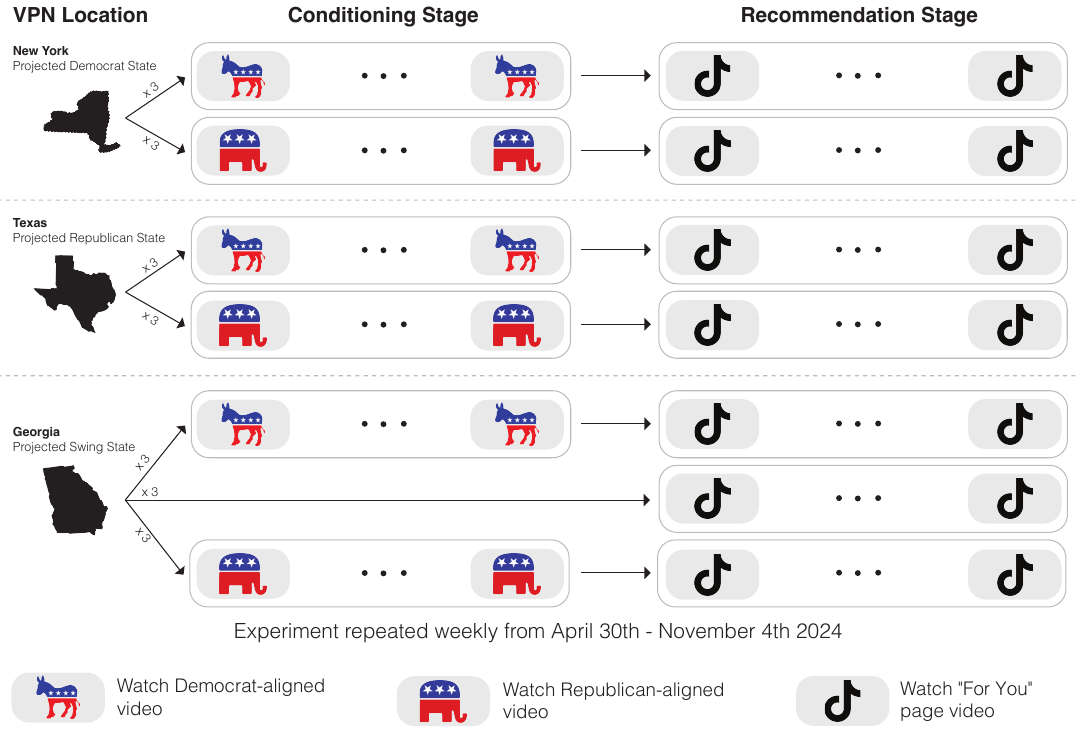}
    \caption{\fontsize{10}{10}\selectfont{\textbf{Experimental Setup.} 21 experiment runs are performed each week between April 30th and November 4th 2024. Bots are assigned an experimental condition based on two attributes, namely, their VPN and GPS location (New York, Texas, or Georgia), as well as the political leaning (Democratic or Republican) of the videos they watch during the conditioning stage. Each bot initially passes through a conditioning stage where it views up to 400 videos of a given political leaning, and then transitions to the recommendation stage, where it views up to 1200 videos on their TikTok ``For You'' page.}}
    \label{fig:experimental_setup}
\end{figure}

Through 323 experiment runs over 27 weeks, our bots watched 114,522 total (3,181 unique) videos in the conditioning stage and an additional 284,093 (176,252 unique) videos in the recommendation stage. Figure~\ref{fig:experimental_setup} illustrates our experimental design and data collection procedure; more details are available in the \hyperref[methods:experiment_setup]{Experimental Setup} section in Methods. The study was preregistered shortly before the end of the data collection period, and before the data was analyzed (asPredicted link: \url{https://aspredicted.org/mjj2-vxrb.pdf}).

\subsection*{Measuring Political Content}
Having collected 176,252 unique recommended TikTok videos, we next detail our measurement process for capturing the political content recommended by TikTok's algorithm.

We first download the English transcripts of each video from the subtitle URLs provided in their metadata, which were available in 40,264 videos, constituting 22.8\% of the unique videos and 26.2\% of all videos recommended to our bots. To categorize a given video's partisan content, we use a human-validated pipeline utilizing three large language models (LLMs)---GPT-4o~\cite{openai_gpt4o}, Gemini-Pro~\cite{deepmind_gemini_pro}, and GPT-4~\cite{openai_gpt4}---to answer the following questions about a given video: (Q1) Is the video political in nature?, (Q2) Is the video concerned with the 2024 U.S. elections or major U.S. political figures?, and (Q3) What is the ideological stance of the video in question (Pro Democratic, Anti Democratic, Pro Republican, Anti Republican, or Neutral)? For each video, we prompt each LLM to answer Q1, and if the answer is Yes, we ask Q2 and Q3. For each question, our outcome measure is the majority vote of the three LLMs' answers. A majority is guaranteed for Q1 and Q2 as they have binary outcomes. As for Q3, despite having five outcome categories, 89.4\% of videos reached a majority label; our analysis focuses on videos with an LLM majority vote. In the \hyperref[methods:data_representativeness]{Data Representativeness} in Methods, we show that the partisan distributions of videos with transcripts does not differ significantly from a large sample of videos without transcripts, indicating that the videos analyzed in this study are representative of the entire set of recommendations made to the bots. 

The above classification pipeline improves upon prior work characterizing video-level political content on TikTok. Tjaden et al., who study TikTok political content in the context of German elections~\cite{tjadenautomated}, associate a video with a party if it includes a party-affiliated hashtag in its description. This approach is not applicable in the U.S., where videos which include hashtags containing ``trump'' are often published by Democratic-aligned creators. Notably, ``\#donaldtrump'' and ``\#trump'' are among the six most common hashtags used by Democratic-aligned channels (Supplementary Table~9). We provide more details on our measurement, validation, and prompting in the \hyperref[methods:stance_classification]{Ideological Stance Classification} section in Methods, and we present descriptive statistics about our complete dataset in Supplementary Table~2.

\section*{Partisan Presence on TikTok}
What is the supply and partisan distribution of political content creators on TikTok? To categorize channels as Democratic-aligned or Republican-aligned, we calculate the proportion of each creator's videos labeled as Pro-Democratic or Anti-Republican versus Pro-Republican or Anti-Democratic, supplementing our dataset with up to 30 additional pre-election videos from the TikAPI~\cite{tikapi} for channels with fewer than 10 labeled videos in our sample. We label a channel as Democratic-aligned if at least 75\% of its videos are either Pro-Democratic or Anti-Republican, and Republican-aligned if at least 75\% of its videos are Pro-Republican or Anti-Democrat. This process yielded 56 Democratic-aligned channels and 75 Republican-aligned channels, which we manually validated following best practices on channel-level classification tasks ~\cite{hosseinmardi2020evaluating, ledwich2019algorithmic}. Table~\ref{tab:channel_classification} in the Methods section summarizes the average proportion of party-aligned videos across these channels. 

Table~\ref{tab:top_20_accounts_combined} details the 20 largest channels classified as either Democratic-aligned or Republican-aligned ranked by follower count, including cumulative engagement metrics such as the total likes received and the number of videos published. The left columns show that the top 20 Democratic channels in our sample are dominated by known political figures (e.g., Kamala Harris, Tim Walz, Bernie Sanders, and Alexandria Ocasio-Cortez), talk shows (e.g., The View, Jimmy Kimmel Live, and The Colbert Late Show), and news-media outlets (e.g., New York Times, MSNBC, and Courier News Room). In contrast, the top Republican-aligned channels in the right  columns of Table~\ref{tab:top_20_accounts_combined}
feature more independent ``influencer'' channels (e.g., Adam Calhoun, Charlie Kirk, Ben Shapiro, Brandon Tatum, and Patrick Bet-David). The bottom of Table~\ref{tab:top_20_accounts_combined} summarizes engagement metrics for the top 50 accounts in both partisan categories. Republican channels saw higher median engagement across followers and likes per channel compared to their Democratic-aligned counterparts, despite fewer videos published on the platform.

\begin{table}[p]

{\fontsize{8}{8}\selectfont{
\begin{tabularx}{\columnwidth}{|Lccc|Lccc|}
\hline
{\cellcolor{blue!20}\rotatebox{0}{\textbf{\makecell[c]{Account name (ID) }}}} &  {\cellcolor{blue!20}\rotatebox{0}{\textbf{\makecell[l]{Number \\ of followers}}}} & 
{\cellcolor{blue!20}\rotatebox{0}{\textbf{\makecell[l]{Number \\ of likes}}}} & 
{\cellcolor{blue!20}\rotatebox{0}{\textbf{\makecell[l]{Number \\ of videos}}}} & 
{\cellcolor{red!20}\rotatebox{0}{\textbf{\makecell[c]{Account name (ID) }}}} &  
{\cellcolor{red!20}\rotatebox{0}{\textbf{\makecell[l]{Number \\ of followers}}}} & 
{\cellcolor{red!20}\rotatebox{0}{\textbf{\makecell[l]{Number \\ of likes}}}} &  
{\cellcolor{red!20}\rotatebox{0}{\textbf{\makecell[l]{Number \\ of videos}}}} \\
\endhead 
\endfoot
\hline
           Kamala Harris (\texttt{kamalaharris}) &          9.3M &     149.5M &        219 & President Donald J Trump (\texttt{realdonaldtrump}) &         14.7M &     107.2M &         58  \\\hline
                        TizzyEnt (\texttt{tizzyent}) &          6.7M &     270.9M &       2.6K & Team Trump (\texttt{teamtrump}) &          8.3M &     204.3M &        447 \\\hline
                       Kamala HQ (\texttt{kamalahq}) &          5.7M &     308.9M &       1.1K & The Charlie Kirk Show (\texttt{thecharliekirkshow}) &          5.3M &     134.7M &        638  \\\hline
                              MSNBC (\texttt{msnbc}) &          4.1M &     265.9M &       4.3K & Date Right Stuff (\texttt{daterightstuff}) &          3.4M &      197.6M &        787 \\\hline
             NowThis Impact (\texttt{nowthisimpact}) &          4.0M &     257.7M &       2.9K & Candace Owens Show (\texttt{candaceoshow}) &          3.3M &     35.4M &        342 \\\hline
                      The View (\texttt{theviewabc}) &          3.7M &      37.7M &       1.8K & Robert F. Kennedy Jr (\texttt{robertfkennedyjrofficial}) &          3.2M &     63.7M &        886 \\\hline
Late Show With Stephen Colbert (\texttt{colbertlateshow}) &          2.2M &     113.3M &       1.4K & Adam Calhoun (\texttt{adamcalhoun1}) &          3M &     36.1M &       153 \\\hline
               Jeff Jackson (\texttt{jeffjacksonnc}) &          2.2M &      38.7M &        121 & Ben Shapiro (\texttt{real.benshapiro}) &          2.7M &     59.9M &        1056  \\\hline
                         Tim Walz (\texttt{timwalz}) &          2.1M &      22.0M &        104 & J.D. Vance (\texttt{jd}) &          2.3M &     12.3M &        39 \\\hline
        Jimmy Kimmel Live (\texttt{jimmykimmellive}) &          2.0M &      76.2M &        587 & Piers Morgan Uncensored (\texttt{piersmorganuncensored}) &          2.2M &     32M &         630 \\\hline
                    Bernie Sanders (\texttt{bernie}) &          1.6M &     14.4M &        331 & Tucker Carlson (\texttt{tuckercarlson}) &          2.2M &     26.2M &        224 \\\hline
                Harry Sisson (\texttt{harryjsisson}) &          1.5M &     136.2M &       3.7K & Fox News (\texttt{foxnews}) &          1.9M &     55.5M &        1138  \\\hline
               The Democrats (\texttt{thedemocrats}) &          1.4M &      56.7M &       1.3K & The Comments Section (\texttt{thecommentssectiondw}) &          1.9M &     65.6M &        999 \\\hline
         Courier Newsroom (\texttt{couriernewsroom}) &          1.3M &     156.4M &       3.8K & Jeff Mead (\texttt{the\_jefferymead}) &          1.9M &     45.7M &        1.5K \\\hline
                Aaron Parnas (\texttt{aaronparnas1}) &          1.3M &      70.1M &       2.7K & Donald Trump Jr. (\texttt{donaldjtrumpjr}) &          1.7M &     14.4M &        166 \\\hline
               The New York Times (\texttt{nytimes}) &          1.3M &      31.6M &       1.5K & Charlie Kirk (\texttt{charliekirkdebateclips}) &          1.6M &     17.5M &        147 \\\hline
                  MeidasTouch (\texttt{meidastouch}) &          1.2M &      91.6M &       3.3K & Make America Great Again (\texttt{maga}) &          1.4M &     44.7M &        408 \\\hline
   Alexandria Ocasio-Cortez (\texttt{aocinthehouse}) &          1M &      7.5M &         74 & The Officer Tatum (\texttt{theofficertatum}) &          1.4M &     41.6M &       671 \\\hline
                     Robert Reich (\texttt{rbreich}) &        938.8K &     17.8M &       1.0K & Jesse Watters (\texttt{jessebwatters}) &          1.3M &     36.2M &       1.3K \\\hline
Late Night with Seth Meyers (\texttt{latenightseth}) &        887.9K &      72.8M &       1.3K & Patrick Bet-David (\texttt{patrickbetdavid}) &          1.3M &     27.5M &       3.4K \\
\hline  
& & \multicolumn{3}{c}{\footnotesize\textbf{\makecell[c]{Top 50 Accounts}}} & & & \\ \hline
{\cellcolor{blue!20}Sum} & {\cellcolor{blue!20}66.8M} & {\cellcolor{blue!20}2.6B} & {\cellcolor{blue!20}83703} & {\cellcolor{red!20}Sum} & {\cellcolor{red!20}87.3M} & {\cellcolor{red!20}1.78B} & {\cellcolor{red!20}48642} \\ 
{\cellcolor{blue!20}Median} & {\cellcolor{blue!20}709K} & {\cellcolor{blue!20}17.9M} & {\cellcolor{blue!20}1038}  &  {\cellcolor{red!20}Median} &  {\cellcolor{red!20}1.2M} & {\cellcolor{red!20}18.5M} & {\cellcolor{red!20}794} \\
{\cellcolor{blue!20}IQR} & {\cellcolor{blue!20}1M} & {\cellcolor{blue!20}61.5M} & {\cellcolor{blue!20}2180.5} & {\cellcolor{red!20}IQR} & {\cellcolor{red!20}1.3M} & {\cellcolor{red!20}33.9M} & {\cellcolor{red!20}739} \\ 
\hline
    \caption{Top 20 Democrat-aligned channels (left) and Republican-aligned channels (right) by follower count, and the cumulative engagement metrics for the 50 Democrat and Republican-aligned channels with the largest number of followers (bottom).}
    \label{tab:top_20_accounts_combined}
\end{tabularx}
}}

\end{table}

\newpage

\begin{figure}
    \centering

    \includegraphics[width=\linewidth]{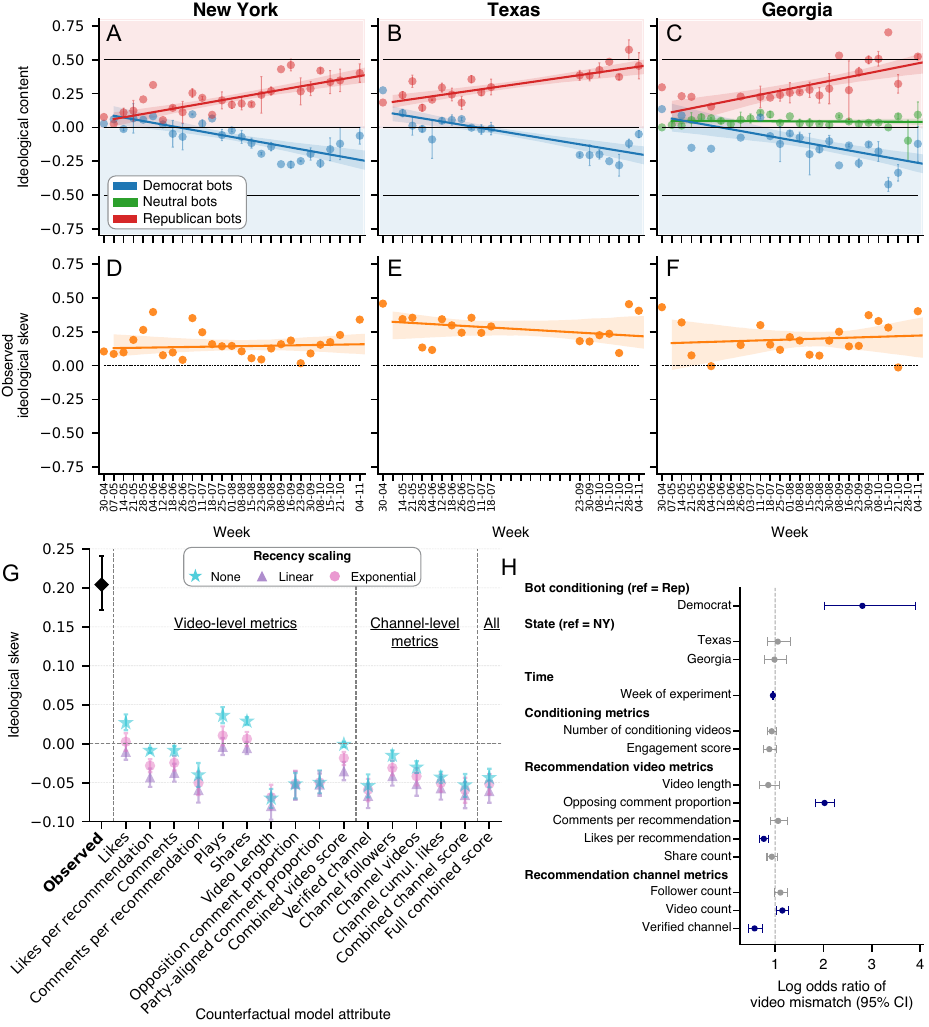}
    \caption{\fontsize{10}{10}\selectfont{\textbf{Differences in recommendation rates over time.} (\textbf{A}-\textbf{C}) The mean ideological alignment for bots of a given conditioning-leaning in New York, Texas and Georgia, measured as the difference between the proportion of recommended videos which are Republican aligned minus the proportion of Democratic aligned videos. OLS regression lines are plotted with 95\% confidence intervals. (\textbf{D}-\textbf{F}) The observed ideological skew for each state over time; positive values indicate a Republican skew. (\textbf{G}) The observed ideological skew, and the expected ideological skew based on counterfactual models built on different engagement metrics. (\textbf{H}) Log odds ratios computed through logistic regression on the likelihood of a video being a mismatch/cross-partisan. Statistically significant coefficients are highlighted in navy.}}
    \label{fig:recommendation_rates}
\end{figure}

\section*{Political recommendation rates}
Does TikTok's recommendation algorithm favor one party over the other? Arguably, one could expect the algorithm's behavior to fall into one of the following broad categories: stasis, symmetric polarization, asymmetric polarization, and partisan bias. A static recommendation algorithm would identify a user's partisan content mix, and serve them exactly that mix at every time point. A symmetrically polarizing algorithm would identify a user's preferred partisan content and serve them more of that content over time, gradually segmenting users into partisan echo chambers. An asymmetrically polarizing algorithm feeds users from one party more partisan content than the users from another party, creating stronger echo chambers for users of that partisanship. Finally, a partisan-biased algorithm shifts all users' partisan content mixes toward one party over time, regardless of their starting mix. Which of these does TikTok's algorithm most resemble? Recall that our experimental design conditions some bots to Democratic-aligned content, some to Republican-aligned content, and others to no partisan content at all; by comparing the partisan content of TikTok's recommended videos to the partisan content of the conditioning videos, we can begin to answer this question.

We measure the ``ideological content'' of a given bot's recommendations as the proportion of political recommended videos that are Republican-aligned (Pro Republican or Anti Democratic) minus the proportion of recommendations that are Democratic-aligned (Pro Democratic or Anti Republican). This allows us to score each bot's video recommendations from -1 (all videos are Democratic-aligned) to +1 (all videos are Republican-aligned). We plot the bots' recommendations across experimental conditions and over time in Figure~\ref{fig:recommendation_rates}A, B, C for the states of New York, Texas, and Georgia, respectively. Each point represents the mean ideological content of the recommendations received by the bots in a given week and experimental condition. Here, bars indicate 95\% confidence intervals, while color indicates the conditioned partisanship of the bots. Missing points represent weeks without any successful runs for bots located in a given state (see Supplementary Table~1 for the number of successful runs per condition over time). We also plot OLS regression lines of ideological content by experimental condition over time, showing a clear increase in polarization in all states: TikTok served partisans more co-partisan content over time. However, while both Democratic and Republican-conditioned bots received more ideologically aligned content over time, the magnitude is significantly greater for Republicans.

We confirm that the TikTok algorithm exhibits an ideological skew towards Republican-aligned content in Figure~\ref{fig:recommendation_rates}D, E, and F. In these figures, the Y axis shows the overall partisan content of all TikTok's recommended videos regardless of partisan alignment, where negative values indicate a Democratic bias and positive values indicate a Republican bias. On average, Republican bots saw approximately 11.8\% more co-partisan recommendations than their Democratic bot counterparts while simultaneously seeing approximately 7.5\% less cross-partisan recommendations. Supplementary Table~3 details the mean monthly ideological content of Republican and Democratic bots in each state, as well as Neutral bots in Georgia. Supplementary Table~4 details results from independent t-tests comparing the partisan content of bots of each type of conditioning, revealing that Republican bots received significantly more co-partisan content across all three states (NY: $t = 5.82$, $p < 0.001$; Texas: $t = 9.21$, $p < 0.001$; Georgia: $t = 5.83$, $p < 0.001$), and both Republican and Democratic bots received more co-partisan content than Neutral bots (Rep vs. Neutral: $t = 13.22$, $p < 0.001$; Dem vs. Neutral: $t = 4.70$, $p < 0.001$). Supplementary Figures 2-5 illustrate rolling averages of the amount of political content, election-related content, ideological matches, and ideological mismatches viewed by different bots during a given experimental run. As shown in these figures, for both Democratic- and Republican-conditioned bots, the conditioning stage results in an immediate spike in political content recommended to bots, decaying as bots are recommended more videos but never approaching the control group.

\subsection*{Robustness Checks}
Figure~\ref{fig:recommendation_rates} establishes that TikTok shows more Republican-aligned content than Democratic-aligned content, and that Democrats receive more Republican-aligned content than Republicans receive Democratic-aligned content. However, if there are more Republican users or content creators on TikTok, if Republican-aligned content is more engaging, or if Democrats are more \textit{interested in} Republican-aligned content than Republicans are in Democratic-aligned videos, then this ideological skew is exactly what we might expect TikTok's algorithm to produce. In this section we show that these mechanisms cannot fully explain the observed difference in ideological content. First, we use a weighted random sampling approach to show that observed engagement metrics like channel size, video likes, and commenting, or linear combinations of the above, would not produce the observed ideological skew. Second, we show that the number of Democratic-aligned comments on a Republican-aligned video does not predict whether that video is shown more to Democrats. Neither of these two explanations, however, preclude the possibility that TikTok has developed a latent engagement metric that \textit{would} explain the observed ideological gap; our third robustness test develops a sensitivity analysis to show that for any hypothetical hidden engagement metric, the difference in that metric between Republican and Democratic videos would have to be 98 times bigger than the gap in likes to explain the ideological content gap we observe on TikTok in our experiments.

\subsubsection*{Sampling by Engagement Metrics}
To verify that our results are not due to differences in the engagement metrics of Republican and Democratic videos or channels, we consider counterfactual scenarios where video recommendations are functions of these engagement metrics. We showed above that the TikTok algorithm recommends more Republican-aligned content than Democratic-aligned content, but this may not be surprising if there is more Republican content overall on TikTok, or if that content is more popular. In other words, Figure~\ref{fig:recommendation_rates} assumes a counterfactual where Republican-aligned and Democratic-aligned videos appear in equal proportion; we consider alternative counterfactuals below. Our first robustness test answers the question: how big of an ideological skew in content should we expect under different counterfactual scenarios, and how does the observed skew compare?

To confirm that the skew towards Republican-aligned content exists even after accounting for potential differences in video engagement metrics that might produce baseline differences, we take a weighted-random sample of \textit{N} videos (\textit{N} being the number of videos watched by pairs of bots in a given week and experimental condition) with weights proportional to that video's \textit{engagement}, and calculate the proportion of Republican- and Democratic-aligned videos in that sample. We then compare these proportions to the observed proportion of recommended Republican and Democratic-aligned videos, and show that our bots received more such Republican-aligned videos than we would expect if recommendations were only a function of video engagement. To this end, we consider 48 scenarios where recommendation rates are a function of TikTok engagement metrics: video-level metrics such as a video's likes, likes per recommendation, comments, comments per recommendation, plays, and share count, as well as channel-level metrics such as verification status, follower count, and cumulative like and video counts. We then aggregate the above metrics using principal component analysis (PCA) and extract the video loadings to produce a combined video engagement score, a combined channel engagement score, and a combined overall engagement score. For each of the 16 models described (nine video-level models, four channel-level models, and three combined models), we also compute two alternative versions, penalizing old videos linearly and exponentially, to up-weight more recent content.

We show the results from this counterfactual analysis in Figure~\ref{fig:recommendation_rates}G. The leftmost point represents the observed ideological skew (the proportion of Republican-aligned videos minus that of Democratic-aligned ones recommended to the bots).  Each of the remaining points presents the expected ideological skew as computed by the 48 counterfactuals, shown along the x-axis; plot point markers represent the alternative methods for up-weighting more recent content. If the observed engagement metrics fully explain the partisan gap, we would expect the simulated gaps when we sample proportionate to those metrics to be close to the observed gap of approximately 0.2. We do not observe that. Instead, across all 48 counterfactuals, the observed bias toward Republican-aligned content substantially exceeded the counterfactual proportion based on engagement, with the majority of engagement metrics yielding a Democratic bias rather than a Republican one. Supplementary Table~5 lists the expected ideological skews as computed by each model, as well as independent t-tests comparing these values to that observed by the bots; see the ~\hyperref[methods:counterfactual]{Video~Recommendation~Counterfactual~Models} section in Methods for further details. 

\subsubsection*{Asymmetric Homophily}
In addition to these public engagement metrics, we also consider bias which may stem from co-viewership networks~\cite{hosseinmardi2024causally}. Republican-aligned videos may garner more engagement from Democrats than vice-versa, which would explain why Republican-aligned videos are recommended to Democrats more frequently; we refer to this possibility as \textit{asymmetric homophily}. A key implication of asymmetric homophily is that among Republican-aligned videos, those with more Democratic-aligned comments will be shown to Democrats more than those with fewer Democratic-aligned comments. We show that is not the case, suggesting that asymmetric homophily cannot explain the ideological skew we observe.

To test this, we collect a random sample of 200 comments from all political videos examined in our study and classify the content of each comment in the context of that video. From this we compute the proportion of comments ideologically aligned with the video (see the \hyperref[methods:stance_classification:comment_classification]{Comment~Classification} section in Methods).

We find meaningful differences between Republican-aligned and Democratic-aligned videos in the composition of comments, but not in the expected direction. We find that Democratic-aligned videos have a somewhat higher proportion of Republican-aligned comments than vice versa (13.4\% versus 10.3\%, $\chi^2 = 0.299$, $p = 0.59$), as well as a higher proportion of copartisan comments than Republican-aligned videos (64.7\% versus 52.9\%, $\chi^2 = 0.544$, $p = 0.46$). However, Republican-aligned videos have a much higher proportion of ``neutral'' comments that are not explicitly partisan (45.6\% versus 27\%, $\chi^2 = 2.07$, $p = 0.15$). We also include this metric in our weighted sampling robustness check (Figure~\ref{fig:recommendation_rates}G), finding similar results to our other observed engagement metrics. These results suggest that, at least with regards to commenting behavior, Democrats are not engaging with Republican-aligned videos more than Republicans are engaging with Democratic-aligned videos: we find no asymmetric homophily that would explain the ideological skew. 

We formalize these analyses in a logistic regression framework (Figure~\ref{fig:recommendation_rates}H). We conduct a regression at the bot-video recommendation level where the outcome variable indicates whether a video is an ideological mismatch as a function of that bot's conditioned partisanship, controlling for that bot's state, the week in which the experiment was conducted, as well as the engagement metrics of the videos and channels watched during the recommendation stage, in addition to those watched in the conditioning stage. We isolate the independent variables with a Variance Inflation Factor (VIF) value less than five (Supplementary Table~11 lists the VIF scores for each of the independent variables). Considering that all conditioning video- and channel-level engagement metrics had a VIF score greater than five, we compute a combined engagement score for those videos using PCA. In Figure~\ref{fig:recommendation_rates}H, the top-most point shows the logistic regression coefficient on bot conditioning party. While opposition-comment proportion meaningfully predicts a video being mismatched, compared to Republican-conditioned bots, Democratic-conditioned bots remain 2.8 times as likely to receive ideologically mismatched videos even when controlling for opposition comment proportion. We show the full regression table for this model in Supplementary Table~13. Additionally, Supplementary Tables~16-19 include descriptive statistics of the engagement metrics of videos and channels seen by bots during the conditioning and recommendation stages.

\subsubsection*{Sensitivity Analysis}
While we have shown evidence that observable engagement metrics and viewership patterns cannot explain the ideological skew in Republican-aligned content on TikTok, we cannot rule out the possibility that there are unobserved or latent engagement metrics or viewership patterns, known to the TikTok algorithm but unknown to us, that would explain the gap we observe. Instead, we conduct a sensitivity analysis to calculate how strong such an unobserved factor would have to be to fully explain TikTok's apparent ideological skew.

Across our 14 observed metrics, there are four where Republican-aligned videos outperform Democratic-aligned videos: comments, likes, plays, and shares. If we consider only the optimal linear combination of those four in our weighted sampling approach, they explain 5.7\% of the observed ideological gap. Therefore, for any latent or unobserved metric to fully explain that gap, Republican-aligned videos would need to outperform Democratic-aligned videos by 9.4 times as strongly as they do on the aggregation of comments, likes, plays, and shares (see Table~\ref{tab:sensitivity_analysis} in Methods for the required scaling needed on the various engagement metrics analyzed). We argue that such a metric is unlikely to exist. We formalize the assumptions behind this in the \hyperref[methods:sensitivity]{Sensitivity Analysis} component of the Methods section, in which we experiment with sampling from a variety of distributions.

\begin{figure}
    \centering
    \includegraphics[width=\linewidth]{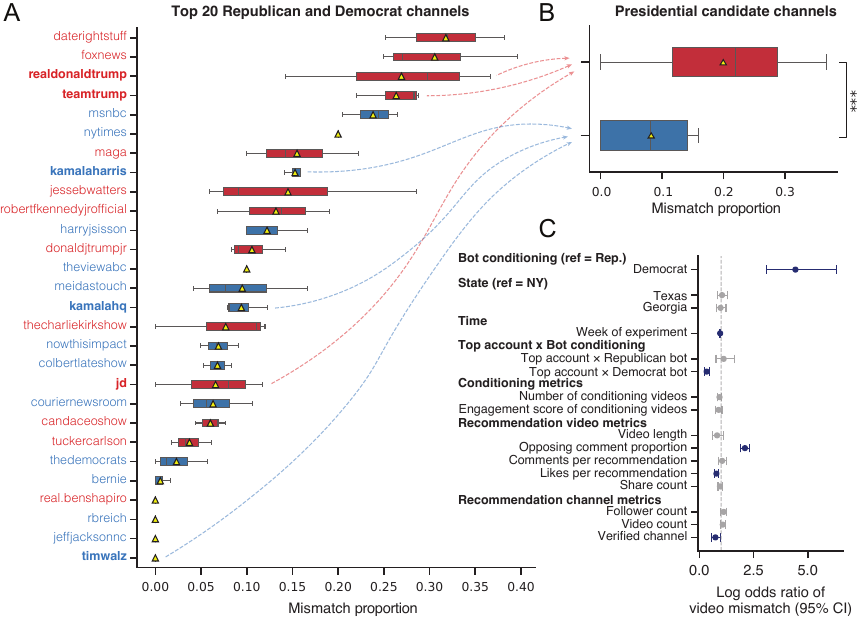}
    \caption{\fontsize{10}{10}\textbf{Mismatch proportions of top Democratic and Republican channels.} \selectfont{(\textbf{A}) The mismatch proportion, or the proportion of video watches by bots of an opposite conditioning-leaning for top Democratic and Republican TikTok channels by follower count. (\textbf{B}) The mean mismatch proportions of the TikTok accounts for the main political figures in the 2024 U.S. elections (Trump, JD Vance, Kamala Harris, and Tim Walz) (\textbf{C}) Logistic regression estimates on the log-odds ratio of recommended video mismatch for top Republican and Democratic channels (Statistically significant coefficients are highlighted in navy).}}
    \label{fig:channel-analysis}
\end{figure}

\subsection*{Top Democratic and Republican channels}
Having examined ideological mismatches at the video-level, we next examine such mismatches at the channel-level. We compute a channel's mismatch proportion as the proportion of that channel's videos shown to bots conditioned with the opposite partisanship. Here, we focus on the top Democratic and Republican channels by follower count who were watched at least 10 times by bots in each state. The mismatch proportions for these channels are illustrated in Figure~\ref{fig:channel-analysis}A. Videos published by Donald Trump's official TikTok channel (\texttt{realdonaldtrump}) had an average mismatch proportion of 0.269, meaning that nearly 27\% of the time his videos were recommended to our bots, they were recommended to Democratic-conditioned bots. In contrast, Kamala Harris's average mismatch proportion was only 0.153, despite being the sitting Vice President during our experiment. 

Indeed, of the top Republican and Democratic channels, the highest four mismatch proportions were of Republican channels (\texttt{daterightstuff}, \texttt{foxnews}, \texttt{realdonaldtrump}, and \texttt{teamtrump}). In contrast, four channels were never shown to a bot of an opposite ideological alignment during our experiments (\texttt{real.benshapiro}, \texttt{rbreich}, \texttt{jeffjacksonnc}, and \texttt{timwalz}).
Overall, Republican channels had a significantly higher mismatch proportion than Democratic channels (Chi-squared test; $\chi^2 = 36.5$, $p < 0.001$). Figure~\ref{fig:channel-analysis}B shows that the official channels of the Republican candidates, Donald Trump and JD Vance, had a significantly higher mismatch proportion than those of Democratic candidates Kamala Harris and Tim Walz (Chi-squared test; $\chi^2 = 28.1$, $p < 0.001$). To test whether it is specifically mismatches within top accounts that contribute to the biases observed overall, we compute a logistic regression similar to that shown in Figure~\ref{fig:recommendation_rates}H, while including another binary channel-level metric designating whether a channel is amongst one of the top 20 channels by follower count. The results of this analysis are in Figure~\ref{fig:channel-analysis}C. Here, we again find that the partisan discrepancy observed is most strongly associated with the bot's partisan conditioning, as well as the opposing comment proportion of a given video. In contrast, we find that top channels were less likely to be recommended as a mismatch, despite the discrepancies seen in Figure~\ref{fig:channel-analysis}A and ~\ref{fig:channel-analysis}B, suggesting that less popular channels were even more likely to be ideologically mismatched. We again considered variables with a VIF value less than 5, as listed in Supplementary Table~12. See Supplementary Table~14 for the logistic regression table containing these results.

\begin{figure}
    \centering
    \includegraphics[width=\linewidth]{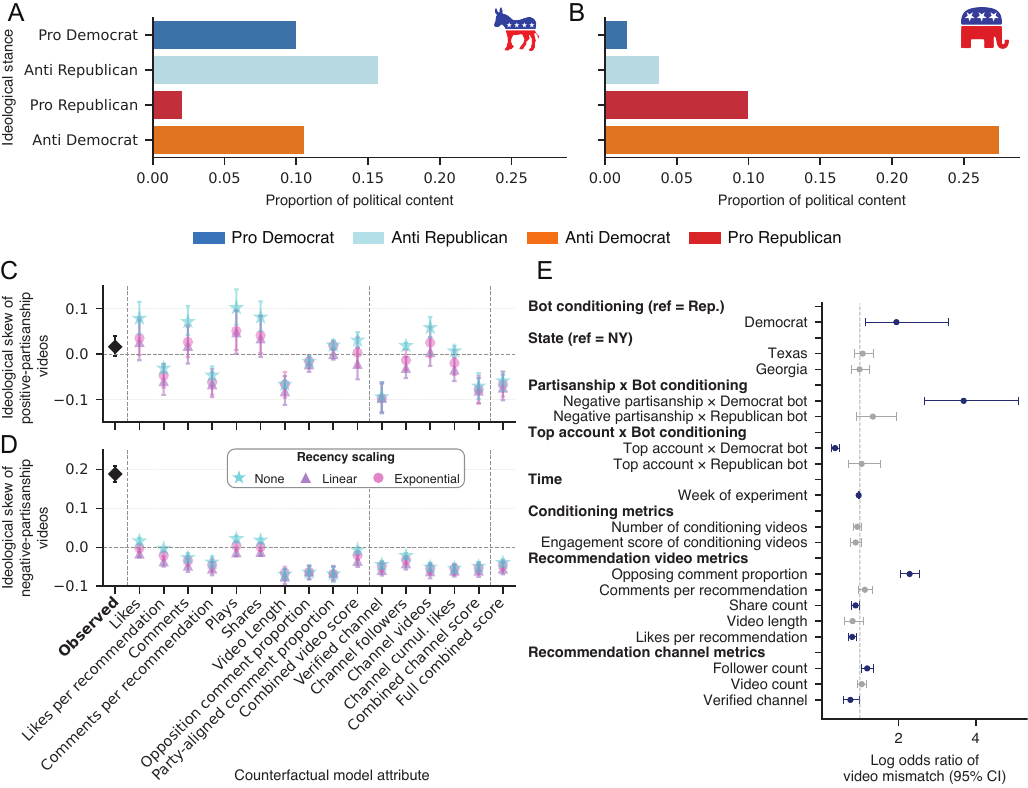}
    \caption{\fontsize{10}{10}\selectfont{\textbf{Ideological stance recommendation rates over time.} The mean proportion of videos of a given partisan (non-neutral) ideological stance recommended to Democratic bots (\textbf{A}) and Republican bots (\textbf{B}). (\textbf{C}, \textbf{D}) Observed ideological skew and the expected ideological skew based on counterfactual models when considering only Pro-party videos or Anti-opposition videos, respectively. (\textbf{E}) Log odds ratios computed through logistic regression on the likelihood of a video being a mismatch, accounting for the type of partisanship. Statistically significant coefficients are highlighted in navy.}}
    \label{fig:props}
\end{figure}

\subsection*{Negative Partisanship}
Recall that we previously grouped Pro-Democratic and Anti-Republican videos as Democratic-aligned and Pro-Republican and Anti-Democratic videos as Republican-aligned. Next, we unpack the observed partisan bias into positive partisanship, or favor toward one's own party, and negative partisanship, which is animosity toward the out-party. 

We plot the distribution of individual partisan (non-neutral) stances recommended to Democratic-conditioned bots and Republican-conditioned bots in Figures~\ref{fig:props}A and ~\ref{fig:props}B, respectively. As shown in Panel~A, Democratic-conditioned bots were recommended Anti-Republican videos (light blue) more frequently than Pro-Democratic videos (dark blue). Similarly, Panel~B shows that Republican-conditioned bots were recommended more Anti-Democratic videos (orange) than Pro Republican ones (red), indicating a general negative partisanship bias~\cite{abramowitz2018negative}. However, this bias was greater for Republican bots than Democratic ones (Republican bots: $\chi^2 = 507.3$, $p < 0.001$; Democratic bots:  $\chi^2 = 85.6$, $p < 0.001$).

Considering this gap, we recompute the ideological skew measured in Figure~\ref{fig:recommendation_rates}G, once while only considering positive-partisan videos (Pro Democratic and Pro Republican), and again while only considering negative-partisan videos. For each of these sets of videos, we also recompute the expected ideological skew based on the different engagement metrics these videos receive (Supplementary Tables~6 and 7 detail the ideological skews as measured by each of these models for positive-partisan and negative-partisan videos, respectively). The results of these analyses are presented in Figures~\ref{fig:props}C and ~\ref{fig:props}D for positive- and negative-partisan videos, respectively. Broadly speaking, the Republican skew observed in positive-partisan videos is markedly smaller than that in negative-partisan videos. Formalizing this analysis using a logistic regression (Figure~\ref{fig:props}E) with video mismatch as an outcome variable suggests that it is specifically Anti Democrat videos which were most likely to be recommended as an ideological mismatch relative to other types of videos. See Supplementary Table~15 for the logistic regression table containing these results.

\subsection*{Survey on TikTok algorithm changes}
\begin{figure}
    \centering
    \includegraphics[width=\linewidth]{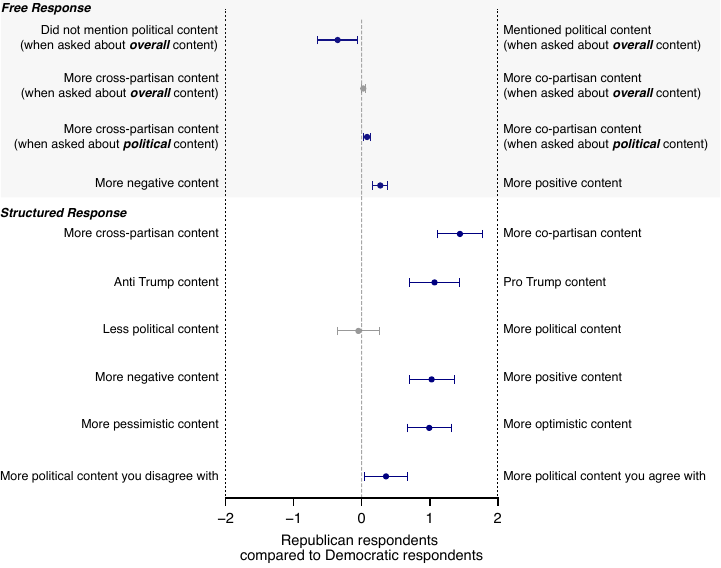}
    \caption{\fontsize{10}{10}\selectfont{\textbf{Survey responses}. Regression coefficients of Republican respondents relative to Democratic respondents for different survey questions regarding content on their TikTok feeds. The top four rows are with regards to responses to text-entry questions, while the bottom six rows are with regards to responses to structured scale-based questions. Statistically significant coefficients are highlighted in navy.}}
    \label{fig:survey}
\end{figure}

While the experimental results above reveal systematic partisan biases in TikTok’s recommendation algorithm, it remains an open question whether these changes are perceptible to real users. Do TikTok users themselves notice shifts in political content on the platform, and do perceptions of these changes vary along partisan lines? To address this question, we conducted a pre-registered survey of 1,000 U.S.-based TikTok users to examine whether they had observed changes in their TikTok feeds over the past year. Our goal was to assess the public’s awareness of algorithmic shifts and determine whether those shifts were interpreted differently based on political affiliation.

Participants were first asked a series of three open-ended, text-entry questions about perceived changes to their TikTok feed. The first question asked generally whether they had noticed any changes in content over the past year; the second focused specifically on changes to political content; and the third asked whether political content had become more positive or more negative. Following these open-ended questions, participants responded to six structured items using a 0–10 scale, measuring their perception of whether their feed had shifted toward more political or less political content, more Democratic or more Republican content, whether it had become more positive or negative, more optimistic or pessimistic, more Pro or Anti Trump, and whether they were seeing more content they agreed or disagreed with politically (see the Survey section of the Methods for more details).

Figure~\ref{fig:survey} presents the regression coefficients of Republican respondents relative to Democratic respondents across all questions. Specifically, each of the coefficients presented is taken from a separate regression, each of which controls for participant demographics (see Supplementary Table~20 and 21 for the full set of regression results and Supplementary Table~22 for a breakdown of participant demographics). In the free response portion, Republican respondents were less likely than Democrats to mention political content when asked about overall feed changes, and no significant partisan differences were observed when respondents were asked whether they noticed a shift toward Republican or Democratic content in their feed overall. However, when asked about political content specifically, Republicans were significantly more likely to mention increases in co-partisan content and more positive political content.

In the structured response section, there were clear partisan differences as well. Republicans were significantly more likely than Democrats to report that they had seen an increase in party-aligned content, and were more likely to characterize the political content on their feed as positive, optimistic, agreeable, and pro-Trump.

\subsection*{Topic analysis}
\begin{figure}
    \centering
    \includegraphics[width=\linewidth]{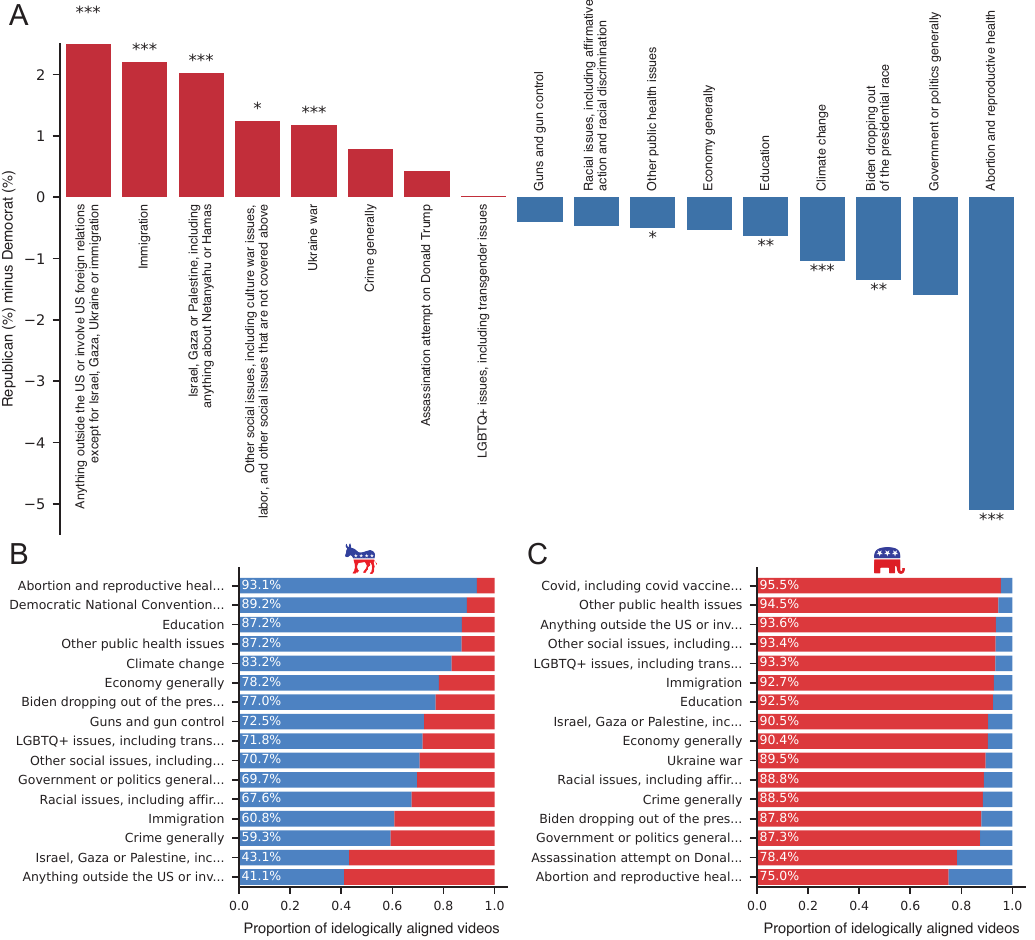}
    \caption{\fontsize{10}{10}\selectfont{\textbf{Topic analysis of political TikTok videos.} (\textbf{A}) The percentage of Republican-aligned videos on a given topic minus the percentage of Democratic-aligned videos on that topic. Statistically significant difference between Republican and Democratic percentages are highlighted with stars (Chi-squared tests; $*: p < 0.05$, $**: p < 0.01$, $***: p < 0.001$). (\textbf{B}, \textbf{C}) The proportion of videos on a given topic which are ideologically-aligned and ideologically-opposing seen by Democratic and Republican-conditioned bots, respectively. For each plot, topics are listed in descending order of ideological-alignment.}}
    \label{fig:topics}
\end{figure}

What do partisan videos discuss, and what topics are most likely to be recommended as an ideological mismatch? To answer these questions, we use methods from~\cite{pewresearchAmericasNews} to identify the substantive topics covered in a given political video (see details in the Topic Analysis section of the Supplementary Materials). To start, we first isolate topics that were covered in at least 100 unique videos. We then compute the difference between the proportion of Republican- and Democratic-aligned videos on a given topic. Figure~\ref{fig:topics}A illustrates differences in Republican and Democratic coverage of different topics, ordered from most proportionally Republican to most proportionally Democratic. As can be seen, topics stereotypically associated with the Democratic party (e.g., climate change, as well as abortion and reproductive health) emerge as topics with significantly more proportional coverage from Democratic-aligned videos. On the other hand, topics such as immigration, foreign policy, or the Ukraine war were topics more frequently covered by Republican-aligned videos. See Supplementary Table~10 for additional details regarding the distribution of topics within Democratic- and Republican-aligned videos, as well as chi-squared tests comparing these distributions.

Having established that Democratic-conditioned bots viewed significantly more ideologically mismatched videos in previous sections, next, we analyze what topics such videos discussed. To this end, we focus on topics with a given partisan stance (Democratic- or Republican-aligned) that were viewed at least 100 times. For each such topic, Figure~\ref{fig:topics}B illustrates the proportion of watches by Democratic-conditioned bots that were Democratic-aligned (blue) and Republican-aligned (red). Similarly, Figure~\ref{fig:topics}C illustrates these proportions but for Republican-conditioned bots. Videos seen by Democratic-conditioned bots on the topics of immigration, crime, the Gaza conflict, or foreign policy broadly, were most likely to be of an opposing ideological alignment, with ideological mismatches representing the 56.9\% and 58.9\% of the latter two topics. In contrast, videos on the topics of abortion and reproductive health were proportionally the most commonly mismatched videos for Republican-conditioned bots, although these only account for 25.0\% of all abortion-related videos seen by Republican bots. Supplementary Figures~7A and 7B mirror Figures~\ref{fig:topics}B and ~\ref{fig:topics}C while including videos classified as ``Neutral'' as well. Supplementary Figure~6 illustrates the difference in the proportion of videos on a topic with a given stance and the proportion of videos of that stance out of all political videos. 

Lastly, we also conduct an exploratory analysis of misinformation on TikTok. Using 20,950 ``fake news'' headlines collected from Snopes~\cite{snopes} and Politifact~\cite{politifact}, we compute the cosine similarity of the vector embeddings of these headlines, and those of the video transcripts. As shown in Table~\ref{table:misinfo}, at similarity thresholds above 0.7, less than 1\% of videos of each ideological stance include misinformation as computed by this metric. See the \hyperref[methods:misinfo]{Misinformation Classification} section of the Methods for more details.

\section*{Discussion} 

Our longitudinal analysis of TikTok's recommendation algorithm during the 2024 U.S. presidential elections reveals substantial asymmetries in partisan content recommendation rates. Through controlled experiments across three states, we document several key patterns that characterize political content recommendations on the platform. Most notably, we find persistent ideological segregation that manifests differently across partisan lines, with Republican-conditioned accounts receiving approximately 11.5\% more ideologically aligned content compared to Democratic-conditioned accounts. Moreover, we observe asymmetric cross-party exposure, where Democratic-conditioned accounts encounter significantly more opposite-party content than their Republican counterparts, particularly from high-profile Republican figures and channels. We also show that negative-partisan content, or content criticizing the opposing party rather than promoting one's own, formed the majority recommendations across both partisan alignments, and it is this type of content specifically that is more likely to be recommended as an ideological mismatch.

These findings have broad partisan implications, especially since TikTok is a critical news source for young voters---a demographic that shifted 10 percentage points toward Trump from 2020 to 2024~\cite{tuftsYoungVoters}. Compared to legacy social media platforms like Facebook and Twitter, TikTok's recommendation is more central to users' experience due to how users engage with the platform, which is almost entirely dominated by recommendations. Taken together, this suggests that TikTok may have a greater impact on political discourse worldwide than any of the platforms to come before. 

\subsection*{Limitations}
While we made every effort to scrutinize our results and ensure their validity, we list some empirical and technical limitations below. First, our experiments may not fully capture the complexity of real user behavior or the full range of factors influencing content recommendations (e.g., non-public engagement metrics only known to TikTok). Second, our classification of political and topical content relies on LLM-based assessments of political stance. While such use-cases of LLMs have been validated in our study and in prior work~\cite{ziems2024can, gambini2024evaluating}, our classification pipeline does not have perfect accuracy. Having said that, we aimed to mitigate potential biases from classifications made by a single LLM by employing an ensemble method to take the majority vote of three different models. Third, given our reliance on video transcripts, our annotation process does not capture any potential visual or audio cues included in the videos. Lastly, after November 11th, we were unable to conduct additional experimental runs due to TikTok increasing its bot-detection efforts, preventing us from collecting data for the weeks after the election; as a result, it may be difficult to replicate this analysis or measure changes in the partisan bias of TikTok's recommendation algorithm .

Future research could address these limitations in several ways. First, studies combining sock puppet experiments with real user data could provide a more comprehensive understanding of how TikTok's algorithm responds to true user behavior. Second, developing methods to analyze visual content at scale alongside transcripts could capture political messaging that relies primarily on images or non-verbal cues. Third, longitudinal studies extending beyond the election period could help distinguish between election-specific effects and more persistent patterns in content recommendation. Fourth, comparative studies across different social media platforms could help isolate TikTok-specific effects from broader patterns in social media political communication. Lastly, there is a clear need for future work to study the extent to which misinformation exists on TikTok, both in general and across partisan lines. While we perform a relatively small exploration into this question using cosine similarity matching of sentence embeddings, further efforts that rely on larger datasets of misinformation relating to the 2024 U.S.\ elections specifically, may be needed in an effort to advance the literature relating to misinformation spread on social media. This is true both on TikTok and on other social media platforms as well, considering, for instance, Meta reportedly intending to end its third-party fact-checking program~\cite{nbcnewsMetaEnding}.

\section*{Ethics statement}
Throughout this study, we took care to follow relevant ethical standards. The use of sock puppet accounts is an established research technique for investigating personalization and bias on Internet platforms~\cite{Sandvig2014AuditingA, ibrahim2023youtube, haroon2023auditing}, provided it is strictly for noncommercial, public-interest purposes and does not compromise user privacy. Prior work and legal precedent has indicated that social media platforms’ Terms of Service, which may prohibit automated access to content, do not necessarily conflict with collecting publicly available data for research aims~\cite{golla2020webscraping, apnews2024twitterlawsuit}. Given our focus on potential impacts on democratic processes and political discourse, this project falls under well-recognized academic exceptions for studying online information ecosystems. Moreover, we minimized the risk of privacy breaches or commercial harm by restricting our data collection to publicly available content. While the experiment itself did not include any human subjects, we obtained informed consent from the human annotators for our LLM classification validation tasks. 

For the survey component of this study, the survey was deemed exempt by the authors’ institutional review board (IRB Protocol Number: HRPP-2025-69)

\section*{Methods}
\label{methods}
\subsection*{Experimental Setup}
\label{methods:experiment_setup}
In this experiment, we aimed to understand the rate at which TikTok's recommendation algorithm recommends videos of a certain political leaning, in the context of U.S. politics. To do so, we employ bots which simulate TikTok users by watching both predefined sequences of videos of a given political leaning (the ``conditioning'' stage), and then subsequently watching recommended videos on a given bot's ``For You Page'' (FYP) (the ``recommendation'' stage). Each experimental run, comprising these two stages, lasts for a one-week period.

Over the duration of 27 weeks, 567 experiments were conducted. Specifically, each week, 21 new TikTok accounts are created by randomly combining the most common American first and last names from~\cite{ssa_babynames_2010s} and assigning an age between 22 and 24. 
This allowed each account to impersonate a potential voter for the U.S. presidential election likely to be active on TikTok. The 21 accounts created each week are split into one of nine experimental conditions, which are defined by two attributes. The first attribute is the state in which the bot is manually geo-located to, which is either New York, Texas, or Georgia\footnote{Georgia was largely regarded as a key swing state for 2024 U.S. presidential elections, with the narrow 0.23\% margin in favor of Joe Biden in the previous 2020 elections.}. These states were chosen specifically as projected Democrat, Republican, and Swing states in the 2024 U.S. presidential elections, respectively.  
The second variable is the political leaning of the videos the bot watches in the ``conditioning'' stage. The videos watched in the conditioning stage are either published by known Democrat supporting channels or Republican supporting channels. Lastly, each week, three bots geo-located to Georgia bypass the conditioning stage of the experiment, and move directly to the recommendation stage. This is done to collect recommendations made to users who do not have a particular interest in politics. A summary of the experimental conditions can be seen in Table~\ref{table:experimental_conditions}.
Supplementary Figure~1 illustrates a more detailed timeline of a bot during a given experimental run.

\begin{table}[htbp!]
\centering
\begin{tabular}{l|ccc}
    & \multicolumn{3}{c}{State}   \\ \hline
Conditioning-Leaning & New York  & Georgia   & Texas     \\ \hline
Democrat   & 3x weekly & 3x weekly & 3x weekly \\
Republican & 3x weekly & 3x weekly & 3x weekly \\
Neutral &     & 3x weekly &    \\\hline
\end{tabular}
\caption{Summary of experimental conditions}
\label{table:experimental_conditions}
\end{table}

\subsubsection*{Pre- and Post-Experiment Protocols}
\label{methods:experiment_setup:pre_post_protocol}
TikTok infers user location through the device's GPS or network (IP) geo-location~\cite{tiktok_location_services}. This required us to dedicate an Android smartphone, namely Samsung Galaxy A34 5G, to each of the 21 accounts created every week. Before each experiment, we control device geo-location across three target states using a combined approach of GPS mocking and VPN tunneling. Specifically, we employed AnyTo~\cite{imyfone_location_changer} for GPS coordinate spoofing, setting New York bots to $<$40.7308, -73.9976$>$ in Manhattan, New York City, Texas bots to $<$33.148, -96.638$>$ in Collin County, and Georgia bots to $<$33.961, -84.537$>$ in Cobb County. These specific locations were chosen as counties which voted strongly Democrat, Republican, or was a close call in the 2020 U.S. presidential elections, respectively. 
Furthermore, to align each bot's network identity with the intended state's geo-location, we tunnel each phone's public IP address to one of three custom VPN servers we hosted on third-party cloud providers. We avoided commercial VPN services to minimize the risk of TikTok identifying the IPs as virtual. We install TikTok from Google PlayStore only after each phone's GPS and IP address had been appropriately modified. At the conclusion of a weekly experiment, we factory-reset every phone before beginning the next round of experiment. This step ensures that any TikTok-related cache is cleared and does not influence the subsequent experiments conducted on the same phones. Finally, all phones operated on Android 13 which re-randomizes the MAC address every 24 hours~\cite{android_mac_randomization}, precluding the possibility of TikTok's device-level tracking or bot detection throughout a weekly experiment.         

\subsubsection*{Conditioning Stage}
\label{methods:experiment_setup:conditioning_stage}
In the conditioning stage of the experiment, bots watch a sequence of videos published by TikTok channels aligned with either Democratic or Republican political leanings. To collect the channels that the bots would watch in this stage, channels were compiled iteratively by searching for politically charged keywords, such as ``Trump'', ``Biden'', or ``Kamala'' on TikTok's search bar, in addition to the terms ``Democrat'' and ``Republican''. From there, the channels published videos supporting each candidate were compiled and verified by the authors to fit the following criteria: (1) the majority of videos published by the channel concern political content; (2) all the political videos published by the channel are aligned with either the Democratic or Republican political parties. 

In total, 54 candidate channels were compiled across both political party affiliations. Each channel associated with a given political party was then matched with a channel from the opposite political party based on its number of followers and cumulative number of likes using Euclidean distance matching. This is done to account for the possibility that the strength of a given bot's political conditioning may be partially attributed to the popularity of the channels it watches in the conditioning stage. After matching, the 10 pairs of channels with the smallest Euclidean distance were selected. Table~\ref{table:conditioning_stage_accounts} below details the channels used in the conditioning stage, showing 12 pairs in total, as two channels (``donaldtrumpwasright'', ``kayetriots'') became no longer available (deleted or made private) during our experiments. As such, those two pairs were replaced with the new pairs that had the smallest Euclidean distance. The accounts of the main political candidates (\texttt{kamalaharris} and \texttt{realdonaldtrump}) did not exist when the experiment began in May, and hence, were not included as conditioning channels. Kamala Harris officially joined TikTok in July, while Donald Trump joined the platform in June.

In each experiment, a bot is conditioned to lean either Democratic or Republican by watching up to 50 most recent videos from 8 randomly selected channels aligned with the respective political party.  
The target of each bot watching 400 conditioning videos in total was not always achieved, as shown in Table~\ref{table:persona_counts}. On average, Republican bots viewed less conditioning videos due to several of the Republican accounts having less than 50 total published videos, particularly during the first few months of the experiment. To account for this, the number of videos watched during the conditioning stage was controlled for during all analyses that compared recommendation rates across experimental conditions.   
Every video during the conditioning stage is watched for one minute, ensuring consistent exposure to the content from the selected channels. 
After completing the predefined set of videos, the bot would then ``sleep'' for 24 hours, where it would not open or interact with TikTok. This pause was implemented to simulate realistic human viewing patterns, minimizing the risk of bot detection that could result from consuming an excessive number of videos in rapid succession.

\begin{table}[htbp!]
\centering
\begin{tabular}{l|cc}
State    & Democrat bots & Republican bots \\ \hline
Georgia  & 376.96        & 317.44          \\
New York & 341.30        & 311.878         \\
Texas    & 376.58        & 311.85         \\ \hline
\end{tabular}
\caption{The mean number of conditioning videos watched per experimental condition.}
\label{table:persona_counts}
\end{table}

\begin{table}[htbp!]
\centering
\begin{tabular}{ccc}
\hline
Left-leaning Account & Right-leaning Account & Euclidean Distance \\ 
\hline
indianahousedemocrats & republicantoks             & 3.83K \\           
repbowman             & elsakurt\_official         & 13.90K  \\         
repstansbury          & republicanfamily           & 13.94K    \\       
repsumnerlee          & republican\_army\_         & 20.73K      \\     
iampoliticsgirl       & therepublicanjournal       & 209.84K       \\   
bernie                & tuckercarlson              & 608.28K \\          
theproblem            & donaldtrumpwasright         & 636.73K \\         
reprokhanna           & kayetriots                  & 707.32K  \\        
aocinthehouse         & restoringamerica           & 920.82K   \\       
thedemocrats          & thecamhigby                & 1.6M      \\      
biancagraulau         & conservativeant2.0         & 2.4M     \\       
jeffjacksonnc         & real.benshapiro            & 13.5M \\
\hline
\end{tabular}
\caption{Left- and right-leaning TikTok accounts used during the conditioning phase, paired based on the minimum Euclidean distance with respect to the number of followers and likes.}
\label{table:conditioning_stage_accounts}
\end{table}

\subsubsection*{Recommendation Stage}
\label{methods:experiment_setup:recommendation_stage}
Following the conditioning stage, each bot in the experiment transitions to the ``recommendation stage'' where they watch videos that appear on their ``For You'' page. The For You page is the default interface on TikTok, where users can watch videos recommended to them based on their interests as implicitly determined by TikTok's algorithm. 
In this stage, each video is watched for up to 10 seconds, after which the video's URL was retrieved. To circumvent TikTok's bot detection mechanism, only the first 10 recommended videos were watched per hour, followed by a 60-minute sleep. Additionally, the TikTok app was reloaded before every hourly session to avoid watching videos pre-loaded during the previous iteration.     

All URLs collected in a given experimental run are then used to retrieve the video's metadata, including the author of the video, the video's description, and its embedded transcript if available. Of the 176,252 unique videos watched, 40,264 had a transcript available, and it is this set of 40,264 videos which we analyze in the remainder of this study. In the Data Representativeness section, we show that this sample is representative of the entire dataset of recommended videos.

\subsubsection*{Experimental run validation}
\label{methods:experiment_setup:validation}
Naturally with audit experiments such as this one, experimental failures are inevitable due to issues such TikTok classifying the account as a bot and subsequently suspending the account, or internet outages. As such, to ensure the same amount of recommendation exposure for Democrat- and Republican-conditioned bots, we match bots of opposite conditioning in each state during a given experiment week. Consequently, for each pair of bots, we only consider the first \textit{N} recommendations made to each bot, where \textit{N} is the lesser of the total number of recommendations made to either bot. Furthermore, we only consider pairs of bots which watched at least 150 videos each. 

There were a handful of weeks in which the bots failed to meet our inclusion threshold. Firstly, during the end of July and the month of August, bots with a Texas geo-location lost internet connectivity while the authors were not available to tend to the bots. As such, we were unable to collect data from Texas during this time. In another weeks, some accounts were recognized as bots by TikTok's bot detection algorithm and subsequently suspended before reaching 150 video threshold. Of the 567 experiments conducted, 323 met our inclusion criteria. Supplementary Table~1 lists the total number of bots which met the inclusion criteria on a weekly basis across the different bot conditioning and geo-location classes.

\subsection*{Ideological Stance Classification}
\label{methods:stance_classification}
To analyze the ideological stances present in the video content, we implemented a three-step classification approach using an ensemble of Large Language Models (LLMs) comprising GPT-4o, Gemini-Pro, and GPT4. First, each video transcript was evaluated for political content using the prompt:
\begin{quote}
    "Given the following video transcript, do you think the topic is political? "
\end{quote}

For transcripts identified as political, we conducted two additional classification steps. The first assessed election relevance through the prompt
\begin{quote}
    "Given the following video transcript, do you think the topic is related to the 2024 US election or related to Donald Trump, Kamala Harris, Joe Biden, JD Vance, or Tim Walz? Answer with only Yes or No."
\end{quote}

The second step determined the partisan stance using the prompt
\begin{quote}
    "Given the following video transcript, classify the transcript into one of the following categories:
    
    Anti Democrat
    
    Anti Republican
    
    Pro Democrat
    
    Pro Republican
    
    Neutral." 
\end{quote}

The distribution of classifications across these categories is presented in Table~\ref{table:question_counts} below.

\begin{table}[htbp!]
\centering
\begin{tabular}{lcc}
\textbf{Question}                                                                                                                                                                                                                                       & \textbf{Majority Vote} & \textbf{Count} \\ \hline
\multirow{2}{*}{\begin{tabular}[c]{@{}l@{}}Question 1: Given the following video transcript,\\ do you think the topic is political?\end{tabular}}                                                                                                       & No                     & 31520          \\
                                                                                                                                                                                                                                                        & Yes                    & 8744          \\ \hline
\multirow{4}{*}{\begin{tabular}[c]{@{}l@{}}Question 2: Given the following video transcript,\\ do you think the topic is related to the 2024 US\\ election or related to Donald Trump, Kamala Harris,\\ Joe Biden, JD Vance, or Tim Walz?\end{tabular}}  \\ & No                    & 4392           \\
                                                                                                                                                                                                                                                        & Yes                     & 4229           \\ \\ \hline
\multirow{4}{*}{\begin{tabular}[c]{@{}l@{}}Question 3: Given the following video transcript,\\ classify the transcript into one of\\ the following categories:\end{tabular}}                                                                            & Neutral                & 3389           \\
                                                                                                                                                                                                                                                        & Anti Democrat          & 1949           \\
                                                                                                                                                                                                                                                        & Anti Republican        & 1177           \\
                                                                                                                                                                                                                                                        & Pro Democrat           & 667           \\
                                                                                                                                                                                                                                                        & Pro Republican         & 586       \\ \hline   
\end{tabular}
\caption{The number of videos classified as political, pertaining to the U.S. presidential election, and of each ideological stance.}
\label{table:question_counts}
\end{table}

Throughout our study, we often group the categories ``Anti Democrat'' and ``Pro Republican'' under the broad category of ``Republican-aligned'', and similarly group the categories ``Anti Republican'' and ``Pro Democrat'' under the category of ``Democrat-aligned''. 

To ensure classification reliability, we employed a consensus-based approach where GPT-4 served as the tiebreaker in cases of disagreement between GPT-4o and Gemini-Pro. The inter-model agreement rates for each classification task are detailed in Table~\ref{table:stance_validation} below. This ensemble method was chosen to mitigate individual model biases and enhance the robustness of our ideological stance classifications.

\begin{table}[htbp!]
\centering
\begin{tabular}{l|cc}
           & Fleiss Kappa & Krippendorf's Alpha \\ \hline
Question 1 & 0.715        & 0.615               \\ 
Question 2 & 0.696        & 0.626               \\ 
Question 3 & 0.631        & 0.560              \\ \hline
\end{tabular}
\caption{Inter-LLM agreement metrics for the three prompts.}
\label{table:stance_validation}
\end{table}

Lastly, after all videos had been classified, we exclude the ten advertisement videos encountered by the bots during the entire duration of the experiment which were labeled as political. Of these, only one had a non-neutral ideological stance (``Anti Democrat''). Furthermore, none of the ten advertisements were published by an official political candidate channel.

\subsubsection*{Ideological stance validation}
\label{methods:stance_classification:validation}
To validate our automated classification approach, we conducted a human annotation study on a subset of 500 randomly selected transcripts. Three independent annotators, all political science undergraduate students, were recruited to perform the classification tasks. The annotators were permitted to use web searches to research unfamiliar topics or references within the transcripts, ensuring informed labeling decisions. Each transcript received independent classifications from all three annotators following the same categorical framework used in the LLM classification. The results of this validation study, including inter-rater reliability metrics, human majority-LLM majority classification accuracy, Cohen's kappa coefficient comparing human majority and LLM majority decisions, and the LLM ensemble's F1-scores, are presented in Table~\ref{table:stance_valdiation}, while Table~\ref{table:indv_llms} details these scores for each model separately.

\begin{table}[htbp!]
\centering
\begin{tabular}{lcccc}
           & \begin{tabular}[c]{@{}c@{}}Inter-Rater Reliability\\ (Krippendorf's Alpha)\end{tabular} & Accuracy & Cohen Kappa & F1 Score \\ \hline
Question 1 & 0.886                                                                                   & 0.965    & 0.930       & 0.965    \\
Question 2 & 0.851                                                                                   & 0.970    & 0.940       & 0.970    \\
Question 3 & 0.886                                                                                   & 0.957    & 0.936       & 0.784   \\ \hline
\end{tabular}
\caption{Human validation accuracy and agreement metrics for LLM-ensemble classification tasks.}
\label{table:stance_valdiation}
\end{table}

\begin{table}[htbp!]
\centering
\small
\begin{tabular}{ll|ccc}
                             &            & \textbf{Question 1} & \textbf{Question 2} & \textbf{Question 3} \\ \hline
\multirow{4}{*}{Accuracy}    & GPT 4o     & 0.710               & 0.965               & 0.854               \\
                             & Gemini-Pro & 0.915               & 0.955               & 0.890               \\
                             & GPT 4      & 0.955               & 0.945               & 0.909               \\
                             & Majority   & 0.965               & 0.970               & 0.957               \\\hline
\multirow{4}{*}{Cohen Kappa} & GPT 4o     & 0.410               & 0.930               & 0.767               \\
                             & Gemini-Pro & 0.831               & 0.910               & 0.838               \\
                             & GPT 4      & 0.910               & 0.888               & 0.859               \\
                             & Majority   & 0.930               & 0.940               & 0.936               \\ \hline
\multirow{4}{*}{F1 Score}    & GPT 4o     & 0.683               & 0.965               & 0.652               \\
                             & Gemini-Pro & 0.915               & 0.955               & 0.707               \\
                             & GPT 4      & 0.955               & 0.944               & 0.728               \\
                             & Majority   & 0.965               & 0.970               & 0.784     \\ \hline         
\end{tabular}
\caption{The accuracy, Cohen kappa, and F1-scores of each LLM model individually, as well as the LLM majority vote, when compared with human ground truth labels.}
\label{table:indv_llms}
\end{table}

\subsubsection*{Channel classification}
\label{methods:stance_classification:channel_classification}
To identify channels potentially leaning toward either the Democratic or Republican ideologies, we focused on channels that had published videos previously labeled with a specific non-neutral ideological stance (e.g., ``Anti-Democrat,'' ``Pro-Republican,'' ``Anti-Republican,'' or ``Pro-Democrat''). From this process, we identified 170 unique channels for analysis. For each channel, we calculated the proportion of videos aligned with a particular party’s ideology as a fraction of the total videos collected for that channel.
To ensure robustness in our classification, channels were labeled as Democrat-aligned or Republican-aligned if more than 75\% of their analyzed videos were ideologically consistent with one party. 
For 85 channels with fewer than 10 unique videos collected during the experiment phase, we used TikAPI~\cite{tikapi} to retrieve up to 30 additional videos published prior to the election. These additional videos were processed through the same transcript classification pipeline to determine their ideological stance. This step was necessary to avoid potential misclassification caused by a limited number of channels’ unique videos watched by the bots. Table~\ref{tab:channel_classification} presents a summary of the classified channels, including the total number of channels in each category (Republican-aligned, Democrat-aligned, and neutral) as well as the average proportion of partisan content for each group. 

\begin{table}[htbp!]
    \centering
    \begin{tabular}{l|cc}\hline
         Category &  Count & Proportion ($M \pm SD$) \\ \hline
         \hline
         Democrat-aligned & 56 & 0.925 $\pm$ 0.0733\\
         Republican-aligned & 75 & 0.94 $\pm$ 0.0568\\ 
         \hline
    \end{tabular}
    \caption{Classification of TikTok channels based on political videos, including channel counts and average proportions of partisan content.}
    \label{tab:channel_classification}
\end{table}

\subsubsection*{Comment Classification}
\label{methods:stance_classification:comment_classification}
To identify the ideological stance of the comments on a given TikTok video, we follow a similar methodology to that of classifying a video. Specifically, we pass both the transcript of the video, as well as the comment in question in a prompt to GPT-4o, which can be seen below:

\begin{quote}
    "Given the following TikTok video transcript and comment, classify the comment into one of the following categories.
    
    Transcript of the video: {{ \textbf{TRANSCRIPT }}}
    
    Comment: {{ \textbf{COMMENT} }}
    
    Categories:
    
    Anti Democrat
    
    Anti Republican
    
    Pro Democrat
    
    Pro Republican
    
    Neutral" 
\end{quote}

The outputs of the model were validated by three independent annotators, again all political science undergraduate students. The results of this validation process are shown in Table~\ref{tab:comment_validation} below.

\begin{table}[!htbp]
\centering
\begin{tabular}{lcccc}
                       & \begin{tabular}[c]{@{}l@{}}Inter-Rater Reliability\\ Krippendorf's Alpha\end{tabular} & Accuracy & Cohen Kappa & F1 Score \\ \hline
Comment Classification & 0.89371                                                                               & 0.95016  & 0.93366     & 0.95025 \\ \hline
\end{tabular}
\caption{Human validation accuracy and agreement metrics for comment classification task.}
\label{tab:comment_validation}
\end{table}

\subsection*{Sensitivity Analysis}
\label{methods:sensitivity}
To effectively test the strength of the bias observed in our analysis, we conduct a sensitivity analysis with regards to the various engagement metrics analyzed in our study. Specifically, here, we ask the following question. If the observed bias was due to some unobserved internal engagement metric only available to TikTok, how different would Republican and Democrat videos need to be in this engagement metric to explain the observed biases measured in our study. To answer this question, we formulate this unobserved metric as a random variable sampled from a set of different distributions, namely, a binomial, lognormal, normal, or poisson distribution. Democratic videos and Republican videos then sample from two distributions of a given type which differ based only on their means. As an example, Democratic videos would sample from a normal distribution with a mean of 1, while Republican videos would sample from a normal distribution with a mean of 2. We can then ask, how large of a difference in means between these two distributions would result in a bias consistent with that seen in our study. The results of this analysis can be seen in Table 1 below.

Specifically, in Table~\ref{tab:sensitivity_analysis} below, we show in the first column the true difference between the Republican and Democratic videos with regards to their normalized values for a given engagement metric. In the remaining four columns, we show the ratio between the difference needed to match the observed bias and that of the true difference in mean between Republican and Democratic videos in a given engagement metric. As can be seen, for the majority of engagement metrics, it is Democratic videos which tend to have higher engagement on average, with the exceptions being in the video comment count, like count, play count, share count, and composite score. Although the combination of these metrics produces the largest normalized Republican skew in engagement gap, this gap would need to be 9.4 times as large in order to full explain the observed bias in recommendation rates seen in our experiments.

\begin{table}[htbp!]
\centering
\footnotesize
\begin{tabular}{lc|cccc}
\hline
\multicolumn{2}{c}{}                                                                                                                                       & \multicolumn{4}{l}{\textbf{Required Scaling with different distributions}}           \\ \hline
\multicolumn{1}{c}{\textbf{Engagement metric}} & \textbf{\begin{tabular}[c]{@{}c@{}}True diff.\\ in normalized\\ metric values\\ (Rep - Dem)\end{tabular}} & \textbf{Binomial} & \textbf{Lognormal} & \textbf{Normal} & \textbf{Poisson} \\ \hline
(Comments, Likes, Shares, Plays) Score & 0.0209 & 31.2593 & 9.4323 & 34.0259 & 29.03865 \\ 
Video share count                              & 0.0028                                                                                                    & 169.4288          & 51.1241            & 184.4241        & 157.3927         \\
Video play count                               & 0.0024                                                                                                    & 177.4398          & 53.5413            & 193.1442        & 164.8346         \\
Comment count                                  & 0.0006                                                                                                    & 782.0713          & 235.9851           & 851.2887        & 726.5135         \\
Video Like count                               & 0.0009                                                                                                    & 325.8119          & 98.3117            & 354.6479        & 302.6664         \\
Comments per recommendation                    & -0.0005                                                                                                   & -77.8383          & -23.4872           & -84.7274        & -72.3087         \\
Video engagement rate                          & -0.0016                                                                                                   & -70.573           & -21.295            & -76.8191        & -65.5596         \\
Channel follower count                         & -0.0017                                                                                                   & -122.0704         & -36.834            & -132.8742       & -113.3986        \\
Likes per recommendation                       & -0.0026                                                                                                   & -73.7079          & -22.2409           & -80.2314        & -68.4717         \\
Channel cumul. likes                           & -0.0116                                                                                                   & -48.4503          & -14.6196           & -52.7384        & -45.0084         \\
Video length                                   & -0.014                                                                                                    & -31.1608          & -9.4026            & -33.9187        & -28.9472         \\
Channel video count                            & -0.0143                                                                                                   & -26.1214          & -7.882             & -28.4333        & -24.2657        \\
Opposing comment proportion                    & -0.0543                                                                                                   & -15.3969          & -4.6459            & -16.7596        & -14.3031         \\
Party-aligned comment proportion               & -0.1043                                                                                                   & -4.8802           & -1.4726            & -5.3121         & -4.5335          \\
Verified account                               & -0.1556                                                                                                   & -3.6068           & -1.0883            & -3.926          & -3.3505          \\

\hline
\end{tabular}
\caption{The true difference between Republican and Democratic videos with regards to their normalized values for a given engagement metric as well as the ratio between the difference needed to match the observed bias and that of the true difference in mean between Republican and Democratic videos for a given engagement metric.}
\label{tab:sensitivity_analysis}
\end{table}

\subsection*{Data Representativeness}
\label{methods:data_representativeness}

Given that our primary analysis throughout this study focuses on videos with transcripts, we must ensure that this sample is representative of videos throughout the entire dataset. To do so, we annotate a random sample of videos recommended to each of the set of Republican and Democrat during the experiment. Specifically, for each of these sets, we analyze a random sample of 2000 videos which do not include transcripts to measure the following features of a given video: (1) whether the video is political in nature, (2) whether the video is concerned with the 2024 U.S. elections or major U.S. political figures, and (3) the ideological stance of the video in question (Pro-Democrat, Anti-Democrat, Pro-Republican, Anti-Republican, or Neutral). This annotation task was again completed by three political science undergraduates, with high agreement between annotators (Krippendorf's alpha; Q1: $\alpha = 0.858$, Q2: $\alpha = 1.0$, Q3: $\alpha = 0.989$). Indeed, we find that the inter-rater agreement when viewing the videos directly, rather than relying on the transcript alone (the results of which are in Table~\ref{table:stance_valdiation}), yielded higher agreement between the raters. 

\begin{figure}
    \centering
    \includegraphics[width=\linewidth]{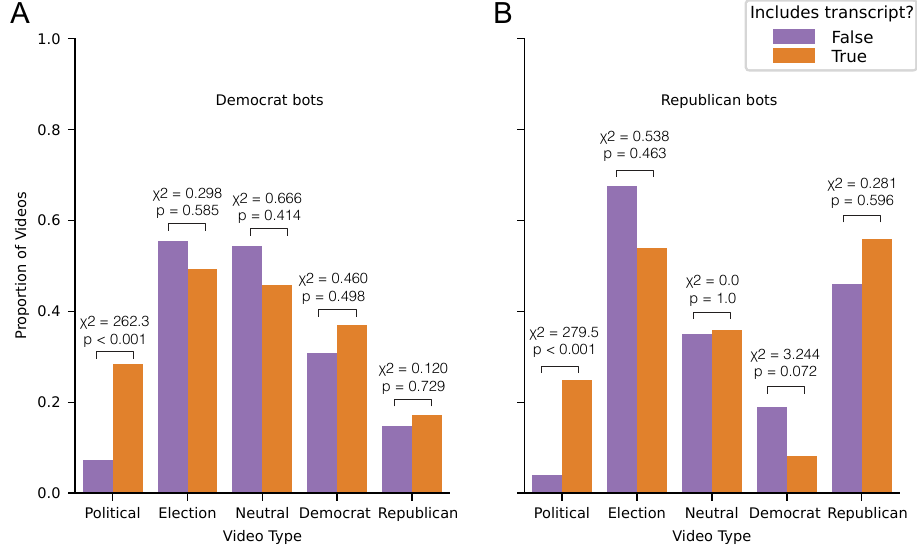}
    \caption{(\textbf{A}, \textbf{B}) A comparison between videos with and without transcripts seen by Democrat and Republican bots, respectively.}
    \label{fig:no_transcript}
\end{figure}

This process allows us to compare the distribution of political content in videos with transcripts against those without transcripts. The results of this comparison is illustrated in Figure~\ref{fig:no_transcript}. As can be seen, videos with no transcripts contained significantly fewer political videos on average relative to those that did include transcripts. This result is perhaps unsurprising given that videos which do not include a transcript are also less likely to include dialogue (e.g., videos of landscapes, pets, etc.). However, of those which were political, we do not find significant differences (computed through chi-squared tests) between transcript and non-transcript videos with regards to the proportion of political videos pertaining to the election, nor their distribution of Democrat-aligned, Republican-aligned, or Neutral videos.

\subsection*{Misinformation classification}
\label{methods:misinfo}

We also conduct an exploratory analysis into the transcripts which may contain misinformation. To this end, we start by collecting all ``fake news'' headlines from Snopes~\cite{snopes} and Politifact~\cite{politifact}. For Snopes, we collect all headlines with a rating of ``False'', ``Mostly False'', ``Unproven'', ``Fake'', or ``Unfounded'', while for Politifact, we collect headlines with a rating of ``False'' or ``Pants-on-fire''. This amounted to a total of 20,950 headlines. Next, we compute the vector embeddings of these headlines, as well as the vector embeddings of all video transcripts collected using OpenAI's ``text-embedding-3-large'' embedding model~\cite{embeddings}. 

Using these vector embeddings, we compute cosine similarities across all possible headline-transcript pairs. Table~\ref{table:misinfo} lists the percentage of transcripts of a given ideological stance which had a headline with a cosine similarity greater than a certain threshold. As can be seen, at thresholds greater than 0.8, less than 1\% of Anti Democrat and Neutral videos include misinformation, and no Anti Republican, Pro Democrat, or Pro Republican videos included misinformation as computed by this metric. While this process mirrors approaches taken by prior work analyzing misinformation on social media~\cite{ibrahim2023youtube, vosoughi2018spread}, further efforts that rely on larger datasets of misinformation, especially those related to the 2024 U.S.\ elections,  may be needed to fully capture the extent of misinformation on TikTok.

\begin{table}[htbp!]
\centering
\footnotesize
\begin{tabular}{lcccc}
Ideological Stance & \begin{tabular}[c]{@{}c@{}}\% of transcripts\\ with similarity $\geq$ 0.6\end{tabular} & \begin{tabular}[c]{@{}c@{}}\% of transcripts\\ with similarity $\geq$ 0.7\end{tabular} & \begin{tabular}[c]{@{}c@{}}\% of transcripts\\ with similarity $\geq$ 0.8\end{tabular} & \begin{tabular}[c]{@{}c@{}}\% of transcripts\\ with similarity $\geq$ 0.9\end{tabular} \\ \hline
Anti Democrat      & 2.874                                                                                        & 0.779                                                                                        & 0.146                                                                                        & 0.0                                                                                          \\
Anti Republican    & 2.375                                                                                        & 0.085                                                                                        & 0.0                                                                                          & 0.0                                                                                          \\
Neutral            & 1.908                                                                                        & 0.66                                                                                         & 0.281                                                                                        & 0.012                                                                                        \\
Pro Democrat       & 2.399                                                                                        & 0.15                                                                                         & 0.0                                                                                          & 0.0                                                                                          \\
Pro Republican     & 1.193                                                                                        & 0.17                                                                                         & 0.0         & 0.0                                                        \\ \hline                                            
\end{tabular}
\caption{The percentage of transcripts that had a headline cosine similarity score above a certain threshold.}
\label{table:misinfo}
\end{table}

\subsection*{Topic Analysis}
\label{methods:topic}

To identify the topics discussed in each political video, we utilize the methodology employed by Stocking et al. ~\cite{pewresearchAmericasNews}. Specifically, the authors utilized GPT-4 to classify a video in one of a set of topics, and thoroughly validated their approach, finding accuracies greater than 0.9 for all topics.

We slightly modify the prompt used by Stocking et al.\ to allow the LLM to select more than one topic option. The prompt used can be seen below:

\begin{quote}
    You are an AI assistant trained to look at social media posts and determine what the post is about.
    
    You will receive the transcript of the video in question.
    
    You will be given a list of topics. Please tell us what this post is about.
    
    Transcript : \textbf{[TRANSCRIPT]}

    Some other things to keep in mind:

    – There is an election in November, so many posts will be about that. The Democratic candidates were Joe Biden and Kamala Harris, but Joe Biden dropped out and it’s Kamala Harris and Tim Walz. The Republican candidates are Donald Trump and JD Vance. The independent candidates are Robert F Kennedy Jr (RFK Jr), Cornel West, and Jill Stein.

    – If there are multiple topics, pick all the relevant topics.

    – If a post seems political, first see if it goes into another category. For example a post about politics and race would go into the race category. Fall back on politics if none other fits, as long as there is something political.

    – With the exception of immigration, most posts that reference a foreign country will go into the last category.

    Categories : 

    1. Crime
    
  1.1. Crime generally
    
    2. Environment
    
  2.1. Climate change
    
  2.2. Other environmental issues
    
    3. Immigration
    
  3.1. Immigration generally
    
    4. Social issues
    
  4.1. Abortion and reproductive health
    
  4.2. Guns and gun control
    
  4.3. LGBTQ+ issues, including transgender issues
    
  4.4. Racial issues, including affirmative action and racial discrimination
    
  4.5. Education
    
  4.6. Other social issues, including culture war issues, labor, and other social issues that are not covered above
    
    5. Public health
    
  5.1. Covid, including covid vaccines
    
  5.2. Other vaccines
    
  5.3. Other public health issues
    
    6. Economy
    
  6.1. Economy generally
    
    7. Technology
    
  7.1. AI, LLMs
    
  7.2. Crypto
    
  7.3. Other technology issues
    
    8. Government, politics and elections
    
  8.1. Assassination attempt on Donald Trump
    
  8.2. Republican National Convention (RNC)
    
  8.3. Democratic National Convention (DNC)
    
  8.4. Biden dropping out of the presidential race
    
  8.5. Other political or government related posts that do not fit into other categories
    
    9. International issues
    
  9.1. Israel, Gaza or Palestine, including anything about Netanyahu or Hamas
    
  9.2. Ukraine war
    
  9.3. Anything outside the US or involve US foreign relations except for Israel, Gaza, Ukraine, or immigration
    
    10. No topic

  10.1. None of the above topics

\end{quote}

Using the above prompt, the topics discussed in the 7,767 ideologically stanced TikTok videos in our dataset were identified. Table~\ref{table:topic_counts} below describes the number of videos falling under each category, split by the video's ideological stance.

\begin{table}[]
\centering
\scriptsize
\begin{tabular}{lccccc}
Topic                                                                                                                                                    & Anti Democrat & Anti Republican & Neutral & Pro Democrat & Pro Republican \\ \hline
AI, LLMs                                                                                                                                                 & 9           & 1             & 29    & 4          & 0            \\
Abortion and reproductive health                                                                                                                         & 82          & 148           & 96    & 148        & 8            \\
\begin{tabular}[c]{@{}l@{}}Anything outside the US or involve US foreign\\ relations except for Israel, Gaza, Ukraine,\\ or immigration\end{tabular}     & 235         & 72            & 1217  & 8          & 40           \\
Assassination attempt on Donald Trump                                                                                                                    & 38          & 46            & 57    & 3          & 53           \\
Biden dropping out of the presidential race                                                                                                              & 288         & 68            & 153   & 227        & 25           \\
Climate change                                                                                                                                           & 33          & 33            & 64    & 40         & 2            \\
Covid, including covid vaccines                                                                                                                          & 52          & 29            & 35    & 7          & 7            \\
Crime generally                                                                                                                                          & 236         & 141           & 1257  & 19         & 25           \\
Crypto                                                                                                                                                   & 4           & 5             & 11    & 0          & 4            \\
Democratic National Convention (DNC)                                                                                                                     & 38          & 6             & 28    & 43         & 1            \\
Economy generally                                                                                                                                        & 260         & 147           & 496   & 133        & 82           \\
Education                                                                                                                                                & 54          & 44            & 176   & 36         & 15           \\
Environment generally                                                                                                                                    & 18          & 15            & 117   & 12         & 6            \\
Government or politics generally                                                                                                                         & 1926        & 1254          & 2466  & 715        & 609          \\
Guns and gun control                                                                                                                                     & 55          & 38            & 227   & 35         & 18           \\
Immigration                                                                                                                                              & 286         & 123           & 244   & 34         & 56           \\
\begin{tabular}[c]{@{}l@{}}Israel, Gaza or Palestine, including anything\\ about Netanyahu or Hamas\end{tabular}                                         & 239         & 88            & 1014  & 13         & 17           \\
LGBTQ+ issues, including transgender issues                                                                                                              & 132         & 61            & 160   & 48         & 15           \\
Other                                                                                                                                                    & 63          & 6             & 2561  & 3          & 7            \\
Other public health issues                                                                                                                               & 39          & 28            & 350   & 37         & 18           \\
\begin{tabular}[c]{@{}l@{}}Other social issues, including culture war\\ issues, labor, and other social issues that\\ are not covered above\end{tabular} & 355         & 164           & 2082  & 92         & 61           \\
Other technology issues                                                                                                                                  & 28          & 12            & 136   & 2          & 7            \\
Other vaccines                                                                                                                                           & 7           & 0             & 6     & 1          & 2            \\
\begin{tabular}[c]{@{}l@{}}Racial issues, including affirmative action\\ and racial discrimination\end{tabular}                                          & 299         & 175           & 408   & 87         & 23           \\
Republican National Convention (RNC)                                                                                                                     & 5           & 18            & 5     & 0          & 18           \\
Ukraine war                                                                                                                                              & 77          & 14            & 135   & 2          & 15          \\ \hline
\end{tabular}
\caption{The number of videos of each political classification on a certain topic.}
\label{table:topic_counts}
\end{table}

\subsection*{Video Recommendation Counterfactual Models}
\label{methods:counterfactual}

To verify that the Republican skew observed in our experiment are not due to differences in the engagement metrics of Republican and Democrat videos or channels, we build a series of counterfactual-models to predict expected video recommendations based on these engagement metrics. Specifically, we aim to compute differences between observed recommendation rates and predicted recommendation rates when sampling from the dataset of videos based on a series of different metrics. The metrics of interest include several video-level metrics, such as the number of plays/views a video receives, as well as its number of shares, comments, and likes. In addition to video-level metrics, we also collect channel level metrics such as a channel's cumulative number of likes, followers, videos, as well as whether the channel is verified or not. Given that channel verification status is a binary attribute (True or False), we apply a value of 0.1 to channels who are not verified, and 0.9 to channels who are verified. For robustness, we also test all values between 0.5 and 1.0 in increments of 0.05 for verified channels (and inversely, 0.5 to 0.0 for unverified channels), and find similar results (see Supplementary Table~8). 

To retrieve predicted recommendation rates, for each week throughout the duration of the experiment, we first isolate the set of videos seen by the bots during and prior to the week in question. We then compute normalized values of the metric in question using min-max normalization. Next, we compute a bootstrapped measure of differences by sampling \textit{N} videos (where \textit{N} is equivalent to the number of videos seen by a bot during a given week), weighted by the normalized metric in question. The proportion of Republican and Democrat videos is then computed from the sample of \textit{N} videos collected. This allows us to compute the difference between the observed recommendation rates of Republican and Democrat videos by the bots and those of the sampled videos. This sampling process is repeated 100 times for robustness to retrieve the mean and standard error of the differences per week for each set of partisan bots.

In addition to the single value metrics mentioned above, we compute three additional metrics. The first is a linear combination of a video's normalized play, share, like, and comment counts. The second is a linear combination of a channel's cumulative likes, followers, and videos, as well as its verification status. The third is a linear combination of all video- and channel-level metrics. For each of these additional metrics, the weight of each component is derived using Principal Component Analysis (PCA), which identifies the linear combination of components that captures the maximum variance in the data. We use the loadings (coefficients) from the first principal component as weights, normalizing them to sum to 1. The sampling process described above is repeated for each of these three combined metrics. 

The above models that consider the number of likes or comments disregard the fact that these metrics are not independent of the recommendation algorithm. Indeed, videos that receive more recommendations end up being viewed more, which in turn increases their likelihood of receiving likes and/or comments. Thus, to develop counterfactual models that capture the video's intrinsic ``quality'' while isolating the effect of the choices made by the algorithm, we normalize these metrics as follows. First, we take the number of ``recommendations'' a video receives as the number of plays minus the number of shares. This subtraction yields the number of views received by the video specifically via the recommendation algorithm and not by users sharing the video. With this recommendations metric, we additionally develop counterfactual models that account for a video's number of likes or comments per recommendation. 

Lastly, for each of the aforementioned metrics, we repeat the sampling processes while taking into account the potential favorability of recent content. Specifically, we inversely scale each metric in question with the time since its publishing date, once linearly, and once exponentially. In total, this amounts to 39 different counterfactual models with which we compare observed recommendation rates and predicted recommendation rates. Supplementary Table~5 details the ideological skew computed by each model, as well as t-test results between the observed skew and expected skew based on each model. Supplementary Tables~6 and ~7 detail these results for models using only positive-partisan or negative-partisan videos, respectively.

\subsection*{Survey}
We administered a pre-registered survey with a sample of 1,008 U.S.-based TikTok users to assess whether individuals had noticed changes to the content of their TikTok feeds, particularly political content, over the past year. The survey was preregistered on OSF (\url{https://osf.io/udywb/}) and was deemed exempt by the authors’ institutional review board (IRB Protocol Number: HRPP-2025-69).

Participants were recruited via the online platform Prolific and screened to ensure they resided in the United States and were active users of TikTok. The survey consisted of two parts: (1) a series of open-ended text entry questions, and (2) a series of structured, scale-based questions. Open-ended items asked participants whether they had noticed any changes to the content on their TikTok feed in general, any changes to political content specifically, and whether the tone of political content had become more positive or negative. Responses to these questions were manually coded by the first author to determine whether participants explicitly referenced changes to political content and, if so, whether they described seeing more Republican-aligned or Democratic-aligned content.

Structured questions asked participants to rate, on a 0–10 scale, the extent to which their feed had shifted toward Democratic or Republican content, become more positive or negative in tone, or featured more political content they agreed or disagreed with.

For each survey item, we conducted separate linear regression analyses that included participant political affiliation and demographic covariates (age, gender, race, and education level) as predictors. Full regression results are presented in Supplementary Tables~20 and 21 for the open-ended and structured questions, respectively. Participant demographic characteristics are summarized in Supplementary Table~22, and the complete survey instrument is available in the Supplementary Materials under the Survey Materials section.

\section*{Author Contributions}
Y.Z. and T.R. conceived and supervised the study. Y.Z., T.R., H.I. and H.J. designed the experimental procedure. H.I. and H.J. built the data collection pipeline, ran the experiments, collected the data, and built the annotation frameworks. Y.Z., T.R., H.I. and H.J. analyzed the data. N.A. built the misinformation detection protocol and analyzed the results. Y.Z., T.R., H.I. and H.J. produced the visualizations. Y.Z., T.R., H.I., H.J., and A.K. discussed the results and wrote the manuscript. 

\section*{Acknowledgments}
We gratefully acknowledge Jason Greenfield, Kevin Munger, Manoel Horta Ribeiro, Chris Schwartz, Josh Tucker, and the Center for Social Media and Politics' Data Science Reading Group for invaluable feedback during the preparation and revision of this manuscript.

\section*{Competing Interests}
The authors declare no competing interests.

\section*{Data and Code Availability}
The full dataset of political videos on TikTok, in addition to all code to replicate the findings can be found at \url{https://github.com/comnetsAD/politics_tiktok}.

\section*{Additional Information}
Supplementary Information is available for this paper.\\
Correspondence and requests for materials should be addressed to Yasir Zaki.

\clearpage

\bibliographystyle{naturemag}
\bibliography{sample}

\renewcommand{\figurename}{Supplementary Figure}
\renewcommand{\tablename}{Supplementary Table}

\setcounter{figure}{0}
\setcounter{table}{0}

\section*{Survey Materials}

{\large \textbf{Informed Consent}}

\vspace{1em}

You are being asked to provide consent to participate in a research study. Participation is voluntary. You can say yes or no. If you say yes now, you can still change your mind later.

\vspace{1em}

\textbf{Purpose of Research:} This research is being conducted to better understand changes in your TikTok feed over the past 12 months.

\vspace{0.5em}
\textbf{Procedures:} You will be asked a series of open-ended and multiple-choice questions. \textbf{Please do not use ChatGPT or any large language model to generate answers. If use is detected, you will NOT be compensated for your participation.}

\vspace{0.5em}
\textbf{Duration:} Participation will involve approximately 6 minutes of your time.

\vspace{0.5em}
\textbf{Risks:} We believe there are no known risks associated with this research study. Your participation is voluntary, and you may withdraw at any time without penalty or loss of benefits.

\vspace{0.5em}
\textbf{Benefits:} We aim to understand TikTok's recommendation algorithm in the context of U.S. politics, which may be beneficial in the future.

\vspace{0.5em}
\textbf{Compensation:} You will be compensated approximately \$1.15 for your time.

\vspace{0.5em}
\textbf{Privacy and Confidentiality:} Study records will be kept in a secure location. Electronic files will be password-protected. Only the research team will have access. Shared data will be anonymized. Results may be published in summary form without identifying information.

The NYU Abu Dhabi Institutional Review Board (IRB) may inspect study records, but these reviews will only focus on researchers, not participants.

If you have questions about this study, contact the principal investigator, Yasir Zaki (yasir.zaki@nyu.edu). For questions about your rights as a participant, contact the IRB at \href{mailto:irbnyuad@nyu.edu}{irbnyuad@nyu.edu}.

\vspace{1em}

\textbf{Consent Acknowledgment:}
\begin{enumerate}
  \item Your participation is voluntary.
  \item You are 18 years of age or older.
  \item You may withdraw at any time.
\end{enumerate}

\textbf{Options:}
\begin{itemize}
  \item I consent, begin the study.
  \item I do not consent, I do not wish to participate.
\end{itemize}

\vspace{1em}
{\large \textbf{Survey Questions}}

\vspace{1em}
\textbf{Free Response Questions}
\begin{enumerate}[label=Q\arabic*)]
  \item Over the past 12 months, from March 2024 to today, how has the \textbf{overall content} of your TikTok feed changed?\\[1ex]
  \textit{[Text Entry]}

  \item Over the past 12 months, from March 2024 to today, how has the \textbf{political content} of your TikTok feed changed?\\[1ex]
  \textit{[Text Entry]}

  \item Over the past 12 months, from March 2024 to today, has the \textbf{political content} of your TikTok feed become \textbf{more positive or more negative}?\\[1ex]
  \textit{[Text Entry]}
\end{enumerate}

\vspace{1em}
\textbf{Structured Scale Questions}
\begin{enumerate}[label=Q\arabic*), start=4]
  \item Over the past 12 months, your TikTok feed has become:\\
  \textit{0 (Less Political) --- 10 (More Political)}

  \item Over the past 12 months, your TikTok feed has become:\\
  \textit{0 (More political content you disagree with) --- 10 (More political content you agree with)}

  \item Over the past 12 months, the \textbf{political content} in your TikTok feed has become:\\
  \textit{0 (More Democratic) --- 10 (More Republican)}

  \item Over the past 12 months, the \textbf{political content} in your TikTok feed has become:\\
  \textit{0 (More Pro-Trump) --- 10 (More Anti-Trump)}

  \item Over the past 12 months, the \textbf{political content} in your TikTok feed has become:\\
  \textit{0 (More Positive) --- 10 (More Negative)}

  \item Over the past 12 months, the \textbf{political content} in your TikTok feed has become:\\
  \textit{0 (More Pessimistic) --- 10 (More Optimistic)}
\end{enumerate}

\vspace{1em}
\textbf{Demographic Questions}
\begin{enumerate}[label=Q\arabic*), start=10]
  \item How old are you?
  \begin{itemize}
    \item Under 18
    \item 18--24
    \item 25--34
    \item 35--44
    \item 45--54
    \item 55--64
    \item 65+
  \end{itemize}

  \item How do you describe yourself?
  \begin{itemize}
    \item Male
    \item Female
    \item Non-binary / third-gender
    \item Prefer to self-describe: \textit{[Text Entry]}
    \item Prefer not to say
  \end{itemize}

  \item What racial or ethnic group best describes you?
  \begin{itemize}
    \item White
    \item Black
    \item Hispanic
    \item Asian
    \item Native American
    \item Middle Eastern
    \item Two or more races
    \item Other: \textit{[Text Entry]}
    \item Prefer not to answer
  \end{itemize}

  \item What is the highest level of education you have completed?
  \begin{itemize}
    \item Some high school or less
    \item High school diploma or GED
    \item Some college, but no degree
    \item Associates or technical degree
    \item Bachelor's degree
    \item Graduate or professional degree
    \item Prefer not to say
  \end{itemize}

  \item What is your US Zip Code?\\
  \textit{[Text Entry]}

  \item Which political party do you most identify with?
  \begin{itemize}
    \item Democratic Party
    \item Republican Party
    \item Independent
    \item Other: \textit{[Text Entry]}
  \end{itemize}

  \item[Q16)] If Democratic Party (from Q15): Would you call yourself a:
  \begin{itemize}
    \item Strong Democrat
    \item Not so strong Democrat
    \item Prefer not to answer
  \end{itemize}

  \item[Q17)] If Republican Party (from Q15): Would you call yourself a:
  \begin{itemize}
    \item Strong Republican
    \item Not so strong Republican
    \item Prefer not to answer
  \end{itemize}

  \item Do you think of yourself as closer to the Democratic or the Republican party?
  \begin{itemize}
    \item Democratic Party
    \item Republican Party
    \item Neither
    \item Not sure
    \item Prefer not to answer
  \end{itemize}

  \item In general, how would you describe your own political viewpoint?
  \begin{itemize}
    \item Very liberal
    \item Liberal
    \item Moderate
    \item Conservative
    \item Very conservative
    \item Not sure
    \item Prefer not to answer
  \end{itemize}

  \item Did you vote in the 2024 presidential election?
  \begin{itemize}
    \item Yes
    \item No
    \item Prefer not to answer
  \end{itemize}

  \item What party did you vote for?
  \begin{itemize}
    \item Democrat
    \item Green
    \item Independent
    \item Libertarian
    \item Republican
    \item Other
  \end{itemize}

  \item Do you plan to vote in the 2026 midterm elections?
  \begin{itemize}
    \item Yes
    \item No
    \item Not sure
    \item Prefer not to answer
  \end{itemize}

  \item Which party do you plan to vote for?
  \begin{itemize}
    \item Democrat
    \item Green
    \item Independent
    \item Libertarian
    \item Republican
    \item Other
  \end{itemize}

  \item Please tell us anything else you would like to share about your experience on TikTok, politics, or this survey?\\[1ex]
  \textit{[Text Entry]}
\end{enumerate}

\clearpage
\section*{Supplementary Tables}
\label{tables}

\begin{table}[htbp!]
\centering
\footnotesize
\begin{tabular}{l|ccc|cc|cc}
\textbf{Week} & \multicolumn{3}{c}{\textbf{Georgia}} & \multicolumn{2}{c}{\textbf{New York}} & \multicolumn{2}{c}{\textbf{Texas}} \\ \hline
              & Democrat   & Neutral   & Republican  & Democrat         & Republican         & Democrat        & Republican       \\\hline
30-04-2024    & 1        & 1       & 1         & 1              & 1                & 1             & 1              \\
07-05-2024    & 0        & 2       & 0         & 1              & 1                & 0             & 0              \\
14-05-2024    & 1        & 2       & 1         & 2              & 2                & 2             & 2              \\
21-05-2024    & 1        & 2       & 1         & 2              & 2                & 2             & 2              \\
28-05-2024    & 0        & 2       & 0         & 1              & 1                & 2             & 2              \\
04-06-2024    & 1        & 2       & 1         & 1              & 1                & 2             & 2              \\
12-06-2024    & 0        & 2       & 0         & 1              & 1                & 2             & 2              \\
18-06-2024    & 0        & 2       & 0         & 2              & 2                & 2             & 2              \\
26-06-2024    & 1        & 2       & 1         & 2              & 2                & 2             & 2              \\
03-07-2024    & 0        & 2       & 0         & 2              & 2                & 3             & 3              \\
11-07-2024    & 2        & 2       & 2         & 2              & 2                & 3             & 3              \\
18-07-2024    & 1        & 2       & 1         & 1              & 1                & 3             & 3              \\
25-07-2024    & 2        & 2       & 2         & 2              & 2                & 0             & 0              \\
01-08-2024    & 2        & 1       & 2         & 3              & 3                & 0             & 0              \\
08-08-2024    & 2        & 2       & 2         & 3              & 3                & 0             & 0              \\
15-08-2024    & 2        & 2       & 2         & 3              & 3                & 0             & 0              \\
23-08-2024    & 2        & 2       & 2         & 2              & 2                & 0             & 0              \\
30-08-2024    & 2        & 2       & 2         & 3              & 3                & 0             & 0              \\
08-09-2024    & 1        & 2       & 1         & 1              & 1                & 0             & 0              \\
16-09-2024    & 2        & 2       & 2         & 3              & 3                & 0             & 0              \\
23-09-2024    & 3        & 2       & 3         & 3              & 3                & 3             & 3              \\
30-09-2024    & 3        & 2       & 3         & 3              & 3                & 3             & 3              \\
08-10-2024    & 3        & 2       & 3         & 3              & 3                & 3             & 3              \\
15-10-2024    & 2        & 2       & 1         & 2              & 2                & 1             & 1              \\
21-10-2024    & 3        & 2       & 3         & 3              & 3                & 3             & 3              \\
28-10-2024    & 0        & 2       & 0         & 0              & 0                & 2             & 2              \\
04-11-2024    & 3        & 2       & 3         & 2              & 2                & 3             & 3             \\ \hline
\end{tabular}
\caption{The number of successful experimental runs per condition per week.}
\label{table:experiment_counts}
\end{table}

\begin{table}[htbp!]
\centering
\footnotesize
\begin{tabular}{llccccc}
\textbf{State}             & \textbf{Leaning}            & \textbf{\begin{tabular}[c]{@{}c@{}}Is content\\ poltical?\end{tabular}} & \textbf{\begin{tabular}[c]{@{}c@{}}If political, is\\ content about\\ election and/or\\ candidates?\end{tabular}} & \textbf{\begin{tabular}[c]{@{}c@{}}Ideological\\ stance\end{tabular}} & \textbf{\begin{tabular}[c]{@{}c@{}}If political,\\ Stance \%\end{tabular}} & \textbf{\begin{tabular}[c]{@{}c@{}}If political and\\ about election\\ and/or candidates,\\ Stance \%\end{tabular}} \\ \hline
\multirow{15}{*}{Georgia}  & \multirow{5}{*}{Democrat}   & \multirow{5}{*}{32.0\%}                                                 & \multirow{5}{*}{38.9\%}                                                                                           & Neutral                                                               & 43.0\%                                                                     & 14.7\%                                                                                                              \\
                           &                             &                                                                         &                                                                                                                   & Anti Republican                                                       & 22.9\%                                                                     & 36.8\%                                                                                                              \\
                           &                             &                                                                         &                                                                                                                   & Pro Democrat                                                          & 16.8\%                                                                     & 27.2\%                                                                                                              \\
                           &                             &                                                                         &                                                                                                                   & Anti Democrat                                                         & 13.8\%                                                                     & 16.0\%                                                                                                              \\
                           &                             &                                                                         &                                                                                                                   & Pro Republican                                                        & 3.5\%                                                                      & 5.3\%                                                                                                               \\ \cline{2-7}
                           & \multirow{5}{*}{Neutral}    & \multirow{5}{*}{3.8\%}                                                  & \multirow{5}{*}{5.3\%}                                                                                            & Neutral                                                               & 68.6\%                                                                     & 23.8\%                                                                                                              \\
                           &                             &                                                                         &                                                                                                                   & Anti Democrat                                                         & 17.9\%                                                                     & 29.5\%                                                                                                              \\
                           &                             &                                                                         &                                                                                                                   & Pro Republican                                                        & 6.1\%                                                                      & 22.9\%                                                                                                              \\
                           &                             &                                                                         &                                                                                                                   & Anti Republican                                                       & 4.4\%                                                                      & 13.3\%                                                                                                              \\
                           &                             &                                                                         &                                                                                                                   & Pro Democrat                                                          & 2.9\%                                                                      & 10.5\%                                                                                                              \\\cline{2-7}
                           & \multirow{5}{*}{Republican} & \multirow{5}{*}{29.1\%}                                                 & \multirow{5}{*}{39.4\%}                                                                                           & Anti Democrat                                                         & 42.0\%                                                                     & 51.6\%                                                                                                              \\
                           &                             &                                                                         &                                                                                                                   & Neutral                                                               & 33.6\%                                                                     & 11.5\%                                                                                                              \\
                           &                             &                                                                         &                                                                                                                   & Pro Republican                                                        & 16.7\%                                                                     & 26.1\%                                                                                                              \\
                           &                             &                                                                         &                                                                                                                   & Anti Republican                                                       & 5.6\%                                                                      & 7.6\%                                                                                                               \\
                           &                             &                                                                         &                                                                                                                   & Pro Democrat                                                          & 2.1\%                                                                      & 3.1\%                                                                                                               \\\hline
\multirow{10}{*}{New York} & \multirow{5}{*}{Democrat}   & \multirow{5}{*}{26.5\%}                                                 & \multirow{5}{*}{34.1\%}                                                                                           & Neutral                                                               & 46.7\%                                                                     & 16.6\%                                                                                                              \\
                           &                             &                                                                         &                                                                                                                   & Anti Republican                                                       & 23.2\%                                                                     & 40.2\%                                                                                                              \\
                           &                             &                                                                         &                                                                                                                   & Pro Democrat                                                          & 13.8\%                                                                     & 23.3\%                                                                                                              \\
                           &                             &                                                                         &                                                                                                                   & Anti Democrat                                                         & 13.5\%                                                                     & 14.5\%                                                                                                              \\
                           &                             &                                                                         &                                                                                                                   & Pro Republican                                                        & 2.9\%                                                                      & 5.3\%                                                                                                               \\\cline{2-7}
                           & \multirow{5}{*}{Republican} & \multirow{5}{*}{21.7\%}                                                 & \multirow{5}{*}{32.4\%}                                                                                           & Neutral                                                               & 39.3\%                                                                     & 14.0\%                                                                                                              \\
                           &                             &                                                                         &                                                                                                                   & Anti Democrat                                                         & 38.3\%                                                                     & 50.4\%                                                                                                              \\
                           &                             &                                                                         &                                                                                                                   & Pro Republican                                                        & 13.3\%                                                                     & 20.8\%                                                                                                              \\
                           &                             &                                                                         &                                                                                                                   & Anti Republican                                                       & 6.2\%                                                                      & 10.3\%                                                                                                              \\
                           &                             &                                                                         &                                                                                                                   & Pro Democrat                                                          & 2.9\%                                                                      & 4.5\%                                                                                                               \\\hline
\multirow{10}{*}{Texas}    & \multirow{5}{*}{Democrat}   & \multirow{5}{*}{28.1\%}                                                 & \multirow{5}{*}{31.8\%}                                                                                           & Neutral                                                               & 47.3\%                                                                     & 16.8\%                                                                                                              \\
                           &                             &                                                                         &                                                                                                                   & Anti Republican                                                       & 21.8\%                                                                     & 38.2\%                                                                                                              \\
                           &                             &                                                                         &                                                                                                                   & Anti Democrat                                                         & 15.8\%                                                                     & 17.8\%                                                                                                              \\
                           &                             &                                                                         &                                                                                                                   & Pro Democrat                                                          & 12.4\%                                                                     & 22.1\%                                                                                                              \\
                           &                             &                                                                         &                                                                                                                   & Pro Republican                                                        & 2.6\%                                                                      & 5.1\%                                                                                                               \\\cline{2-7}
                           & \multirow{5}{*}{Republican} & \multirow{5}{*}{24.9\%}                                                 & \multirow{5}{*}{35.3\%}                                                                                           & Anti Democrat                                                         & 41.5\%                                                                     & 54.1\%                                                                                                              \\
                           &                             &                                                                         &                                                                                                                   & Neutral                                                               & 34.9\%                                                                     & 11.3\%                                                                                                              \\
                           &                             &                                                                         &                                                                                                                   & Pro Republican                                                        & 15.9\%                                                                     & 24.2\%                                                                                                              \\
                           &                             &                                                                         &                                                                                                                   & Anti Republican                                                       & 5.6\%                                                                      & 6.9\%                                                                                                               \\
                           &                             &                                                                         &                                                                                                                   & Pro Democrat                                                          & 2.1\%                                                                      & 3.6\%   \\ \hline                                                                                                           
\end{tabular}
\caption{For each experimental condition, the proportion of videos that are political, and of those, the proportion of videos that pertain to the US elections or major political candidates. For each set of videos, the distribution of their ideological stances.}
\label{tab:sup_table_12}
\end{table}

\begin{table}[htbp!]
\centering
\begin{tabular}{l|ccc|cc|cc}
\textbf{Month}   & \multicolumn{3}{c}{\textbf{Georgia}} & \multicolumn{2}{c}{\textbf{New York}} & \multicolumn{2}{c}{\textbf{Texas}} \\ \hline
                 & Neutral      & Dem.       & Rep.     & Dem.               & Rep.             & Dem.             & Rep.            \\ \hline
May              & 0.04         & -0.07      & 0.23     & 0.03               & 0.12             & 0.02             & 0.23            \\
June             & 0.05         & -0.11      & 0.21     & -0.05              & 0.15             & 0.03             & 0.24            \\
July             & 0.08         & -0.06      & 0.24     & 0.03               & 0.21             & -0.01            & 0.3             \\
August           & 0.04         & -0.13      & 0.28     & -0.1               & 0.2              & $\sim$               & $\sim$              \\
September        & 0.05         & -0.2       & 0.44     & -0.25              & 0.36             & -0.21            & 0.39            \\
October          & 0.04         & -0.28      & 0.44     & -0.2               & 0.38             & -0.2             & 0.44            \\
November         & 0.09         & -0.12      & 0.52     & -0.06              & 0.4              & -0.05            & 0.46            \\ \hline
\textbf{Overall} & 0.05         & -0.13      & 0.37     & -0.08              & 0.26             & -0.06            & 0.33    \\ \hline       
\end{tabular}
\caption{The mean ideological content seen by bots of different conditioning in each state over time.}
\label{tab:sup_table_13}
\end{table}

\begin{table}[htbp!]
\centering
\small
\begin{tabular}{l|ccc|c|c}
                 & \multicolumn{3}{c}{\textbf{Georgia}}                         & \textbf{New York}      & \textbf{Texas}         \\ \hline
                 & Rep. vs. Neutral & Dem. vs. Neutral & Rep. vs. Dem. & Rep. vs. Dem. & Rep. vs. Dem. \\ \hline
May              & 5.87***          & -0.16            & 1.65          & 3.39**        & 5.62***       \\
June             & 4.41**           & 2.05             & 1.34          & 2.07          & 5.89***       \\
July             & 5.41***          & -0.87            & 2.93*         & 6.29***       & 9.99***       \\
August           & 8.48***          & 2.44*            & 3.73**        & 3.35**        & $\sim$        \\
September        & 7.01***          & 4.81***          & 4.1***        & 2.25*         & 4.19**        \\
October          & 6.27***          & 4.69***          & 1.98          & 3.06**        & 4.91***       \\
November         & 5.81*            & 0.15             & 3.07*         & 3.71          & 4.06*         \\ \hline
\textbf{Overall} & 13.22***         & 4.7***           & 5.83***       & 5.83***       & 9.21***      \\ \hline
\end{tabular}
\caption{Independent t-test results comparing ideological content of bots of different conditioning in each state over time. ($*$ : $p < 0.05$, $**$ : $p < 0.01$, $***$ : $p < 0.001$)}
\label{tab:sup_table_14}
\end{table}

\begin{table}[htbp!]
\centering
\tiny
\begin{tabular}{lcc}
\textbf{Counterfactual model attribute (Recency scaling)} & \textbf{Ideological skew} & \textbf{\begin{tabular}[c]{@{}c@{}}Independent t-test\\ Expected vs. Observed\end{tabular}} \\ \hline
Comment Count (Exponential)                               & -0.024                    & -11.922***                                                                                  \\
Comment Count (Linear)                                    & -0.037                    & -12.479***                                                                                  \\
Comment Count (No scaling)                                & -0.009                    & -11.191***                                                                                  \\
Comments per recommendation (Exponential)                 & -0.05                     & -12.972***                                                                                  \\
Comments per recommendation (Linear)                      & -0.059                    & -13.373***                                                                                  \\
Comments per recommendation (No scaling)                  & -0.04                     & -12.403***                                                                                  \\
Combined Author Score (Exponential)                       & -0.059                    & -13.343***                                                                                  \\
Combined Author Score (Linear)                            & -0.065                    & -13.607***                                                                                  \\
Combined Author Score (No scaling)                        & -0.052                    & -13.067***                                                                                  \\
Combined Likes, Shares, Plays Score (Exponential)         & 0.008                     & -10.162***                                                                                  \\
Combined Likes, Shares, Plays Score (Linear)              & -0.005                    & -10.884***                                                                                  \\
Combined Likes, Shares, Plays Score (No scaling)          & 0.032                     & -8.949***                                                                                   \\
Combined Video Score (Exponential)                        & -0.018                    & -11.63***                                                                                   \\
Combined Video Score (Linear)                             & -0.035                    & -12.36***                                                                                   \\
Combined Video Score (No scaling)                         & -0.001                    & -10.801***                                                                                  \\
Like Count (Exponential)                                  & 0.002                     & -10.438***                                                                                  \\
Like Count (Linear)                                       & -0.01                     & -11.126***                                                                                  \\
Like Count (No scaling)                                   & 0.027                     & -9.238***                                                                                   \\
Video Length (Exponential)                                & -0.068                    & -13.926***                                                                                  \\
Video Length (Linear)                                     & -0.08                     & -14.356***                                                                                  \\
Video Length (No scaling)                                 & -0.07                     & -14.111***                                                                                  \\
(Likes + Comments + Shares) / Plays (Exponential)         & -0.029                    & -12.142***                                                                                  \\
(Likes + Comments + Shares) / Plays (Linear)              & -0.043                    & -12.788***                                                                                  \\
(Likes + Comments + Shares) / Plays (No scaling)          & -0.009                    & -11.263***                                                                                  \\
Channel followers (Exponential)                           & -0.031                    & -12.282***                                                                                  \\
Channel followers (Linear)                                & -0.042                    & -12.672***                                                                                  \\
Channel followers (No scaling)                            & -0.015                    & -11.479***                                                                                  \\
Full Combined score (Exponential)                         & -0.051                    & -13.12***                                                                                   \\
Full Combined score (Linear)                              & -0.061                    & -13.55***                                                                                   \\
Full Combined score (No scaling)                          & -0.044                    & -12.815***                                                                                  \\
Channel cumul. likes (Exponential)                        & -0.05                     & -13.15***                                                                                   \\
Channel cumul. likes (Linear)                             & -0.057                    & -13.321***                                                                                  \\
Channel cumul. likes (No scaling)                         & -0.042                    & -12.925***                                                                                  \\
Likes per recommendation (Exponential)                    & -0.028                    & -12.114***                                                                                  \\
Likes per recommendation (Linear)                         & -0.042                    & -12.796***                                                                                  \\
Likes per recommendation (No scaling)                     & -0.009                    & -11.243***                                                                                  \\
Video plays (Exponential)                                 & 0.01                      & -9.976***                                                                                   \\
Video plays (Linear)                                      & -0.004                    & -10.773***                                                                                  \\
Video plays (No scaling)                                  & 0.036                     & -8.714***                                                                                   \\
Video shares (Exponential)                                & 0.006                     & -10.341***                                                                                  \\
Video shares (Linear)                                     & -0.005                    & -10.972***                                                                                  \\
Video shares (No scaling)                                 & 0.029                     & -9.242***                                                                                   \\
Channel video count (Exponential)                         & -0.042                    & -12.728***                                                                                  \\
Channel video count (Linear)                              & -0.052                    & -13.077***                                                                                  \\
Channel video count (No scaling)                          & -0.03                     & -12.294***                                                                                  \\
Channel verified (Exponential)                            & -0.059                    & -11.519***                                                                                  \\
Channel verified (Linear)                                 & -0.065                    & -11.803***                                                                                  \\
Channel verified (No scaling)                             & -0.05                     & -11.138***                                                                                  \\
Party-aligned comment proportion (Exponential)            & -0.05                     & -9.771***                                                                                   \\
Party-aligned comment proportion (Linear)                 & -0.052                    & -9.84***                                                                                    \\
Party-aligned comment proportion (No scaling)             & -0.049                    & -9.735***                                                                                   \\
Opposition comment proportion (Exponential)               & -0.051                    & -8.608***                                                                                   \\
Opposition comment proportion (Linear)                    & -0.051                    & -8.622***                                                                                   \\
Opposition comment proportion (No scaling)                & -0.051                    & -8.623***             \\ \hline                                                                      
\textbf{Observed }                                & \textbf{0.204 }           &    \\ \hline                                                                
\end{tabular}
\caption{The ideological skew observed in the experiments, as well as the ideological skew computed through counterfactual models which sample videos based on a given attribute. The right-most column denotes independent t-test results between the expected ideological skew and that observed by the bots. ($*$ : $p < 0.05$, $**$ : $p < 0.01$, $***$ : $p < 0.001$)}
\label{tab:sup_table_15}
\end{table}

\begin{table}[htbp!]
\centering
\tiny
\begin{tabular}{lcc}
\textbf{Counterfactual model attribute (Recency scaling)} & \textbf{Ideological skew} & \textbf{\begin{tabular}[c]{@{}c@{}}Independent t-test\\ Expected vs. Observed\end{tabular}} \\ \hline
Comment Count (Exponential)                               & 0.027                     & 0.626                                                                                       \\
Comment Count (Linear)                                    & 0.017                     & 0.061                                                                                       \\
Comment Count (No scaling)                                & 0.071                     & 3.364**                                                                                     \\
Comments per recommendation (Exponential)                 & -0.062                    & -5.025***                                                                                   \\
Comments per recommendation (Linear)                      & -0.067                    & -5.312***                                                                                   \\
Comments per recommendation (No scaling)                  & -0.046                    & -4.36***                                                                                    \\
Combined Author Score (Exponential)                       & -0.078                    & -5.762***                                                                                   \\
Combined Author Score (Linear)                            & -0.078                    & -6.041***                                                                                   \\
Combined Author Score (No scaling)                        & -0.069                    & -5.37***                                                                                    \\
Combined Likes, Shares, Plays Score (Exponential)         & 0.043                     & 1.418                                                                                       \\
Combined Likes, Shares, Plays Score (Linear)              & 0.038                     & 1.122                                                                                       \\
Combined Likes, Shares, Plays Score (No scaling)          & 0.089                     & 4.159***                                                                                    \\
Combined Video Score (Exponential)                        & 0.003                     & -0.788                                                                                      \\
Combined Video Score (Linear)                             & -0.022                    & -2.227*                                                                                     \\
Combined Video Score (No scaling)                         & 0.03                      & 1.053                                                                                       \\
Like Count (Exponential)                                  & 0.034                     & 0.965                                                                                       \\
Like Count (Linear)                                       & 0.029                     & 0.652                                                                                       \\
Like Count (No scaling)                                   & 0.079                     & 3.577***                                                                                    \\
Video Length (Exponential)                                & -0.065                    & -5.162***                                                                                   \\
Video Length (Linear)                                     & -0.082                    & -5.796***                                                                                   \\
Video Length (No scaling)                                 & -0.068                    & -6.513***                                                                                   \\
(Likes + Comments + Shares) / Plays (Exponential)         & -0.047                    & -4.074***                                                                                   \\
(Likes + Comments + Shares) / Plays (Linear)              & -0.06                     & -4.636***                                                                                   \\
(Likes + Comments + Shares) / Plays (No scaling)          & -0.031                    & -3.711***                                                                                   \\
Channel followers (Exponential)                           & -0.014                    & -2.082*                                                                                     \\
Channel followers (Linear)                                & -0.028                    & -3.031**                                                                                    \\
Channel followers (No scaling)                            & 0.016                     & 0.015                                                                                       \\
Full Combined score (Exponential)                         & -0.068                    & -5.339***                                                                                   \\
Full Combined score (Linear)                              & -0.073                    & -5.758***                                                                                   \\
Full Combined score (No scaling)                          & -0.058                    & -5.057***                                                                                   \\
Channel cumul. likes (Exponential)                        & -0.019                    & -2.206*                                                                                     \\
Channel cumul. likes (Linear)                             & -0.033                    & -3.2**                                                                                      \\
Channel cumul. likes (No scaling)                         & 0.008                     & -0.591                                                                                      \\
Likes per recommendation (Exponential)                    & -0.046                    & -3.989***                                                                                   \\
Likes per recommendation (Linear)                         & -0.059                    & -4.602***                                                                                   \\
Likes per recommendation (No scaling)                     & -0.034                    & -3.938***                                                                                   \\
Video plays (Exponential)                                 & 0.052                     & 1.87                                                                                        \\
Video plays (Linear)                                      & 0.048                     & 1.575                                                                                       \\
Video plays (No scaling)                                  & 0.102                     & 4.728***                                                                                    \\
Video shares (Exponential)                                & 0.042                     & 1.329                                                                                       \\
Video shares (Linear)                                     & 0.035                     & 0.984                                                                                       \\
Video shares (No scaling)                                 & 0.081                     & 3.841***                                                                                    \\
Channel video count (Exponential)                         & 0.025                     & 0.511                                                                                       \\
Channel video count (Linear)                              & 0.005                     & -0.703                                                                                      \\
Channel video count (No scaling)                          & 0.058                     & 2.9**                                                                                       \\
Channel verified (Exponential)                            & -0.074                    & -5.459***                                                                                   \\
Channel verified (Linear)                                 & -0.081                    & -5.941***                                                                                   \\
Channel verified (No scaling)                             & -0.066                    & -5.011***                                                                                   \\
Party-aligned comment proportion (Exponential)            & 0.018                     & 0.119                                                                                       \\
Party-aligned comment proportion (Linear)                 & 0.001                     & -0.874                                                                                      \\
Party-aligned comment proportion (No scaling)             & 0.02                      & 0.245                                                                                       \\
Opposition comment proportion (Exponential)               & -0.016                    & -1.62                                                                                       \\
Opposition comment proportion (Linear)                    & -0.022                    & -1.963                                                                                      \\
Opposition comment proportion (No scaling)                & -0.016                    & -1.653                                                                                     \\ \hline
\textbf{Observed}                                 & \textbf{0.0159}           &                                                                \\ \hline                   
\end{tabular}
\caption{The ideological skew observed in the experiments, as well as the ideological skew computed through counterfactual models which sample videos based on a given attribute when only considering positive-partisanship videos (Pro Democrat or Pro Republican). The right-most column denotes independent t-test results between the expected ideological skew and that observed by the bots. ($*$ : $p < 0.05$, $**$ : $p < 0.01$, $***$ : $p < 0.001$)}
\label{tab:sup_table_16}
\end{table}

\begin{table}[htbp!]
\centering
\tiny
\begin{tabular}{lcc}
\textbf{Counterfactual model attribute (Recency scaling)} & \textbf{Ideological skew} & \textbf{\begin{tabular}[c]{@{}c@{}}Independent t-test\\ Expected vs. Observed\end{tabular}} \\ \hline
Comment Count (Exponential)                               & -0.033                    & -17.873***                                                                                  \\
Comment Count (Linear)                                    & -0.046                    & -17.981***                                                                                  \\
Comment Count (No scaling)                                & -0.027                    & -17.39***                                                                                   \\
Comments per recommendation (Exponential)                 & -0.045                    & -18.194***                                                                                  \\
Comments per recommendation (Linear)                      & -0.054                    & -18.326***                                                                                  \\
Comments per recommendation (No scaling)                  & -0.039                    & -17.763***                                                                                  \\
Combined Author Score (Exponential)                       & -0.052                    & -18.875***                                                                                  \\
Combined Author Score (Linear)                            & -0.058                    & -19.009***                                                                                  \\
Combined Author Score (No scaling)                        & -0.048                    & -19.066***                                                                                  \\
Combined Likes, Shares, Plays Score (Exponential)         & 0.002                     & -15.192***                                                                                  \\
Combined Likes, Shares, Plays Score (Linear)              & -0.012                    & -16.356***                                                                                  \\
Combined Likes, Shares, Plays Score (No scaling)          & 0.021                     & -14.131***                                                                                  \\
Combined Video Score (Exponential)                        & -0.02                     & -17.043***                                                                                  \\
Combined Video Score (Linear)                             & -0.034                    & -17.445***                                                                                  \\
Combined Video Score (No scaling)                         & -0.006                    & -16.478***                                                                                  \\
Like Count (Exponential)                                  & -0.002                    & -15.511***                                                                                  \\
Like Count (Linear)                                       & -0.015                    & -16.654***                                                                                  \\
Like Count (No scaling)                                   & 0.016                     & -14.501***                                                                                  \\
Video Length (Exponential)                                & -0.065                    & -19.538***                                                                                  \\
Video Length (Linear)                                     & -0.075                    & -19.655***                                                                                  \\
Video Length (No scaling)                                 & -0.068                    & -20.294***                                                                                  \\
(Likes + Comments + Shares) / Plays (Exponential)         & -0.022                    & -17.163***                                                                                  \\
(Likes + Comments + Shares) / Plays (Linear)              & -0.035                    & -17.954***                                                                                  \\
(Likes + Comments + Shares) / Plays (No scaling)          & -0.004                    & -16.426***                                                                                  \\
Channel followers (Exponential)                           & -0.033                    & -18.069***                                                                                  \\
Channel followers (Linear)                                & -0.041                    & -18.175***                                                                                  \\
Channel followers (No scaling)                            & -0.021                    & -17.557***                                                                                  \\
Full Combined score (Exponential)                         & -0.045                    & -18.723***                                                                                  \\
Full Combined score (Linear)                              & -0.054                    & -19.052***                                                                                  \\
Full Combined score (No scaling)                          & -0.04                     & -18.874***                                                                                  \\
Channel cumul. likes (Exponential)                        & -0.054                    & -19.356***                                                                                  \\
Channel cumul. likes (Linear)                             & -0.059                    & -19.191***                                                                                  \\
Channel cumul. likes (No scaling)                         & -0.054                    & -19.995***                                                                                  \\
Likes per recommendation (Exponential)                    & -0.022                    & -17.15***                                                                                   \\
Likes per recommendation (Linear)                         & -0.035                    & -18.002***                                                                                  \\
Likes per recommendation (No scaling)                     & -0.002                    & -16.342***                                                                                  \\
Video plays (Exponential)                                 & 0.003                     & -15.049***                                                                                  \\
Video plays (Linear)                                      & -0.012                    & -16.294***                                                                                  \\
Video plays (No scaling)                                  & 0.022                     & -13.918***                                                                                  \\
Video shares (Exponential)                                & 0.001                     & -15.407***                                                                                  \\
Video shares (Linear)                                     & -0.011                    & -16.346***                                                                                  \\
Video shares (No scaling)                                 & 0.018                     & -14.478***                                                                                  \\
Channel video count (Exponential)                         & -0.054                    & -19.207***                                                                                  \\
Channel video count (Linear)                              & -0.061                    & -18.856***                                                                                  \\
Channel video count (No scaling)                          & -0.051                    & -19.586***                                                                                  \\
Channel verified (Exponential)                            & -0.052                    & -16.929***                                                                                  \\
Channel verified (Linear)                                 & -0.057                    & -17.276***                                                                                  \\
Channel verified (No scaling)                             & -0.044                    & -16.432***                                                                                  \\
Party-aligned comment proportion (Exponential)            & -0.068                    & -15.53***                                                                                   \\
Party-aligned comment proportion (Linear)                 & -0.064                    & -15.275***                                                                                  \\
Party-aligned comment proportion (No scaling)             & -0.068                    & -15.431***                                                                                  \\
Opposition comment proportion (Exponential)               & -0.063                    & -13.608***                                                                                  \\
Opposition comment proportion (Linear)                    & -0.06                     & -13.466***                                                                                  \\
Opposition comment proportion (No scaling)                & -0.064                    & -13.646*** \\ \hline
\textbf{Observed}                        & \textbf{0.1882}  &                                                                                   \\ \hline
\end{tabular}
\caption{The ideological skew observed in the experiments, as well as the ideological skew computed through counterfactual models which sample videos based on a given attribute when only considering negative-partisanship videos (Anti Democrat or Anti Republican). The right-most column denotes independent t-test results between the expected ideological skew and that observed by the bots. ($*$ : $p < 0.05$, $**$ : $p < 0.01$, $***$ : $p < 0.001$)}
\label{tab:sup_table_17}
\end{table}

\begin{table}[htbp!]
\centering
\scriptsize
\begin{tabular}{lcc}
\begin{tabular}[c]{@{}l@{}}Counterfactual model based\\ on weight of Channel Verification status\\(Recency scaling)\end{tabular} & Ideological skew & \begin{tabular}[c]{@{}c@{}}Independent t-test\\ Expected vs. Observed\end{tabular} \\ \hline
Weight = 0.55 (Exponential) & -0.058           & -11.238***                                                                         \\
Weight = 0.55 (Linear)      & -0.065           & -11.549***                                                                         \\
Weight = 0.55 (No scaling)  & -0.049           & -10.881***                                                                         \\ \hline
Weight = 0.6 (Exponential)  & -0.058           & -11.25***                                                                          \\
Weight = 0.6 (Linear)       & -0.065           & -11.553***                                                                         \\
Weight = 0.6 (No scaling)   & -0.05            & -10.888***                                                                         \\\hline
Weight = 0.65 (Exponential) & -0.059           & -11.259***                                                                         \\
Weight = 0.65 (Linear)      & -0.065           & -11.537***                                                                         \\
Weight = 0.65 (No scaling)  & -0.05            & -10.9***                                                                           \\\hline
Weight = 0.7 (Exponential)  & -0.058           & -11.244***                                                                         \\
Weight = 0.7 (Linear)       & -0.065           & -11.558***                                                                         \\
Weight = 0.7 (No scaling)   & -0.049           & -10.87***                                                                          \\\hline
Weight = 0.75 (Exponential) & -0.058           & -11.237***                                                                         \\
Weight = 0.75 (Linear)      & -0.065           & -11.546***                                                                         \\
Weight = 0.75 (No scaling)  & -0.049           & -10.873***                                                                         \\\hline
Weight = 0.8 (Exponential)  & -0.058           & -11.238***                                                                         \\
Weight = 0.8 (Linear)       & -0.065           & -11.557***                                                                         \\
Weight = 0.8 (No scaling)   & -0.05            & -10.908***                                                                         \\\hline
Weight = 0.85 (Exponential) & -0.058           & -11.248***                                                                         \\
Weight = 0.85 (Linear)      & -0.065           & -11.532***                                                                         \\
Weight = 0.85 (No scaling)  & -0.049           & -10.879***                                                                         \\\hline
Weight = 0.9 (Exponential)  & -0.058           & -11.237***                                                                         \\
Weight = 0.9 (Linear)       & -0.065           & -11.552***                                                                         \\
Weight = 0.9 (No scaling)   & -0.049           & -10.87***                                                                          \\\hline
Weight = 0.95 (Exponential) & -0.058           & -11.243***                                                                         \\
Weight = 0.95 (Linear)      & -0.065           & -11.548***                                                                         \\
Weight = 0.95 (No scaling)  & -0.05            & -10.905***                                                                         \\\hline
Weight = 1.0 (Exponential)  & -0.058           & -11.256***                                                                         \\
Weight = 1.0 (Linear)       & -0.065           & -11.548***                                                                         \\
Weight = 1.0 (No scaling)   & -0.05            & -10.89***  \\\hline
\textbf{Observed}                                                                                             & \textbf{0.204}   &          \\ \hline                                                                         
\end{tabular}
\caption{The ideological skew observed in the experiments, as well as the ideological skew computed through counterfactual models which sample videos based on channel verification status with varying weights given to a channels verification status. The right-most column denotes independent t-test results between the expected ideological skew and that observed by the bots. ($*$ : $p < 0.05$, $**$ : $p < 0.01$, $***$ : $p < 0.001$)}
\label{table:ideological_skew_channel_verification}
\end{table}

\begin{table}[htbp!]
\centering
\begin{tabular}{lc|lc}
Republican hashtag      & Count & Democrat hashtag         & Count \\ \hline
foryou                  & 173   & jamaalbowman             & 163   \\
fyp :)                  & 173   & therentistoodamnhigh     & 163   \\
usa                     & 173   & donaldtrump              & 159   \\
trump2024               & 173   & supremecourt             & 158   \\
country                 & 173   & immigration              & 157   \\
america                 & 173   & trump                    & 155   \\
fyp                     & 173   & moreteachinglesstesting  & 155   \\
trump                   & 173   & emmys                    & 155   \\
politics                & 172   & jonstewart               & 155   \\
biden                   & 170   & theproblem               & 155   \\
fypdoesntwork           & 168   & theproblemwithjonstewart & 154   \\
republican              & 168   & transgenderrights        & 154   \\
military                & 168   & fyp                      & 154   \\
backtheblue & 168   & climatechange            & 153   \\
facts                   & 168   & climate                  & 153   \\
opinion                 & 166   & migrantes                & 153   \\
w                       & 166   & scotus                   & 153   \\
president               & 166   & appletv                  & 153   \\
4u                      & 166   & tiktok                   & 153   \\
conservative            & 160   & strictscrutiny           & 153   \\
duet                    & 160   & law                      & 153   \\
army                    & 159   & breakingnews             & 153   \\
police                  & 159   & biden                    & 153   \\
stitch                  & 155   & elections                & 153   \\
donaldtrump             & 154   & outstandingtalkseries    & 153   \\
tucker                  & 153   & peace                    & 153   \\
consantanderconecto     & 153   & podcast                  & 153   \\
tuckercarlson           & 153   & globalwarming            & 153   \\
tiktokban               & 152   & fyc                      & 153   \\
reaction                & 149   & economy                  & 152   \\ \hline
\end{tabular}
\caption{The 20 most common hashtags for Democrat and Republican conditioning videos.}
\label{tab:sup_table_19}
\end{table}

\begin{table}[htbp!]
\centering
\footnotesize
\begin{tabular}{lcccc}
Topic                                                                                                                                                  & \begin{tabular}[c]{@{}c@{}}Prop. of\\ Dem. videos\end{tabular} & \begin{tabular}[c]{@{}c@{}}Prop. of\\ Rep. videos\end{tabular} & Difference & $\chi^2$ test \\ \hline
Abortion and reproductive health                                                                                                                       & 6.6                                                            & 1.5                                                            & 5.1        & 172.0***         \\
Government or politics generally                                                                                                                       & 43.9                                                           & 42.3                                                           & 1.6        & 1.02             \\
Biden dropping out of the presidential race                                                                                                            & 6.6                                                            & 5.2                                                            & 1.4        & 7.41**           \\
Climate change                                                                                                                                         & 1.6                                                            & 0.6                                                            & 1.0        & 25.78***         \\
Education                                                                                                                                              & 1.8                                                            & 1.2                                                            & 0.6        & 6.66**           \\
Economy generally                                                                                                                                      & 6.2                                                            & 5.7                                                            & 0.5        & 1.08             \\
\begin{tabular}[c]{@{}l@{}}Racial issues, including affirmative action\\ and racial discrimination\end{tabular}                                        & 5.8                                                            & 5.4                                                            & 0.5        & 0.87             \\
Other public health issues                                                                                                                             & 1.4                                                            & 1.0                                                            & 0.5        & 4.98*            \\
Democratic National Convention (DNC)                                                                                                                   & 1.1                                                            & 0.7                                                            & 0.4        & 5.39*            \\
Guns and gun control                                                                                                                                   & 1.6                                                            & 1.2                                                            & 0.4        & 2.75             \\
Environment generally                                                                                                                                  & 0.6                                                            & 0.4                                                            & 0.2        & 1.73             \\
LGBTQ+ issues, including transgender issues                                                                                                            & 2.4                                                            & 2.5                                                            & -0.0       & 0.0              \\
Republican National Convention (RNC)                                                                                                                   & 0.4                                                            & 0.4                                                            & 0.0        & 0.0              \\
AI, LLMs                                                                                                                                               & 0.1                                                            & 0.2                                                            & -0.0       & 0.07             \\
Crypto                                                                                                                                                 & 0.1                                                            & 0.1                                                            & -0.0       & 0.0              \\
Other vaccines                                                                                                                                         & 0.0                                                            & 0.2                                                            & -0.1       & 3.15             \\
Covid, including covid vaccines                                                                                                                        & 0.8                                                            & 1.0                                                            & -0.2       & 0.74             \\
Other technology issues                                                                                                                                & 0.3                                                            & 0.6                                                            & -0.3       & 3.48             \\
Assassination attempt on Donald Trump                                                                                                                  & 1.1                                                            & 1.5                                                            & -0.4       & 3.13             \\
Crime generally                                                                                                                                        & 3.6                                                            & 4.4                                                            & -0.8       & 3.62             \\
Other                                                                                                                                                  & 0.2                                                            & 1.2                                                            & -1.0       & 30.38***         \\
\begin{tabular}[c]{@{}l@{}}Other social issues, including culture war issues,\\ labor, and other social issues that are not covered above\end{tabular} & 5.7                                                            & 6.9                                                            & -1.2       & 5.55*            \\
Ukraine war                                                                                                                                            & 0.4                                                            & 1.5                                                            & -1.2       & 33.14***         \\
\begin{tabular}[c]{@{}l@{}}Israel, Gaza or Palestine, including anything\\ about Netanyahu or Hamas\end{tabular}                                       & 2.3                                                            & 4.3                                                            & -2.0       & 29.21***         \\
Immigration                                                                                                                                            & 3.5                                                            & 5.7                                                            & -2.2       & 24.66***         \\
\begin{tabular}[c]{@{}l@{}}Anything outside the US or involve US foreign\\ relations except for Israel, Gaza, Ukraine,\\ or immigration\end{tabular}   & 1.8                                                            & 4.6                                                            & -2.8       & 57.05***      \\ \hline  
\end{tabular}
\caption{The proportion of Democrat and Republican videos on a given topic, the difference in these proportions, and chi-squared tests comparing these proportions. ($*$ : $p < 0.05$, $**$ : $p < 0.01$, $***$ : $p < 0.001$)}
\label{tab:sup_table_20}
\end{table}

\begin{table}[htbp!]
\centering
\begin{tabular}{lc}
Variable                                               & VIF     \\ \hline
Conditioning stage video like count                    & 58.6463 \\
Conditioning stage video play count                    & 37.0054 \\
Conditioning stage video comment count                 & 26.2118 \\
Conditioning stage video share count                   & 13.3406 \\
Recommendation stage video play count                  & 12.2398 \\
Conditioning stage channel verification status         & 11.6664 \\
Conditioning stage channel video count                 & 10.6664 \\
Conditioning stage video like count                    & 10.2772 \\
Conditioning stage channel cumulative like count       & 8.5119  \\
Conditioning stage channel follower count              & 5.2342  \\
Recommendation stage channel cumulative like count     & 5.2128  \\
Recommendation stage video comment count               & 4.4883  \\
Recommendation stage channel verification status       & 4.1009  \\
Week of experiment                                     & 3.7490  \\
Recommendation stage channel video count               & 3.3335  \\
Recommendation stage channel follower count            & 3.1999  \\
Recommendation stage video share count                 & 3.0170  \\
Number of videos watched during conditioning stage     & 1.7269  \\
Recommendation stage video likes per recommendation    & 1.5127  \\
Recommendation stage video comments per recommendation & 1.4689  \\
Opposition comment proportion                          & 1.2853  \\
Video Duration                                         & 1.0427  \\\hline
\end{tabular}
\caption{VIF values for control variables in logistic regression considering all videos viewed during experiment.}
\label{tab:sup_table_21}
\end{table}

\begin{table}[htbp!]
\centering
\begin{tabular}{lc}
Variable                                               & VIF     \\ \hline
Conditioning stage video like count                    & 58.7064 \\
Conditioning stage video play count                    & 47.1637 \\
Recommendation stage video play count                  & 27.6304 \\
Conditioning stage video comment count                 & 25.0417 \\
Recommendation stage channel verification status       & 22.8050 \\
Conditioning stage video share count                   & 16.5838 \\
Recommendation stage video like count                  & 14.3400 \\
Conditioning stage channel verification status         & 13.4786 \\
Conditioning stage channel video count                 & 12.1204 \\
Conditioning stage channel cumulative like count       & 8.8994  \\
Recommendation stage video comment count               & 7.7629  \\
Conditioning stage channel follower count              & 5.7607  \\
Recommendation stage video share count                 & 5.1790  \\
Recommendation stage channel video count               & 2.4859  \\
Recommendation stage channel follower count            & 2.2943  \\
Recommendation stage channel cumulative like count     & 2.1489  \\
Week of experiment                                     & 1.8398  \\
Top account X Republican bot                           & 1.7770  \\
Top account X Democrat bot                             & 1.6151  \\
Number of videos watched during conditioning stage     & 1.6803  \\
Recommendation stage video likes per recommendation    & 1.3990  \\
Recommendation stage video comments per recommendation & 1.3096  \\ 
Opposition comment proportion                          & 1.2853  \\
Video Duration                                         & 1.0427  \\\hline

\end{tabular}
\caption{VIF values for control variables in logistic regression when considering top account status.}
\label{tab:sup_table_22}
\end{table}

\begin{table}[htbp!]
\centering
\footnotesize
\begin{tabular}{lccccc}
Variable                                & Coefficient & Odds Ratio & OR Lower CI & OR Upper CI & P-value                    \\ \hline
\textbf{Bot conditioning (ref = Rep)}   &             &            &             &             & \textit{}                  \\
Democrat                                & 1.0314      & 2.8051     & 2.0177      & 3.8997      & \textit{p \textless 0.001} \\\hline
\textbf{State (ref = NY)}               &             &            &             &             & \textit{}                  \\
Georgia                                 & -0.0148     & 0.9853     & 0.7867      & 1.234       & \textit{p = 0.8975}        \\
Texas                                   & 0.0524      & 1.0538     & 0.839       & 1.3236      & \textit{p = 0.6522}        \\\hline
\textbf{Time}                           &             &            &             &             & \textit{}                  \\
Week of experiment                      & -0.0477     & 0.9534     & 0.9376      & 0.9695      & \textit{p \textless 0.001} \\\hline
\textbf{Conditioning metrics}           &             &            &             &             & \textit{}                  \\
Engagement score of conditioning videos & -0.1296     & 0.8784     & 0.7506      & 1.028       & \textit{p = 0.1062}        \\
Number of conditioning videos           & -0.0747     & 0.928      & 0.8383      & 1.0273      & \textit{p = 0.1497}        \\\hline
\textbf{Recommendation video metrics}   &             &            &             &             & \textit{}                  \\
Share count                             & -0.0727     & 0.9299     & 0.8309      & 1.0406      & \textit{p = 0.2054}        \\
Likes per recommendation                & -0.2692     & 0.764      & 0.6781      & 0.8608      & \textit{p \textless 0.001} \\
Comments per recommendation             & 0.0591      & 1.0609     & 0.9001      & 1.2505      & \textit{p = 0.481}         \\
Opposing comment proportion             & 0.704       & 2.0219     & 1.8348      & 2.2281      & \textit{p \textless 0.001} \\
Video length                            & -0.1549     & 0.8565     & 0.6763      & 1.0848      & \textit{p = 0.1988}        \\\hline
\textbf{Recommendation channel metrics} &             &            &             &             & \textit{}                  \\
Verified channel                        & -0.5568     & 0.5731     & 0.451       & 0.7282      & \textit{p \textless 0.001} \\
Video count                             & 0.1387      & 1.1487     & 1.0366      & 1.273       & \textit{p = 0.0081}        \\
Follower count                          & 0.1028      & 1.1082     & 0.9794      & 1.254       & \textit{p = 0.103}        \\ \hline
\end{tabular}
\caption{Logistic regression coefficients and odds-ratios predicting whether a given video watch is mismatched/cross-party aligned.}
\label{tab:sup_table_23}
\end{table}

\begin{table}[htbp!]
\centering
\footnotesize
\begin{tabular}{lccccc}
Variable                                & Coefficient & Odds Ratio & OR Lower CI & OR Upper CI & P-value                    \\ \hline
\textbf{Bot conditioning (ref = Rep.)}  &             &            &             &             &                            \\
Democrat                                & 1.4891      & 4.4329     & 3.0969      & 6.3453      & \textit{p \textless 0.001} \\\hline
\textbf{State (ref = NY)}               &             &            &             &             & \textit{}                  \\
Georgia                                 & -0.0211     & 0.9791     & 0.7797      & 1.2296      & \textit{p = 0.8561}        \\
Texas                                   & 0.0473      & 1.0484     & 0.8325      & 1.3204      & \textit{p = 0.6878}        \\\hline
\textbf{Time}                           &             &            &             &             & \textit{}                  \\
Week of experiment                      & -0.04       & 0.9608     & 0.9447      & 0.9771      & \textit{p \textless 0.001} \\\hline
\textbf{Top account x Bot conditioning} &             &            &             &             & \textit{}                  \\
Top account × Democrat bot              & -1.0949     & 0.3346     & 0.2494      & 0.4489      & \textit{p \textless 0.001} \\
Top account × Republican bot            & 0.1143      & 1.1211     & 0.7686      & 1.6353      & \textit{p = 0.5527}        \\\hline
\textbf{Conditioning metrics}           &             &            &             &             & \textit{}                  \\
Engagement score of conditioning videos & -0.115      & 0.8914     & 0.7609      & 1.0442      & \textit{p = 0.1545}        \\
Number of conditioning videos           & -0.072      & 0.9306     & 0.8391      & 1.032       & \textit{p = 0.1728}        \\\hline
\textbf{Recommendation video metrics}   &             &            &             &             & \textit{}                  \\
Share count                             & -0.052      & 0.9494     & 0.861       & 1.0468      & \textit{p = 0.297}         \\
Likes per recommendation                & -0.2392     & 0.7873     & 0.6967      & 0.8896      & \textit{p \textless 0.001} \\
Comments per recommendation             & 0.0488      & 1.05       & 0.8887      & 1.2406      & \textit{p = 0.5663}        \\
Opposing comment proportion             & 0.7433      & 2.1029     & 1.9043      & 2.3222      & \textit{p \textless 0.001} \\
Video length                            & -0.2003     & 0.8185     & 0.6097      & 1.0988      & \textit{p = 0.1826}        \\\hline
\textbf{Recommendation channel metrics} &             &            &             &             & \textit{}                  \\
Verified channel                        & -0.2995     & 0.7412     & 0.5673      & 0.9684      & \textit{p = 0.0281}        \\
Video count                             & 0.0768      & 1.0798     & 0.9595      & 1.2152      & \textit{p = 0.2027}        \\
Follower count                          & 0.1158      & 1.1228     & 0.9892      & 1.2745      & \textit{p = 0.0732}       \\ \hline 
\end{tabular}
\caption{Logistic regression coefficients and odds-ratios predicting whether a given video watch is mismatched/cross-party aligned when only considering videos published by top Republican and Democrat channels.}
\label{tab:sup_table_24}
\end{table}

\begin{table}[htbp!]
\centering
\footnotesize
\begin{tabular}{lccccc}
Variable                                 & Coefficient & Odds Ratio & OR Lower CI & OR Upper CI & P-value                    \\ \hline
\textbf{Bot conditioning (ref = Rep.)}   &             &            &             &             & \textit{}                  \\
Democrat                                 & 0.6652      & 1.9449     & 1.144       & 3.3066      & \textit{p = 0.014}         \\\hline
\textbf{State (ref = NY)}                &             &            &             &             & \textit{}                  \\
Georgia                                  & -0.011      & 0.9891     & 0.7844      & 1.2471      & \textit{p = 0.9259}        \\
Texas                                    & 0.0664      & 1.0687     & 0.8464      & 1.3494      & \textit{p = 0.5767}        \\\hline
\textbf{Partisanship x Bot conditioning} &             &            &             &             & \textit{}                  \\
Negative partisanship × Republican bot   & 0.2881      & 1.3339     & 0.9091      & 1.9573      & \textit{p = 0.1408}        \\
Negative partisanship × Democrat bot     & 1.3051      & 3.6881     & 2.6652      & 5.1037      & \textit{p \textless 0.001} \\\hline
\textbf{Top account x Bot conditioning}  &             &            &             &             & \textit{}                  \\
Top account × Republican bot             & 0.0433      & 1.0442     & 0.7113      & 1.5331      & \textit{p = 0.8251}        \\
Top account × Democrat bot               & -1.0339     & 0.3556     & 0.2651      & 0.4771      & \textit{p \textless 0.001} \\\hline
\textbf{Time}                            &             &            &             &             & \textit{}                  \\
Week of experiment                       & -0.0336     & 0.967      & 0.9506      & 0.9836      & \textit{p \textless 0.001} \\\hline
\textbf{Conditioning metrics}            &             &            &             &             & \textit{}                  \\
Engagement score of conditioning videos  & -0.1176     & 0.889      & 0.7566      & 1.0446      & \textit{p = 0.1529}        \\
Number of conditioning videos            & -0.0721     & 0.9304     & 0.8384      & 1.0325      & \textit{p = 0.1746}        \\\hline
\textbf{Recommendation video metrics}    &             &            &             &             & \textit{}                  \\
Likes per recommendation                 & -0.2262     & 0.7975     & 0.7043      & 0.903       & \textit{p \textless 0.001} \\
Video length                             & -0.2116     & 0.8093     & 0.5988      & 1.0938      & \textit{p = 0.1686}        \\
Share count                              & -0.1223     & 0.8849     & 0.7914      & 0.9894      & \textit{p = 0.0317}        \\
Comments per recommendation              & 0.1199      & 1.1274     & 0.9562      & 1.3293      & \textit{p = 0.1537}        \\
Opposing comment proportion              & 0.8272      & 2.287      & 2.0563      & 2.5436      & \textit{p \textless 0.001} \\\hline
\textbf{Recommendation channel metrics}  &             &            &             &             & \textit{}                  \\
Verified channel                         & -0.2846     & 0.7523     & 0.575       & 0.9843      & \textit{p = 0.038}         \\
Video count                              & 0.0449      & 1.046      & 0.9261      & 1.1814      & \textit{p = 0.4693}        \\
Follower count                           & 0.1711      & 1.1866     & 1.0401      & 1.3538      & \textit{p = 0.0109}       \\ \hline
\end{tabular}
\caption{Logistic regression coefficients and odds-ratios predicting whether a given video watch is mismatched/cross-party aligned when accounting for the partisanship of the video.}
\label{tab:sup_table_25}
\end{table}

\begin{table}[htbp!]
\centering
\footnotesize
\begin{tabular}{llccc}
                                                   &               & \textbf{Democrat bots} & \textbf{Neutral bots} & \textbf{Republican bots} \\\hline
                                                   & Unique videos & 1847              & 8177             & 2640                \\\hline
\multirow{7}{*}{Video play count}                  & mean          & 2413648.8843      & 5637526.6721     & 2707925.1944        \\
                                                   & std           & 7506332.0558      & 19249220.6577    & 5683604.0638        \\
                                                   & min           & 239.0             & 48.0             & 516.0               \\
                                                   & 25\%          & 67200.0           & 190250.0         & 103725.0            \\
                                                   & 50\%          & 409400.0          & 970600.0         & 497650.0            \\
                                                   & 75\%          & 1500000.0         & 3700000.0        & 2400000.0           \\
                                                   & max           & 56200000.0        & 177600000.0      & 56600000.0          \\\hline
\multirow{7}{*}{Video share count}                 & mean          & 18692.9357        & 45206.1413       & 21480.1101          \\
                                                   & std           & 64829.2859        & 160370.5541      & 67609.5854          \\
                                                   & min           & 0.0               & 0.0              & 1.0                 \\
                                                   & 25\%          & 256.25            & 572.0            & 363.0               \\
                                                   & 50\%          & 1700.5            & 4151.0           & 2903.0              \\
                                                   & 75\%          & 10475.0           & 21400.0          & 15000.0             \\
                                                   & max           & 586900.0          & 2200000.0        & 1800000.0           \\\hline
\multirow{7}{*}{Video like count}                  & mean          & 250751.5964       & 643048.7369      & 253244.4459         \\
                                                   & std           & 735713.9236       & 2327330.5311     & 560122.1661         \\
                                                   & min           & 21.0              & 2.0              & 2.0                 \\
                                                   & 25\%          & 6954.0            & 14900.0          & 11300.0             \\
                                                   & 50\%          & 39900.0           & 87100.0          & 50950.0             \\
                                                   & 75\%          & 144000.0          & 376050.0         & 214150.0            \\
                                                   & max           & 4700000.0         & 24200000.0       & 8700000.0           \\\hline
\multirow{7}{*}{Video likes per recommendation}    & mean          & 0.1183            & 0.1206           & 0.1172              \\
                                                   & std           & 0.0596            & 0.0821           & 0.0622              \\
                                                   & min           & 0.0125            & 0.0012           & 0.0029              \\
                                                   & 25\%          & 0.0737            & 0.0563           & 0.0695              \\
                                                   & 50\%          & 0.1083            & 0.1045           & 0.1052              \\
                                                   & 75\%          & 0.1567            & 0.169            & 0.1536              \\
                                                   & max           & 0.4994            & 0.7676           & 0.3916              \\\hline
\multirow{7}{*}{Video comment count}               & mean          & 7728.3862         & 7935.9829        & 7235.3002           \\
                                                   & std           & 18115.1643        & 23225.5984       & 13568.1027          \\
                                                   & min           & 0.0               & 0.0              & 0.0                 \\
                                                   & 25\%          & 414.75            & 403.0            & 523.0               \\
                                                   & 50\%          & 2108.5            & 1730.0           & 2384.0              \\
                                                   & 75\%          & 6759.5            & 5733.5           & 8080.5              \\
                                                   & max           & 131200.0          & 773000.0         & 184300.0            \\\hline
\multirow{7}{*}{Video comments per recommendation} & mean          & 0.0078            & 0.0063           & 0.0071              \\
                                                   & std           & 0.0099            & 0.0153           & 0.0088              \\
                                                   & min           & 0.0               & 0.0              & 0.0                 \\
                                                   & 25\%          & 0.0025            & 0.0007           & 0.0019              \\
                                                   & 50\%          & 0.005             & 0.0018           & 0.0043              \\
                                                   & 75\%          & 0.0097            & 0.0049           & 0.0088              \\
                                                   & max           & 0.164             & 0.2166           & 0.1164   \\ \hline          
\end{tabular}
\caption{Engagement metrics of videos watched during recommendation stage.}
\label{tab:sup_table_26}
\end{table}

\begin{table}[htbp!]
\centering
\footnotesize
\begin{tabular}{llccc}
                                               &                 & \textbf{Democrat bots} & \textbf{Neutral bots} & \textbf{Republican bots} \\\hline
                                               & Unique channels & 640               & 5644             & 1197                \\\hline
\multirow{7}{*}{Channel follower count}        & mean            & 3036086.1187      & 3075284.5458     & 3028526.2507        \\
                                               & std             & 3264755.0004      & 5075331.3993     & 4228698.84          \\
                                               & min             & 2102.0            & 32.0             & 746.0               \\
                                               & 25\%            & 665550.0          & 212400.0         & 618200.0            \\
                                               & 50\%            & 1500000.0         & 1000000.0        & 1300000.0           \\
                                               & 75\%            & 4100000.0         & 4100000.0        & 3300000.0           \\
                                               & max             & 16700000.0        & 75300000.0       & 42200000.0          \\\hline
\multirow{7}{*}{Channel cumulative like count} & mean            & 146046830.1019    & 150456706.5605   & 125131927.1811      \\
                                               & std             & 222423644.7813    & 322884691.3661   & 306548628.1349      \\
                                               & min             & 18900.0           & 0.0              & 0.0                 \\
                                               & 25\%            & 11800000.0        & 6100000.0        & 13800000.0          \\
                                               & 50\%            & 63700000.0        & 38750000.0       & 47200000.0          \\
                                               & 75\%            & 265900000.0       & 153400000.0      & 107200000.0         \\
                                               & max             & 1700000000.0      & 1900000000.0     & 1700000000.0        \\\hline
\multirow{7}{*}{Channel video count}           & mean            & 2121.7534         & 1977.6035        & 1852.6265           \\
                                               & std             & 2319.6505         & 2939.3987        & 2635.1561           \\
                                               & min             & 23.0              & 3.0              & 16.0                \\
                                               & 25\%            & 546.5             & 286.0            & 447.0               \\
                                               & 50\%            & 1197.0            & 820.0            & 1025.0              \\
                                               & 75\%            & 3413.0            & 2605.0           & 2605.0              \\
                                               & max             & 24700.0           & 30700.0          & 16300.0             \\ \hline
Channel verification status                    & count           & 70                & 222              & 75      \\ \hline           
\end{tabular}
\caption{Engagement metrics of channels watched during recommendation stage.}
\label{tab:sup_table_27}
\end{table}

\begin{table}[htbp!]
\centering
\footnotesize
\begin{tabular}{llcc}
                                                   &               & \textbf{Democrat bots} & \textbf{Republican bots} \\\hline
                                                   & Unique videos & 1534              & 2561                \\\hline
\multirow{7}{*}{Video play count}                  & mean          & 647921.0081       & 563931.6978         \\
                                                   & std           & 2103402.7393      & 5879274.2821        \\
                                                   & min           & 619.0             & 435.0               \\
                                                   & 25\%          & 12600.0           & 8043.0              \\
                                                   & 50\%          & 58300.0           & 26800.0             \\
                                                   & 75\%          & 366400.0          & 126400.0            \\
                                                   & max           & 174500000.0       & 1100000000.0        \\\hline
\multirow{7}{*}{Video share count}                 & mean          & 4753.1959         & 4515.1846           \\
                                                   & std           & 20217.5614        & 44045.2628          \\
                                                   & min           & 1.0               & 0.0                 \\
                                                   & 25\%          & 142.0             & 70.0                \\
                                                   & 50\%          & 350.0             & 175.0               \\
                                                   & 75\%          & 2180.0            & 698.0               \\
                                                   & max           & 1400000.0         & 4600000.0           \\\hline
\multirow{7}{*}{Video like count}                  & mean          & 93469.9648        & 56878.0389          \\
                                                   & std           & 295319.8032       & 405934.5273         \\
                                                   & min           & 4.0               & 9.0                 \\
                                                   & 25\%          & 1458.0            & 640.0               \\
                                                   & 50\%          & 6865.0            & 2810.0              \\
                                                   & 75\%          & 44400.0           & 13000.0             \\
                                                   & max           & 21800000.0        & 33400000.0          \\\hline
\multirow{7}{*}{Video likes per recommendation}    & mean          & 0.1364            & 0.1059              \\
                                                   & std           & 0.0595            & 0.052               \\
                                                   & min           & 0.0036            & 0.0003              \\
                                                   & 25\%          & 0.0914            & 0.0715              \\
                                                   & 50\%          & 0.133             & 0.0971              \\
                                                   & 75\%          & 0.1791            & 0.1337              \\
                                                   & max           & 0.365             & 0.4419              \\\hline
\multirow{7}{*}{Video comment count}               & mean          & 2580.4743         & 1261.7448           \\
                                                   & std           & 6773.3462         & 4909.7431           \\
                                                   & min           & 0.0               & 0.0                 \\
                                                   & 25\%          & 38.0              & 33.0                \\
                                                   & 50\%          & 244.0             & 106.0               \\
                                                   & 75\%          & 1582.0            & 469.0               \\
                                                   & max           & 109800.0          & 187700.0            \\\hline
\multirow{7}{*}{Video comments per recommendation} & mean          & 0.0054            & 0.0059              \\
                                                   & std           & 0.0044            & 0.0075              \\
                                                   & min           & 0.0               & 0.0                 \\
                                                   & 25\%          & 0.0021            & 0.0018              \\
                                                   & 50\%          & 0.0044            & 0.004               \\
                                                   & 75\%          & 0.0075            & 0.0075              \\
                                                   & max           & 0.0395            & 0.1184             \\ \hline
\end{tabular}
\caption{Engagement metrics of videos watched during conditioning stage.}
\label{tab:sup_table_28}
\end{table}

\begin{table}[htbp!]
\centering
\footnotesize
\begin{tabular}{llcc}
                                               &                 & \textbf{Democrat} & \textbf{Republican} \\\hline
                                               & Unique channels & 12                & 12                  \\\hline
\multirow{7}{*}{Channel follower count}        & mean            & 928811.6177       & 1114729.5803        \\
                                               & std             & 678815.8153       & 982530.0651         \\
                                               & min             & 0.0               & 0.0                 \\
                                               & 25\%            & 384100.0          & 385100.0            \\
                                               & 50\%            & 721900.0          & 720000.0            \\
                                               & 75\%            & 1400000.0         & 2200000.0           \\
                                               & max             & 27700000.0        & 49200000.0          \\\hline
\multirow{7}{*}{Channel cumulative like count} & mean            & 17632176.9303     & 25118611.7036       \\
                                               & std             & 20284093.3183     & 26866917.3477       \\
                                               & min             & 2288.0            & 0.0                 \\
                                               & 25\%            & 7000000.0         & 8600000.0           \\
                                               & 50\%            & 10900000.0        & 14900000.0          \\
                                               & 75\%            & 26500000.0        & 47200000.0          \\
                                               & max             & 1500000000.0      & 1900000000.0        \\\hline
\multirow{7}{*}{Channel video count}           & mean            & 421.0464          & 1218.6955           \\
                                               & std             & 377.7008          & 963.9262            \\
                                               & min             & 0.0               & 0.0                 \\
                                               & 25\%            & 121.0             & 345.0               \\
                                               & 50\%            & 331.0             & 955.0               \\
                                               & 75\%            & 604.0             & 2605.0              \\
                                               & max             & 6377.0            & 30700.0             \\ \hline
Channel verification status                    & count           & 11                & 2          \\ \hline
\end{tabular}
\caption{Engagement metrics of channels watched during conditioning stage.}
\label{tab:sup_table_29}
\end{table}

\begin{table}[htbp!]
\tiny
\centering
\begin{tabular}{lcccc}
                                & \multicolumn{4}{c}{Free response questions}                                                                                                                                                                                                                                                                                                                                                                                                                          \\ \hline
Variable                        & \begin{tabular}[c]{@{}c@{}}Mentioned changes\\ to political content\\ when asked about\\ content overall\end{tabular} & \begin{tabular}[c]{@{}c@{}}Mentioned shifts\\ towards co-partisan\\ content when asked\\ about content overall\end{tabular} & \begin{tabular}[c]{@{}c@{}}Mentioned shifts\\ towards co-partisan\\ content when asked\\ about political content\end{tabular} & \begin{tabular}[c]{@{}c@{}}Mentioned shifts to\\ positive content\end{tabular} \\ \hline
Intercept                       & -1.286**                                                                                                              & 0.011                                                                                                                       & -0.094                                                                                                                        & -0.202                                                                         \\ \hline
Republican Party (ref. Democratic Party)                & -0.353*                                                                                                               & 0.024                                                                                                                       & 0.08**                                                                                                                        & 0.274***                                                                       \\ \hline
Age (ref. 65+ years old)\\
18-24 years old                 & 0.099                                                                                                                 & -0.04                                                                                                                       & 0.071                                                                                                                         & -0.035                                                                         \\
25-34 years old                 & 0.486                                                                                                                 & -0.027                                                                                                                      & 0.036                                                                                                                         & 0.051                                                                          \\
35-44 years old                 & 0.464                                                                                                                 & -0.021                                                                                                                      & 0.024                                                                                                                         & 0.044                                                                          \\
45-54 years old                 & 0.718                                                                                                                 & -0.003                                                                                                                      & -0.001                                                                                                                        & 0.011                                                                          \\
55-64 years old                 & 0.364                                                                                                                 & -0.015                                                                                                                      & 0.062                                                                                                                         & -0.072                                                                         \\ \hline
Race (ref. White) \\
Asian                           & 1.013**                                                                                                               & 0.1                                                                                                                         & 0.152                                                                                                                         & -0.266*                                                                        \\
Black                           & -0.074                                                                                                                & 0.002                                                                                                                       & 0.025                                                                                                                         & 0.001                                                                          \\
Hispanic                        & 0.243                                                                                                                 & -0.03                                                                                                                       & 0.06                                                                                                                          & -0.377**                                                                       \\
Native American                 & -0.558                                                                                                                & -0.005                                                                                                                      & -0.038                                                                                                                        & 0.144                                                                          \\
Two or more races               & 0.891                                                                                                                 & -0.0                                                                                                                        & -0.089                                                                                                                        & -0.456*     \\   
Other                           & -0.123                                                                                                                & -0.003                                                                                                                      & -0.009                                                                                                                        & 0.242                                                                          \\
Prefer not to answer            & -237.16***                                                                                                            & 0.009                                                                                                                       & 0.016                                                                                                                         & 0.15*                                                                          \\ \hline
Female (ref. Male)                          & 0.036                                                                                                                 & 0.007                                                                                                                       & 0.042                                                                                                                         & 0.0                                                                            \\ \hline
Education (ref. Bachelor's degree) \\
Associates or technical degree  & 0.381                                                                                                                 & 0.009                                                                                                                       & 0.07                                                                                                                          & -0.247*                                                                        \\
Graduate or professional degree & -0.345*                                                                                                               & -0.006                                                                                                                      & 0.026                                                                                                                         & 0.078                                                                          \\
High school diploma or GED      & -0.38                                                                                                                 & -0.027                                                                                                                      & -0.071                                                                                                                        & -0.224*                                                                        \\

Some college, but no degree     & 0.133                                                                                                                 & -0.006                                                                                                                      & -0.017                                                                                                                        & -0.119                                                                         \\
Some high school or less        & -0.75                                                                                                                 & -0.013                                                                                                                      & 0.175                                                                                                                         & 0.367                                                                          \\
Prefer not to say               & -240.424***                                                                                                           & -0.021                                                                                                                      & -0.052                                                                                                                        & 0.883*                                                                         \\ \hline
                                                               
\end{tabular}
\caption{Regression coefficients for various models estimating responses to free-response (text-entry) questions based on participant demographics; (Q1 - Q3) of Survey. * : $p < 0.05$; **: $p < 0.01$; ***: $p < 0.001$}
\end{table}

\begin{table}[htbp!]
\tiny
\centering
\begin{tabular}{lcccccc}
                               & \multicolumn{6}{c}{Structured questions}                                                                                                                                                                                                                                                                                                                                                                                                                          \\ \hline
Variable                        & \begin{tabular}[c]{@{}c@{}}More co-partisan/\\ cross-partisan\\ content\end{tabular} & \begin{tabular}[c]{@{}c@{}}More Pro-Trump\\ More Anti-Trump\\ content\end{tabular} & \begin{tabular}[c]{@{}c@{}}More political/\\ Less political\\ content\end{tabular} & \begin{tabular}[c]{@{}c@{}}More positive/\\ More negative\\ content\end{tabular} & \begin{tabular}[c]{@{}c@{}}More optimistic/\\ More pessimistic\\ content\end{tabular} & \begin{tabular}[c]{@{}c@{}}More content \\ you agree\\ /disagree with\end{tabular} \\ \hline
Intercept                       & 0.394                                                                                & 4.755***                                                                           & 7.25***                                                                            & 4.68***                                                                          & 4.575***                                                                              & 6.267***                                                                           \\ \hline
Republican Party (ref. Democratic Party)              & 1.445***                                                                             & 1.071***                                                                           & -0.047                                                                             & 1.029***                                                                         & 0.994***                                                                              & 0.358*                                                                             \\ \hline
Age (ref. 65+ years old) \\
18-24 years old                 & 0.38                                                                                 & 0.174                                                                              & -0.13                                                                              & -0.38                                                                            & 0.114                                                                                 & -0.018                                                                             \\
25-34 years old                 & -0.236                                                                               & -0.172                                                                             & -0.388                                                                             & -0.527                                                                           & -0.091                                                                                & -0.376                                                                             \\
35-44 years old                 & -0.416                                                                               & 0.073                                                                              & -0.29                                                                              & -0.628                                                                           & -0.026                                                                                & -0.386                                                                             \\
45-54 years old                 & -0.677                                                                               & -0.166                                                                             & -0.491                                                                             & -0.698                                                                           & -0.219                                                                                & -0.895*                                                                            \\
55-64 years old                 & -0.228                                                                               & -0.639                                                                             & -0.077                                                                             & -0.531                                                                           & -0.019                                                                                & -0.368                                                                             \\ \hline
Race (ref. White) \\
Asian                           & 0.356                                                                                & -1.156*                                                                            & -1.112*                                                                            & -1.012*                                                                          & -1.726***                                                                             & -0.497                                                                             \\
Black                           & 0.139                                                                                & -0.329                                                                             & 0.164                                                                              & 0.062                                                                            & 0.263                                                                                 & 0.02                                                                               \\
Hispanic                        & -0.751                                                                               & 0.418                                                                              & 0.418                                                                              & -0.516                                                                           & 0.018                                                                                 & -0.251                                                                             \\

Native American                 & 0.715                                                                                & -0.702                                                                             & -0.204                                                                             & 0.619                                                                            & 1.253                                                                                 & 0.043                                                                              \\
Two or more races               & -1.812*                                                                              & 0.317                                                                              & 0.476                                                                              & -1.49*                                                                           & -1.123                                                                                & -2.146**  \\ 
Other                           & 0.583                                                                                & 1.428                                                                              & -1.188                                                                             & 1.182                                                                            & 1.811                                                                                 & 0.657                                                                              \\
Prefer not to answer            & -0.152                                                                               & -1.416                                                                             & 2.507*                                                                             & -2.141                                                                           & -2.587                                                                                & -4.043                                                                             \\
Female (ref. Male)                         & -0.006                                                                               & -0.167                                                                             & -0.369*                                                                            & -0.012                                                                           & 0.103                                                                                 & 0.152       \\ \hline
Education (ref. Bachelor's degree) \\
Associates or technical degree  & 0.337                                                                                & -0.555                                                                             & -0.787                                                                             & -0.325                                                                           & -0.562                                                                                & -0.274                                                                             \\      
Graduate or professional degree & 0.636***                                                                             & 0.172                                                                              & 0.341*                                                                             & 0.424*                                                                           & 0.541**                                                                               & 0.546**                                                                            \\
High school diploma or GED      & -0.332                                                                               & -0.765*                                                                            & -0.134                                                                             & -0.73*                                                                           & -0.367                                                                                & -0.246                                                                             \\
Some college, but no degree     & 0.362                                                                                & -0.364                                                                             & -0.663                                                                             & -0.075                                                                           & -0.414                                                                                & -0.047                                                                             \\
Some high school or less        & 0.66                                                                                 & -0.969                                                                             & -3.007*                                                                            & -0.079                                                                           & -0.567                                                                                & -0.725                                                                             \\
Prefer not to say               & 2.583***                                                                             & 2.269***                                                                           & -5.543*                                                                            & 3.932***                                                                         & -4.645                                                                                & 2.609*                                                                             \\ \hline
  
\end{tabular}
\caption{Regression coefficients for various models estimating responses to structured (scale-based) questions based on participant demographics; (Q4 - Q9) of Survey. * : $p < 0.05$; **: $p < 0.01$; ***: $p < 0.001$}
\end{table}

\begin{table}[htbp!]
\tiny
\centering
\begin{tabular}{lc} \hline
Demographic                                 & Count \\ \hline
25-34 years old                             & 373   \\
35-44 years old                             & 232   \\
45-54 years old                             & 173   \\
18-24 years old                             & 109   \\
55-64 years old                             & 86    \\
65+ years old                               & 35    \\ \hline
Female                                      & 547   \\
Male                                        & 449   \\
Non-binary / third gender                   & 12    \\ \hline
White                                       & 646   \\
Black                                       & 276   \\
Asian                                       & 28    \\
Hispanic                                    & 25    \\
Two or more races                           & 21    \\
Native American                             & 7     \\
Other                                       & 5     \\ \hline
Bachelor’s degree                           & 417   \\
Graduate or professional degree             & 412   \\
Some college, but no degree                 & 74    \\
High school diploma or GED                  & 55    \\
Associates or technical degree              & 46    \\
Some high school or less                    & 3     \\
Prefer not to say                           & 1     \\ \hline
Party                                       &       \\
Republican Party                            & 552   \\
Democratic Party                            & 317   \\
Independent                                 & 139   \\ \hline
Democrat Strong                             &       \\
Strong Democrat                             & 210   \\
Not so strong Democrat                      & 106   \\
Prefer not to answer                        & 1     \\ \hline
Republican Strong                           &       \\
Strong Republican                           & 346   \\
Not so strong Republican                    & 202   \\
Prefer not to answer                        & 4     \\ \hline
Independent close to Democrat or Republican &       \\
Neither                                     & 66    \\
Democratic Party                            & 35    \\
Republican Party                            & 34    \\
Not sure                                    & 3     \\
Prefer not to answer                        & 1     \\ \hline
Liberal/Conservative                        &       \\
Conservative                                & 341   \\
Liberal                                     & 215   \\
Moderate                                    & 178   \\
Very conservative                           & 160   \\
Very liberal                                & 108   \\
Not sure                                    & 5     \\
Prefer not to answer                        & 1     \\ \hline
Voted in 2024 elections                     &       \\
Yes                                         & 924   \\
No                                          & 64    \\
Prefer not to answer                        & 20    \\ \hline
Party in 2024 elections                     &       \\
Republican                                  & 590   \\
Democrat                                    & 307   \\
Independent                                 & 15    \\
Other                                       & 6     \\
Libertarian                                 & 4     \\
Green                                       & 2     \\ \hline
Plans to vote in 2026 elections             &       \\
Yes                                         & 798   \\
Not sure                                    & 152   \\
No                                          & 46    \\
Prefer not to answer                        & 12    \\ \hline
Party in 2026 elections                     &       \\
Republican                                  & 449   \\
Democrat                                    & 290   \\
Independent                                 & 39    \\
Other                                       & 15    \\
Libertarian                                 & 4     \\
Green                                       & 1    \\ \hline
\end{tabular}
\caption{Participant demographic counts}
\end{table}

\clearpage
\section*{Supplementary Figures}
\label{figures}

\begin{figure}[htbp!]
    \centering
    \includegraphics[width=\linewidth]{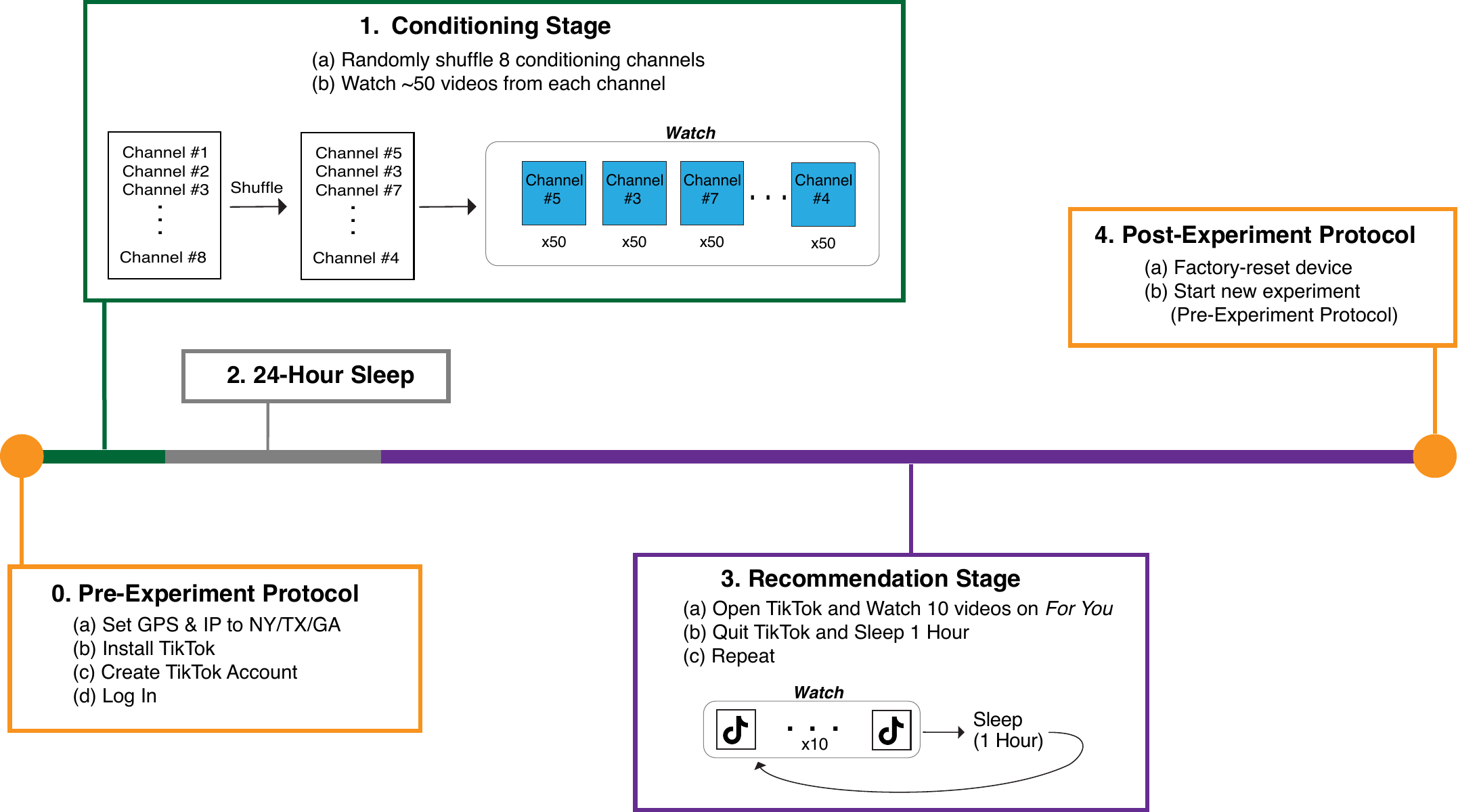}
    \caption{A device's timeline during a weekly experimental run}
    \label{fig:bot_timeline}
\end{figure}

\begin{figure}[htbp!]
    \centering
    \includegraphics[width=\linewidth]{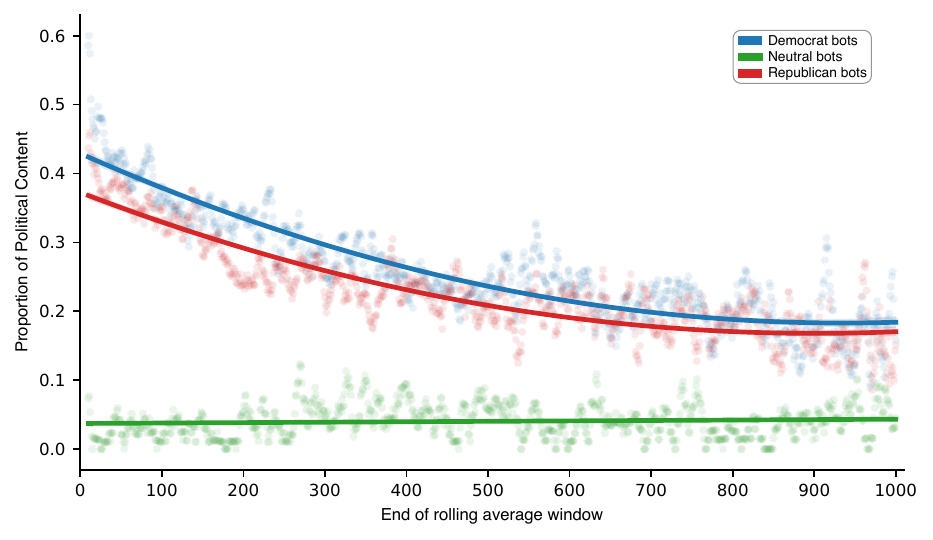}
    \caption{Rolling average of political content viewed by bots of different conditioning in 10 video windows.}
    \label{fig:supp_fig_3}
\end{figure}

\begin{figure}[htbp!]
    \centering
    \includegraphics[width=\linewidth]{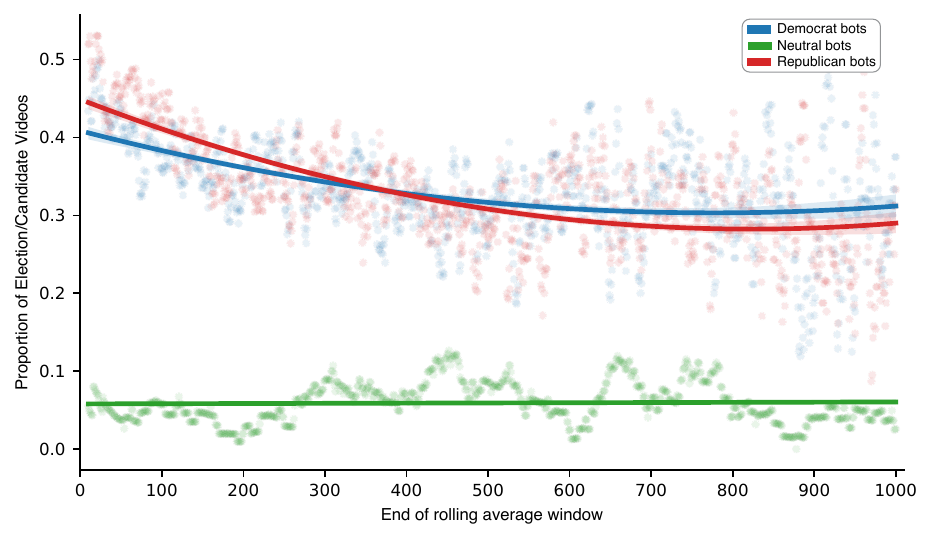}
    \caption{Rolling average of political content about the US elections or about major political candidates (out of all political content) viewed by bots of different conditioning in 10 video windows.}
    \label{fig:supp_fig_4}
\end{figure}

\begin{figure}[htbp!]
    \centering
    \includegraphics[width=\linewidth]{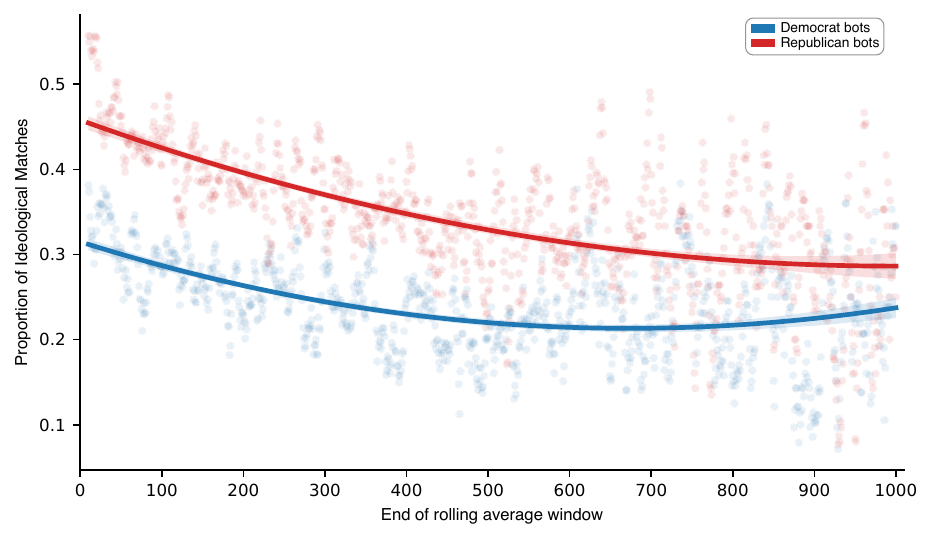}
    \caption{Rolling average of political content that is ideologically matched (out of all political content) to bots of different conditioning in 10 video windows. Pro Democrat or Anti Republican videos are ideologically matched with Democrat bots, while Pro Republican or Anti Democrat videos are ideologically matched with Republican bots.}
    \label{fig:supp_fig_5}
\end{figure}

\begin{figure}[htbp!]
    \centering
    \includegraphics[width=\linewidth]{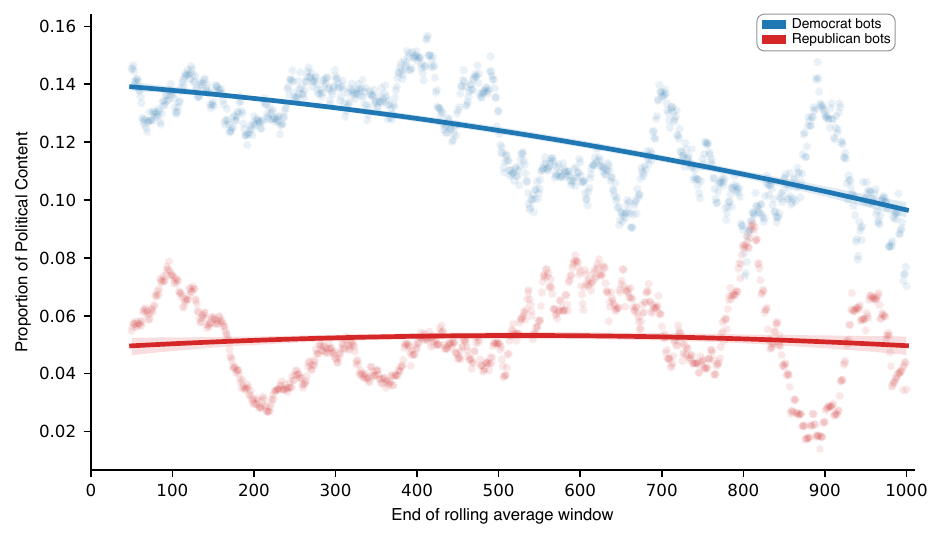}
    \caption{Rolling average of political content that is ideologically mismatched (out of all political content) to bots of different conditioning in 10 video windows. Pro Republican or Anti Democrat videos are ideologically mismatched with Democrat bots, while Pro Democrat or Anti Republican videos are ideologically mismatched with Republican bots.}
    \label{fig:supp_fig_6}
\end{figure}

\begin{figure}[htbp!]
    \centering
    \includegraphics[width=0.9\linewidth]{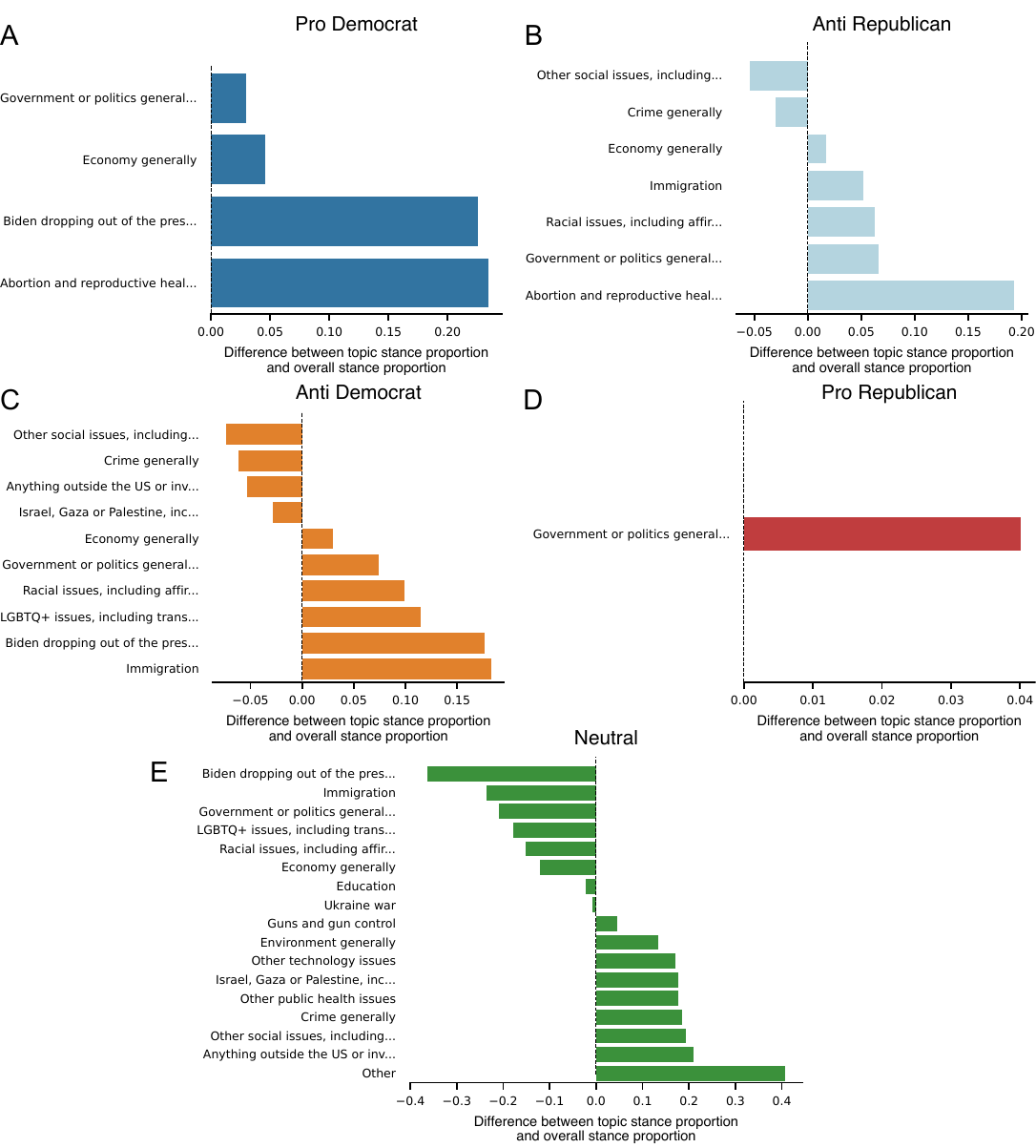}
    \caption{Of topic-stance pairs with at least 100 videos, the difference in the proportion of videos on the topic with the given stance and the proportion of videos of that stance out of all political videos. Positive values indicate a higher than average representation of the stance within a topic, while negative values lower than average representation of the stance within the topic. Plots \textbf{A}, \textbf{B}, \textbf{C}, \textbf{D}, \textbf{E} correspond to the Pro Democrat, Anti Republican, Anti Democrat, Pro Republican, and Neutral stances, respectively.}
    \label{fig:supp_fig_7}
\end{figure}

\begin{figure}[htbp!]
    \centering
    \includegraphics[width=\linewidth]{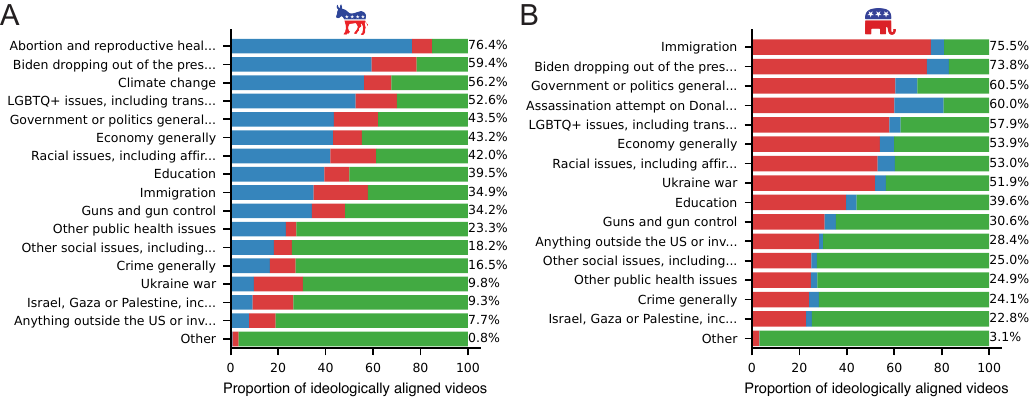}
    \caption{(\textbf{A}, \textbf{B}) The proportion of videos on a given topic which are ideologically-aligned, ideologically-opposing, or neutral, seen by Democrat- and Republican-conditioned bots, respectively. For each plot, topics are listed in descending order of ideological-alignment.}
    \label{fig:supp_fig_8}
\end{figure}

\end{document}